\documentclass[twocolumn,showpacs,preprintnumbers,amsmath,amssymb]{revtex4}
\usepackage{dcolumn}
\usepackage{bm}
\usepackage{graphicx}
\usepackage{amsmath}
\begin{document}

\title{Controlling stable tunneling in a non-Hermitian spin-orbit coupled bosonic junction}
\author{Yunrong Luo$^{1}$\footnote{Corresponding author: lyr\underline{ }1982@hunnu.edu.cn}, \ Xuemei Wang$^{1}$, \ Yuxin Luo$^{1}$, \ Zheng Zhou$^{2}$, Zhao-Yun Zeng$^{3,4}$, and \ Xiaobing Luo$^{3,4}$\footnote{Corresponding author: xiaobingluo2013@aliyun.com}}
\affiliation{$^{1}$Department of Physics and Key Laboratory of
Low-dimensional Quantum Structures and Quantum Control of Ministry
of Education, and Synergetic Innovation Center for Quantum Effects and Applications, Hunan Normal University, Changsha 410081, China\\
$^{2}$Department of Physics, Hunan Institute of Technology, Hengyang 421002, China\\
$^{3}$Department of Physics, Zhejiang Sci-Tech University, Hangzhou, 310018, China\\
$^{4}$School of Mathematics and Physics, Jinggangshan University, Ji'an 343009, China}

\begin{abstract}

In this paper, we study how to apply a periodic driving field to control stable spin tunneling in a non-Hermitian spin-orbit coupled bosonic double-well system. By means of a high-frequency approximation, we obtain the analytical Floquet solutions and their associated quasienergies and thus construct the general non-Floquet solutions of the dissipative spin-orbit coupled bosonic system. Based on detailed analysis of the Floquet quasienergy spectrum, the profound effect of system parameters and the periodic driving field on the stability of spin-dependent tunneling is investigated analytically and numerically for both balanced and unbalanced gain-loss between two wells. Under balanced gain and loss, we find that the stable spin-flipping tunneling is preferentially suppressed with the increase of gain-loss strength. When the ratio of Zeeman field strength to periodic driving frequency $\Omega/\omega$ is even, there is  a possibility that \emph{continuous} stable parameter regions will exist.  When $\Omega/\omega$ is odd, nevertheless, only \emph{discrete} stable parameter regions are found. Under unbalanced gain and loss,
whether $\Omega/\omega$ is even or odd, we can get parametric equilibrium conditions for the existence of stable spin tunneling. The results could be useful for the experiments of controlling stable spin transportation in a non-Hermitian spin-orbit coupled system.

\end{abstract}


\maketitle

\section{Introduction}

Over the last two decades, non-Hermitian systems have attracted increasing interest from both fundamental and application viewpoints\cite{moiseyev2011, hatano77, hatano58, bender70, rotter42, cao87, longhi117, li10, ramy14, ananya31, yuto}, which has  spawned a great deal of research work  in many branches of physics, ranging from atomic and molecular physics\cite{moiseyev2011, lee4} to spin and magnetic systems\cite{giorgi82, galda94}, quantum computing\cite{zhang85, li91}, and mesoscopic solid-state structures\cite{rotter78, rudner82}. Some important applications of non-Hermitian systems have been uncovered in different fields including optics, optomechanics and acoustics\cite{lin106, regen488, xu537, shi7}. As is well known, the quantum dynamics  of a time-dependent non-Hermitian system is usually unstable with the
probability amplitudes either exponentially increasing or decaying. For example, a non-Hermitian Hamiltonian system can possess complex energy eigenvalues whose negative imaginary parts describe an overall probability decrease that can be used to stimulate decay  phenomena\cite{gamow51, dattoli42, okolo374, moiseyev302}. Recently, a scheme on stabilizing non-Hermitian systems by periodic driving has been proposed\cite{gong91}, which is feasible for a continuous range of system parameters and originate from the fact that the eigenvalues of the so-called Floquet Hamiltonian may become all real.

Non-Hermitian transport has long  been studied for fundamental and technological purposes\cite{ramy14, ananya31, yuto, graefe41, graefe82, zhong84, xiao85, yurkevich82, kolesnikov84, rudner102, barontini110, hamazaki123, zeuner115, malzard115, longhi5, longhi95, wang124, valle87, yap96, yap97, zhou98, Zhou100, gong124}. As early as 1996, Hatano and Nelson introduced a paradigmatic example of non-Hermitian transport, in which the localization
transition in a
one-dimensional disordered non-Hermitian lattice  was examined\cite{hatano77}. Progresses on experimental techniques of engineering dissipation have motivated a
renaissance of interest in the non-Hermitian transport and dynamics, an example of which is the hallmark experiment on the implementation of governable open atomic Bose-Einstein condensate\cite{barontini110}. On the theoretical side, many important phenomena of non-Hermitian transport have been unraveled, which include hyperballistic transport on a one-dimensional tight-binding lattice\cite{valle87}, non-Hermitian unidirectional and bidirectional robust transport\cite{longhi5, longhi95}, non-Hermitian many-body localization\cite{hamazaki123}, and so on. In addition, due to the fact that both dissipation and topological mechanisms can be used jointly to control atomic transport, the non-Hermitian transport in topological matters has become a currently developing research topic\cite{yap96, yap97, zhou98, Zhou100, gong124}. In recent years, much effort has been devoted to the specially non-Hermitian
parity-time ($\mathcal{PT}$) symmetric system \cite{bender70}, where
the occurrence of real eigenvalue spectra  in certain parameter regimes enables the stabilization of atomic or wave packet transport.
All studies mentioned above do not involve the problem of  spin-orbit (SO) coupling and spin transport. To date, the connection of the non-Hermitian transport to  spin-orbit coupled atomic system have not yet been discussed in a concrete manner.

In condensed matter physics, SO-coupling arises from the interaction between the motion and spin of a particle. It has led to a number of interesting phenomena like the spin-Hall effect\cite{kato306}, the persistent spin helix\cite{koralek458}, and topological insulators\cite{bernevig314}. Recently, artificial SO-coupling of both bosonic and fermionic ultracold atoms has been realized in experiments\cite{lin471, wang109, cheuk109, zhang109, huang12, wu354}, which provides an ideal platform to study novel quantum phenomena of SO-coupled ultracold atomic systems with an unprecedented level of controllability over all available experimental parameters. There have been some works focusing on the intriguing dynamics of the SO-coupled cold atomic gases, including Josephson dynamics of a SO-coupled Bose-Einstein condensate in a double-well potential\cite{zhang609, garcia89, citro224}, collective dynamics\cite{zhang109, chen86, hu93}, selective coherent spin transportation in a SO-coupled bosonic junction\cite{yu90}, Klein tunneling\cite{zhang628}, nonequilibrium dynamics of SO-coupled lattice bosons\cite{ng92}, tunable Landau-Zener transitions in SO-coupled atomic gases\cite{olson90}, controlling spin-dependent localization and directed transport in a bipartite lattice\cite{luo93}, Bloch oscillations of SO-coupled cold atoms in an optical lattice\cite{ji99}, dynamics of SO-coupled cold atomic gases in a Floquet lattice with an impurity\cite{luo52}, and so on. Nevertheless, all these achievements are hitherto limited in studying the properties of quantum dynamics of SO-coupled cold atoms with Hermitian potential. Thus, many of these works cannot be applied to active systems where gain or loss arises naturally. Overcoming these limitations will not only enrich the conventional investigation in quantum dynamics of SO-coupled cold atoms but also offer new methods for controlling and engineering quantum spin transport. It is therefore highly desirable to investigate quantum spin dynamics of SO-coupled cold atomic systems that incorporate gain and loss mechanisms.

The aim of the present work is to study how to apply periodic driving to stabilize spin transportation in a non-Hermitian SO-coupled bosonic double-well system. Within a high frequency approximation, the Floquet states and quasienergies, as well as non-Floquet states of this non-Hermitian driven system are analytically obtained. Based on the stability analysis, we find the stability of spin-dependent dynamics depends on the competition and balance between the effective coupling parameters (arising from periodic driving) and the gain-loss coefficients.  We have systematically explored how the interplay between periodic driving, dissipation, and other system parameters influences the stability of the spin-dependent tunneling in the SO-coupled bosonic junction, by taking into account balanced/unbalanced gain-loss between two wells.
Under balanced gain and loss, the stable generalized Rabi oscillations with (without) spin-flipping are found, and stable spin-flipping tunneling is shown to be preferentially suppressed with the increase of gain-loss strength. When the ratio of Zeeman field strength to periodic driving frequency $\Omega/\omega$ is even,  \emph{continuous} stable parameter regions is possible to emerge with subtle domain boundaries. When $\Omega/\omega$ is odd, however, only \emph{discrete} stable parameter regions are found. Under unbalanced gain and loss, whether $\Omega/\omega$ is even or odd, we also find the equilibrium conditions of parameters for the existence of stable spin tunneling. The results could be useful for the experiments of manipulating stable spin transportation via a periodic driving field in a non-Hermitian SO-coupled system.

\section{Analytical solutions in the high-frequency approximation}

We consider a single SO-coupled ultracold boson held in
a non-Hermitian driven double-well potential, in which the dynamics is governed by a non-Hermitian Hamiltonian \cite{xiao85,yu90}
\begin{eqnarray}\label{eq1}
\hat{H}(t)&=& -\nu (\hat{a}_{l}^{\dag}e^{-i \pi \lambda \hat{\sigma}_{y}}\hat{a}_{r}+H.c.) + \frac{\Omega}{2} \sum_{j} (n_{j \uparrow}-n_{j \downarrow}) \nonumber\\ &+& \sum_{\sigma} [\varepsilon_{l}(t)n_{l \sigma}-\varepsilon_{r}(t)n_{r \sigma}].
\end{eqnarray}
Here $\hat{a}_{j}=(\hat{a}_{j \uparrow},\hat{a}_{j \downarrow})^{T}$ (the superscript $T$
stands for the matrix transpose) is the two-component vector with elements being the annihilation operators of spin-up and spin-down atoms in the $j$th $(j=l,r)$ well, and $\hat{a}_{j}^{\dag}$ denotes its Hermite conjugation with elements being the creation operators. $\hat{n}_{j\sigma}=\hat{a}_{j \sigma}^{\dag}\hat{a}_{j \sigma}$ denotes the number operator for
spin $\sigma$ $(\sigma=\uparrow, \downarrow)$ in well $j$, $\nu$ denotes the tunneling rate without SO coupling, $\lambda$ characterizes the SO coupling strength,  $\hat{\sigma}_{y}$ is the $y$ component   of Pauli operator, and  $\Omega$ is the effective Zeeman field intensity. The form $\varepsilon_j(t)=\varepsilon\cos(\omega t)+i \beta_j$ is taken, where the first term denotes the periodic driving with driving amplitude $\varepsilon$ and frequency $\omega$, and the latter term  the gain-loss strength. Without loss of generality, we assume  $\beta_j>0$ such that the plus ($+$ is omitted by convention) and minus ($-$) signs in the last summation term of Hamiltonian are used to represent the left well experiencing
gain  while  the right well loss. Throughout this paper, $\hbar=1$ is adopted and the dimensionless parameters $\nu, \Omega, \varepsilon, \omega, \beta_j$ are in units of the reference frequency $\omega_0=0.1 E_r$, with $E_r=k_{L}^2/(2m)$ being the single-photon recoil energy, and time $t$ is normalized in units of $\omega_0^{-1}$. Note that the single-photon recoil energy is $E_r=k_{L}^2/(2m)=22.5$ kHz and the Zeeman field $\Omega$ is set as $-40 \omega_0\sim 40 \omega_0$ in the experiment\cite{lin471}, and the experimentally achievable system parameters can be tuned in a wide range as follows \cite{luo93, kierig100, chen107, ma107, graefe82, xiao85, hai22}: $\nu \sim \omega_0$, $\varepsilon \sim \omega \in [0, 100](\omega_0)$, $\Omega \sim \omega$, $\beta_j \sim \omega_0$.

Using the Fock basis $|0, \sigma\rangle$ ($|\sigma, 0\rangle$) to represent the state of a spin $\sigma$ atom occupying the right (left) well and no atom in the left (right) well, we can expand the quantum state of the SO-coupled system as
\begin{eqnarray}\label{eq2}
|\psi(t)\rangle &=& a_1(t)|0,\uparrow\rangle + a_2(t)|\downarrow,0\rangle + a_3(t)|\uparrow,0\rangle \nonumber\\ &+& a_4(t)|0,\downarrow\rangle,
\end{eqnarray}
where $a_{k}(t)$ ($k=1, 2, 3, 4$) denotes the time-dependent probability amplitude of the atom being in the corresponding Fock state $|0, \sigma\rangle$ or $|\sigma, 0\rangle$ (e.g., $a_{1}(t)$ denotes the time-dependent probability amplitude of the atom being in state $|0, \uparrow\rangle$). The corresponding probability reads $P_{k}(t)=|a_{k}(t)|^2$.
Inserting equations (1) and (2) into Schr\"{o}dinger equation $i\frac{\partial|\psi(t)\rangle}{\partial t}=\hat{H}(t)|\psi(t)\rangle$
results in the coupled equations
\begin{eqnarray}\label{eq3}
i\dot{a}_1(t)&=&-\nu \cos(\pi \lambda) a_3(t)-\nu \sin(\pi \lambda) a_2(t)\nonumber\\&+&[\frac{\Omega}{2}-\varepsilon \cos (\omega t)-i \beta_{r}] a_1(t),\nonumber\\
i\dot{a}_2(t)&=&-\nu \cos(\pi \lambda) a_4(t)-\nu \sin(\pi \lambda) a_1(t)\nonumber\\&+&[-\frac{\Omega}{2}+\varepsilon \cos (\omega t)+i \beta_{l}] a_2(t),\nonumber\\
i\dot{a}_3(t)&=&-\nu \cos(\pi \lambda) a_1(t)+\nu \sin(\pi \lambda) a_4(t)\nonumber\\&+&[\frac{\Omega}{2}+\varepsilon \cos (\omega t)+i \beta_{l}] a_3(t),\nonumber\\
i\dot{a}_4(t)&=&-\nu \cos(\pi \lambda) a_2(t)+\nu \sin(\pi \lambda) a_3(t)\nonumber\\&+&[-\frac{\Omega}{2}-\varepsilon \cos (\omega t)-i \beta_{r}] a_4(t).
\end{eqnarray}
It is hard to obtain the exact analytical solutions of equation (3) because of the periodically varying coefficients, but the quantum
motion of the system  can be investigated analytically in high-frequency approximation\cite{blanes470, thimmel9}.
In the high-frequency regime where
$\omega \gg \nu, \beta_{j}$ and within multiphoton resonance case ($\Omega=n\omega$ with $n$ being integers), we introduce the slowly varying
functions $b_k(t)$ through the transformation
$a_1(t)=b_1(t)e^{-i\int [\frac{\Omega}{2}-\varepsilon \cos(\omega t)]dt}$,
$a_2(t)=b_2(t)e^{-i\int [-\frac{\Omega}{2}+\varepsilon \cos(\omega t)]dt}$,
$a_3(t)=b_3(t)e^{-i\int [\frac{\Omega}{2}+\varepsilon \cos(\omega t)]dt}$,
$a_4(t)=b_4(t)e^{-i\int [-\frac{\Omega}{2}-\varepsilon \cos(\omega t)]dt}$.
Using the Fourier expansions $\exp[\pm i \int
2\varepsilon \cos(\omega t)dt]=\sum_{n'=-\infty}^{\infty}\mathcal
{J}_{n'}(\frac{2\varepsilon}{\omega})\exp(\pm
in'\omega t)$ and $\exp\{\pm i \int [2\varepsilon \cos(\omega t) \pm \Omega]dt\}=\sum_{n''=-\infty}^{\infty}\mathcal
{J}_{n''}(\frac{2\varepsilon}{\omega})\exp[
\pm i(n''\pm \frac{\Omega}{\omega})\omega t]$, and neglecting all the rapidly oscillating terms with $n'\neq 0 $ and $n''\neq \mp \frac{\Omega}{\omega}$ \cite{zou46}, we can arrive at  an
effective non-driving model
\begin{eqnarray}\label{eq4}
i\dot{b}_{1}(t)&=&-J_{0}b_3(t)-J_{\frac{\Omega}{\omega}} b_2(t)-i \beta_{r} b_1(t),\nonumber\\
i\dot{b}_{2}(t)&=&-J_{0}b_4(t)-J_{\frac{\Omega}{\omega}} b_1(t)+i \beta_{l} b_2(t),\nonumber\\
i\dot{b}_{3}(t)&=&-J_{0}b_1(t)+J_{-\frac{\Omega}{\omega}} b_4(t)+i \beta_{l} b_3(t),\nonumber\\
i\dot{b}_{4}(t)&=&-J_{0}b_2(t)+J_{-\frac{\Omega}{\omega}} b_3(t)-i \beta_{r} b_4(t),
\end{eqnarray}
where the effective coupling constants are simply written as $J_{0}=\nu \cos(\pi \lambda)\mathcal
{J}_{0}(\frac{2\varepsilon}{\omega})$ and $J_{\pm\frac{\Omega}{\omega}}=\nu \sin(\pi \lambda)\mathcal
{J}_{\pm\frac{\Omega}{\omega}}(\frac{2\varepsilon}{\omega})$ with $\mathcal{J}_{n}(x)$
being the $n$-order Bessel function of $x$.
It can be apparently seen that the effective coupling constants can be controlled by adjusting the system
parameters, e.g., the SO coupling strength $\lambda$, driving amplitude $\varepsilon$, driving
frequency $\omega$, and the Zeeman field $\Omega$.
When $\lambda=m+\frac{1}{2}$ ($m=0,1,2...$) or/ and $\mathcal
{J}_{0}(\frac{2\varepsilon}{\omega})=0$, which corresponds with $J_{0}=0$ and $J_{\pm\frac{\Omega}{\omega}}\neq 0$, the spin atom will tunnel between two wells with only spin flipping. When $\lambda=m$ or/ and $\mathcal{J}_{\pm \frac{\Omega}{\omega}}(\frac{2\varepsilon}{\omega})=0$, which corresponds with  $J_{\pm \frac{\Omega}{\omega}}=0$ and
$J_{0}\neq 0$, quantum tunneling will be produced without spin flipping. When $\lambda=m+\frac{1}{2}$ ($\lambda=m$) and $\mathcal{J}_{\pm \frac{\Omega}{\omega}}(\frac{2\varepsilon}{\omega})=0$ ($\mathcal{J}_{0}(\frac{2\varepsilon}{\omega})=0$), which corresponds with both $J_{0}=0$ and $J_{\pm\frac{\Omega}{\omega}}=0$,  quantum tunneling between two wells  will be frozen but with either exponential decay of probability amplitude in right well or exponential growth in left well for $\beta_r , \beta_l >0$, while for $\beta_r = \beta_l =0$, conventional coherent destruction of  tunneling (CDT) arises.

Before moving on, we would like to make some remarks on the the validity of the high-frequency expansion employed in the non-Hermitian system. As is well known, the high-frequency expansion is a well-established method which has been commonly used to treat the periodically  driven Hermitian systems. Normally  when the driving frequency is far greater than all energy scales of the system, it is safe to implement the high-frequency expansion approach. The basic idea of this approach is to take a time average of the periodic term  and replace the tunneling constant by the cycle-averaged one, by assuming that the slowly varying variables (after separating out the rapidly oscillating phase) are constants during one driving period. It is apparent that such a high-frequency expansion technique works effectively in analogy to the Hermitian case when the non-Hermitian system evolves to a stable state. Care should be taken with regard to the situation where the time-evolving dynamics of system starts either exponentially growing or decaying. It is common knowledge that  stability of the non-Hermitian system is determined by the imaginary part of quasienergy. When the quasienergies become complex, the probability amplitudes of system will grow or decay exponentially  with time, in which the rate of exponential change is determined by the size of imaginary part of quasienergy.
In our case, the magnitude of the imaginary part of quasienergy is given by the dissipation factor and the tunneling rate without SO-coupling. In the high-frequency limit such that the driving frequency is much larger than the magnitude of the imaginary part of quasienergy, the amount of the exponential change in  probability amplitudes is still very small during a short driving period, and thus we can take the time average of the periodic terms as we have done for the Hermitian problems. It should be noted that the existing literatures have also demonstrated the validity of the high-frequency expansion in the treatment of the periodically driven  non-Hermitian systems\cite{valle87, luo110}.

\subsection{Floquet states and quasienergies}
Applying the well-established  Floquet theorem to the periodic time-dependent equation (3), we can seek the  analytical Floquet states by setting $(b_1(t),b_2(t),b_3(t),b_4(t))^T=(A,B,C,D)^T\exp(-iE t)$, where $(A,B,C,D)^T$ and $E$ are the eigenvector
and eigenvalue of the time-independent version of equation (4) respectively.
In view of the relations between $a_k(t)$ and $b_k(t)$, the analytical solutions of equation (3)  can be constructed as $|\psi(t)\rangle=|\varphi(t)\rangle\exp(-iE t)$, where $|\varphi(t)\rangle=Ae^{-i\int [\frac{\Omega}{2}-\varepsilon \cos(\omega t)]dt}|0,\uparrow\rangle+Be^{-i\int [-\frac{\Omega}{2}+\varepsilon \cos(\omega t)]dt}|\downarrow,0\rangle+Ce^{-i\int [\frac{\Omega}{2}+\varepsilon \cos(\omega t)]dt}|\uparrow,0\rangle+De^{-i\int [-\frac{\Omega}{2}-\varepsilon \cos(\omega t)]dt}|0,\downarrow\rangle$.
According to the  Floquet theorem, it is well known that a Floquet
state should inherit the period of the driving field.
From the expression of $|\varphi(t)\rangle$, we immediately note that $|\varphi(t)\rangle$ is periodic with the driving period
$\tau=2\pi/\omega$, satisfying $|\varphi(t+\tau)\rangle=|\varphi(t)\rangle$.
The result implies that the
solution  $|\varphi(t)\rangle$ can be viewed as the so-called Floquet states and the corresponding
eigenvalues $E$ in equation (4) as the approximate analytical quasienergies \cite{luo2015, luo7}.
Upon inserting the stationary solutions into equation (4), solvability of this equation gives  four analytical  Floquet quasienergies $E_p$ and the corresponding eigenvector components $A_p,B_p, C_p, D_p$ for $p=1,2,3,4$ as follows,
\begin{eqnarray}\label{eq5}
B_{1,2}&=&\pm i A_{1,2} \alpha_{+}(\pm \beta_{r}\pm \beta_{l}+\zeta_{+}), \nonumber\\ C_{1,2}&=&\mp i A_{1,2} \kappa_{+}(\pm \beta_{r}\pm \beta_{l}+\zeta_{+}),\nonumber\\ D_{1,2}&=&-A_{1,2} \eta, E_{1,2}=\frac{1}{2}i(\beta_{l}-\beta_{r}\pm\zeta_{+}),\nonumber\\
B_{3,4}&=&\mp i A_{3,4} \alpha_{-}(\mp \beta_{r}\mp \beta_{l}+\zeta_{-}), \nonumber\\ C_{3,4}&=&\pm i A_{3,4} \kappa_{-}(\mp \beta_{r}\mp \beta_{l}+\zeta_{-}), \nonumber\\ D_{3,4}&=&A_{3,4} \eta, E_{3,4}=\frac{1}{2}i(\beta_{l}-\beta_{r}\mp \zeta_{-}),
\end{eqnarray}
where we have set the following constants $\alpha_{\pm}=\frac{(J_{\frac{\Omega}{\omega}}-J_{-\frac{\Omega}{\omega}})^2\pm2|J_0(J_{\frac{\Omega}{\omega}}-J_{-\frac{\Omega}{\omega}})|}{4(J_{\frac{\Omega}{\omega}}-J_{-\frac{\Omega}{\omega}})(J_0^2+J_{\frac{\Omega}{\omega}}J_{-\frac{\Omega}{\omega}})}$,
$\zeta_{\pm}=\sqrt{(\beta_{r}+\beta_{l})^2-4J_0^2-4J_{\frac{\Omega}{\omega}}^2\pm4|J_0(J_{\frac{\Omega}{\omega}}-J_{-\frac{\Omega}{\omega}})}|$,
$\kappa_{\pm}=\frac{J_0^2 (J_{\frac{\Omega}{\omega}}-J_{-\frac{\Omega}{\omega}})\pm J_{\frac{\Omega}{\omega}}|J_0(J_{\frac{\Omega}{\omega}}-J_{-\frac{\Omega}{\omega}})|}{2J_0(J_{\frac{\Omega}{\omega}}-J_{-\frac{\Omega}{\omega}})(J_0^2+J_{\frac{\Omega}{\omega}}J_{-\frac{\Omega}{\omega}})}$, $\eta=\frac{|J_0(J_{\frac{\Omega}{\omega}}-J_{-\frac{\Omega}{\omega}})|}{J_0 (J_{\frac{\Omega}{\omega}}-J_{-\frac{\Omega}{\omega}})}$,
which can be manipulated by adjusting the related system parameters.
It is noteworthy that the quasienergies given by equation (5) may be complex because the Hamiltonian system is non-Hermitian.
Given equation (5), we can  write down the four Floquet states in the form
\begin{eqnarray}\label{eq6}
|\varphi_{p}(t)\rangle&=&A_{p}e^{-i\int [\frac{\Omega}{2}-\varepsilon \cos(\omega t)]dt}|0,\uparrow\rangle\nonumber \\ &+&B_{p}e^{-i\int [-\frac{\Omega}{2}+\varepsilon \cos(\omega t)]dt}|\downarrow,0\rangle\nonumber \\ &+& C_{p}e^{-i\int [\frac{\Omega}{2}+\varepsilon \cos(\omega t)]dt}|\uparrow,0\rangle\nonumber \\ &+&D_{p}e^{-i\int [-\frac{\Omega}{2}-\varepsilon \cos(\omega t)]dt}|0,\downarrow\rangle
\end{eqnarray}
for $p=1,2,3,4$. Here, we do not bother to normalize
the Floquet states because the normalization factor will have no effect on the final
result. As is well known,  the Floquet state and quasienergy provide two basic concepts
and tools for understanding the tunneling dynamics of periodically
driven system, and all
available information about the system at any time can be deduced from the two basic concepts.

\subsection{General coherent non-Floquet solution}
The Floquet solution  given above, $|\psi(t)\rangle=|\varphi(t)\rangle\exp(-iE t)$, only denotes the simplest fundamental  solution of the periodically driven system, which just acquires a phase $-i
E\tau$ during
the interval of one period $\tau$. In order to study the general spin tunneling dynamics starting from an arbitrary initial state, we have to consider the coherent superposition of the Floquet states. When the superposition states do not obey the well-known definition of the Floquet state, we call them the non-Floquet solutions \cite{lu83}. The superposition principle of quantum mechanics indicates that the non-Floquet states can be constructed by the linear superposition of the Floquet states \cite{jinasundera322}. Directly employing equations (5) and (6) to the linear superposition yields the general non-Floquet solution
\begin{eqnarray}\label{eq7}
|\psi(t)\rangle&=&\sum^{4}_{p=1} \Lambda_p|\varphi_p (t)\rangle e^{-iE_p t}\nonumber\\
&=&d_{1}e^{-i\int [\frac{\Omega}{2}-\varepsilon \cos(\omega t)]dt}|0,\uparrow\rangle\nonumber \\ &+&d_{2}e^{-i\int [-\frac{\Omega}{2}+\varepsilon \cos(\omega t)]dt}|\downarrow,0\rangle\nonumber \\ &+& d_{3}e^{-i\int [\frac{\Omega}{2}+\varepsilon \cos(\omega t)]dt}|\uparrow,0\rangle\nonumber \\ &+&d_{4}e^{-i\int [-\frac{\Omega}{2}-\varepsilon \cos(\omega t)]dt}|0,\downarrow\rangle,
\end{eqnarray}
where $\Lambda_p$ is the superposition coefficient.
In equation (7), the probability amplitudes are rearranged as
$d_1(t)=\sum^{4}_{p=1}\Lambda_pA_pe^{-iE_pt},\
d_2(t)=\sum^{4}_{p=1}\Lambda_pB_pe^{-iE_pt},\
d_3(t)=\sum^{4}_{p=1}\Lambda_pC_pe^{-iE_pt},\
d_4(t)=\sum^{4}_{p=1}\Lambda_pD_pe^{-iE_pt}$ with the undetermined constant $\Lambda_pA_p$ depending on the initial condition. Thus, the occupy probability in each local state is given by $P_{k}=|d_{k}(t)|^{2}$ for $k=1,2,3,4$. The superposition
state, equation (7), implies  quantum interference among the four
Floquet states and may cause dynamical enhancement or suppression
of quantum spin tunneling, whose degree is governed by the values of the
effective coupling parameters \cite{lu45, lu053424} and gain-loss coefficients.

\section{Stability analysis and controlling stable spin tunneling under balanced and unbalanced gain-loss}

We write Floquet quasienergy $E_p$ in the form $E_p$=Re($E_p$)+ \emph{i} Im($E_p$) ($p=1, 2, 3, 4$, hereafter, Re and Im stand for the real part and the imaginary part of a complex value respectively). According to the stability analysis on the linear equations (4)\cite{liu2004, roussel2019}, we know that the system stability is determined by the imaginary part of quasienergy, Im($E_p$), whose value is associated with the following cases.

\emph{Case A}, when all of Im($E_p$) are equal to zero, quasienergies of this system thus become all real. In this case, the evolutions of probabilities are time-periodic and the system is stable\cite{luoxb95}.

\emph{Case B}, when some of Im($E_p$) are equal to zero and the others of Im($E_p$) are less than zero, the system is also stable because its total probability tends to a constant value at $t\rightarrow \infty$ \cite{xiao85, zhou384}.

\emph{Case C}, when all of Im($E_p$) are less than zero, all probabilities will exponentially decay to zero and the atom will be lost asymptotically.

\emph{Case D}, when any one of Im($E_p$) is greater than zero, the total probability  will show exponential growth and the system is unstable.

In this paper, we will mainly present the stability analysis on the non-Hermitian system by employing the  stability criterions of \emph{Case A} and \emph{Case B}, under balanced and unbalanced gain-loss respectively. Here, we shall discuss the simple physics underlying the effective non-driving model (4). As a spin-up (spin-down) particle  tunnels from the left to the right well with spin flipping, there will be a corresponding energy change $\Omega$ ($-\Omega$) respectively. When the Zeeman field strength  is an integer
multiple of the driving frequency, that is, $\Omega=n\omega$ with $n$ being integer, the system is able
to exchange energy with the driving field to
bridge the energy gap $\Omega$ (or $-\Omega$) created by the spin-flipping
tunneling. In a time-averaged sense, the driven SO-coupled system is thus equivalent to an undriven one, where the
 tunneling rate between states $|\downarrow,0\rangle$ and $|0,\uparrow\rangle$  is renormalized by a factor of $\mathcal{J}_{n}(2\varepsilon/\omega)$, whereas the
tunneling rate between $|\uparrow,0\rangle$ and $|0,\downarrow\rangle$ is by $\mathcal{J}_{-n}(2\varepsilon/\omega)$. When $\Omega/\omega$ is even, the effective tunneling rate  between states $|\downarrow,0\rangle$ and $|0,\uparrow\rangle$ is exactly the same as the one between $|\uparrow,0\rangle$ and $|0,\downarrow\rangle$ because of $\mathcal{J}_{-n}(2\varepsilon/\omega)=\mathcal{J}_{n}(2\varepsilon/\omega)$ with even $n$; when $\Omega/\omega$ is odd, the effective tunneling rate  between states $|\downarrow,0\rangle$ and $|0,\uparrow\rangle$ is distinct from the one between $|\uparrow,0\rangle$ and $|0,\downarrow\rangle$  because of $\mathcal{J}_{-n}(2\varepsilon/\omega)\neq\mathcal{J}_{n}(2\varepsilon/\omega)$ with odd $n$. Thus, the system as considered here will show qualitatively different results of both quasienergy spectrums and stable parameter regions, depending on whether $\Omega/\omega$ is even or odd.

\emph{1. Stability analysis under balanced gain and loss}

In this part, we only focus our attention on the balanced gain-loss situation, where the loss (gain) coefficients of two wells take the same values, $\beta_r=\beta_l=\beta$. In such situation, we will analyze  the dependence of the system stability  on different parameters in detail for both even and odd $\Omega/\omega$, by using the stability criterion described by  \emph{Case A}.

(1) even $\Omega/\omega$

 For this case, from equation (5) we easily infer $E_1=-E_2=-E_3=E_4=\frac{1}{2} i \rho$ with $\rho=2 \sqrt{\beta^2-J_{0}^2-J_{\frac{\Omega}{\omega}}^2}$. According to \emph{Case A} of stability analysis, by writing $\rho$=Re($\rho$)+ \emph{i} Im($\rho$), we directly find that the system is stable when Re($\rho$)=0, and is unstable otherwise. In the case of even $\Omega/\omega$, evidently, the boundary between stable (Re($\rho$)=0) and unstable (Re($\rho$)$\neq$ 0) regimes can be parameterized by using the relation $\beta^2-J_{0}^2-J_{\frac{\Omega}{\omega}}^2=0$. In all stability diagrams as illustrated by Figs.~1-3, the boundaries between stable and unstable parameter regimes are plotted as white curves. It is worth mentioning that on the white boundaries are positioned by the non-Hermitian degenerate points (i.e., exceptional points) with $E_1=E_2=E_3=E_4=0$, across which the transition of quasienergies from being real to complex will occur. In Figs.~1 (a)-(c), we take $\nu=1$, $\Omega=100$, $\omega=50$ (hence $\Omega/\omega=2$) to plot Re($\rho$) as a function of $2\varepsilon/\omega$ and $\lambda$ for different gain-loss strengths with (a) $\beta=0.2$, (b) $\beta=0.45$, and (c) $\beta=0.6$.
 As illustrated in these plots, two types of stable parameter regions are possible: when the system is stable for a continuous range of system parameters, we call the parameter regions \emph{continuous} stable parameter regions; when the system is only stable in some disconnected  parameter spaces, we call the parameter regions \emph{discrete} stable parameter regions.
 When $\beta=0.2$, from Fig.~1 (a), it is clearly seen that the system can be stabilized for a continuous range of parameters in the ($ \lambda, 2\varepsilon/\omega$) space, and the width of these continuous stable parameter regions becomes increasingly  narrower with increasing $2\varepsilon/\omega$.
 The \emph{continuous} stable parameter regions are found to possibly occur for the balanced gain-loss situation with $\Omega/\omega$ being even, which provides some convenience for the experimental manipulating stable spin transport in a non-Hermitian spin-orbit coupled cold atomic system.
 As $\beta$ gets stronger, we find that the \emph{continuous} stable parameter regions disappear and the \emph{discrete} stable parameter regions instead emerge, as shown in  Fig.~1 (b).
 Combining Fig.~1 (a) with $\beta=0.2$ and Fig.~1 (b) with $\beta=0.45$, we have another observation: when $2\varepsilon/\omega$ is less than a certain value, stable spin-flipping tunneling with $\lambda=m+0.5 (m=0,1,2...)$ can not occur, but stable non-spin-flipping tunneling with $\lambda=m$ can happen nevertheless.
 This observation implies that the available values of $2\varepsilon/\omega$ for realizing stable spin-flipping tunneling with $\lambda=m+0.5$ are greater than those for stable non-spin-flipping tunneling with $\lambda=m$. In Fig.~1 (c) with $\beta=0.6$, we find that the area of discrete stable parameter regions decreases as compared to that of Fig.~1 (b) with smaller $\beta=0.45$. Specially, it is worth noting that when $\beta=0.6$, the stable spin-flipping tunneling with $\lambda=m+0.5$ vanishes, but the stable tunneling without spin flipping with $\lambda=m$ still exists. This finding shows that with the increase of dissipative strength $\beta$,
 stable spin-flipping tunneling, when/if compared with stable non-spin-flipping tunneling, will be preferentially suppressed.

 Now let us numerically illustrate the above results  by directly integrating the original model equation (3). In Figs.~1 (d)-(e), we arbitrarily pick two parameter sets localized in stable parameter regions of Figs.~1 (a) and 1 (b), $(\beta, 2\varepsilon/\omega,\lambda)=(0.2,3,0.5)$ and $(\beta, 2\varepsilon/\omega,\lambda)=(0.45,1,1)$, as examples to plot the time evolutions of probabilities $P_{k}=|a_{k}(t)|^{2}$ for the particle initially occupying the state $|0, \uparrow\rangle$.
 As we will see in Fig.~1 (d), a stable but non-norm-preserving population oscillation with spin flipping takes place between states $|0, \uparrow\rangle$ and $|\downarrow, 0\rangle$, which represents a type of generalized Rabi oscillation \cite{gong91, luoxb95}. Similarly, a stable generalized Rabi oscillation without spin flipping between states $|0, \uparrow\rangle$ and $|\uparrow, 0\rangle$ is also shown in Fig.~1 (e). In Fig.~1 (f), we choose one parameter set $(\beta, 2\varepsilon/\omega,\lambda)=(0.6,1,0.5)$ in the unstable region of Fig.~1 (c) to illustrate the time evolutions of probabilities with the same initial condition as above. It is shown that the probabilities exponentially grow without bound which marks an unstable spin tunneling.

\begin{figure*}
\includegraphics[height=1.3in,width=2.2in]{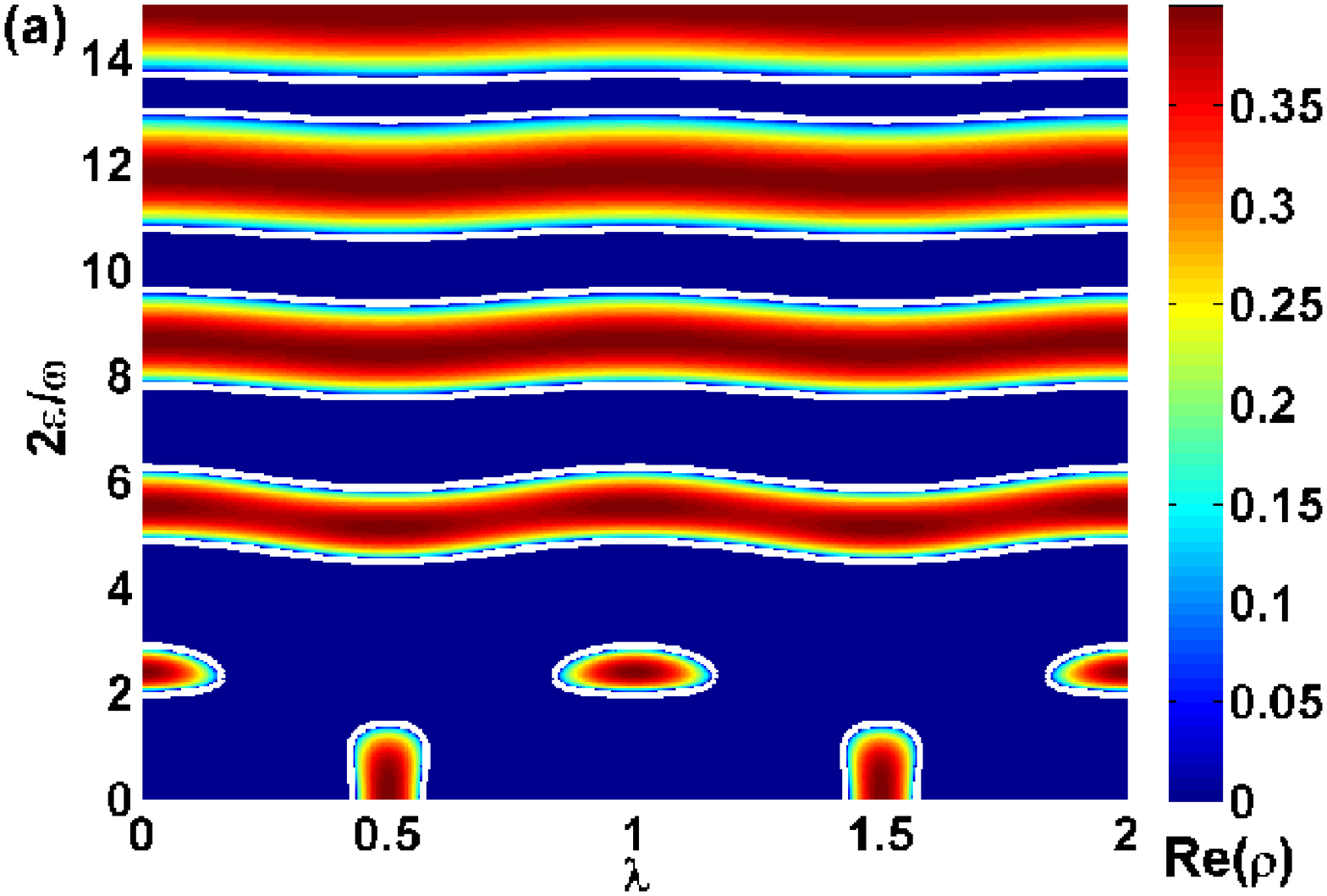}
\includegraphics[height=1.3in,width=2.2in]{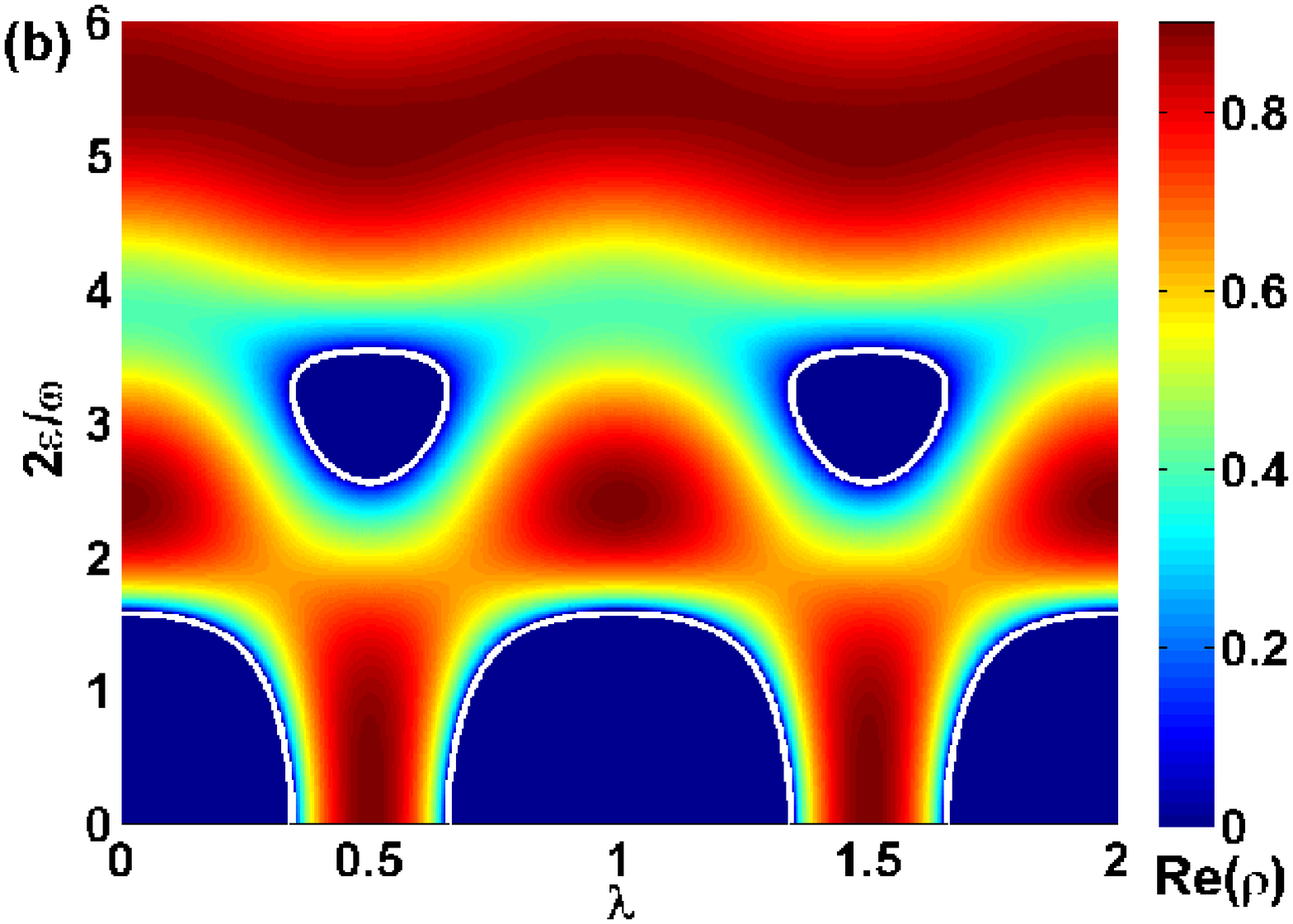}
\includegraphics[height=1.3in,width=2.2in]{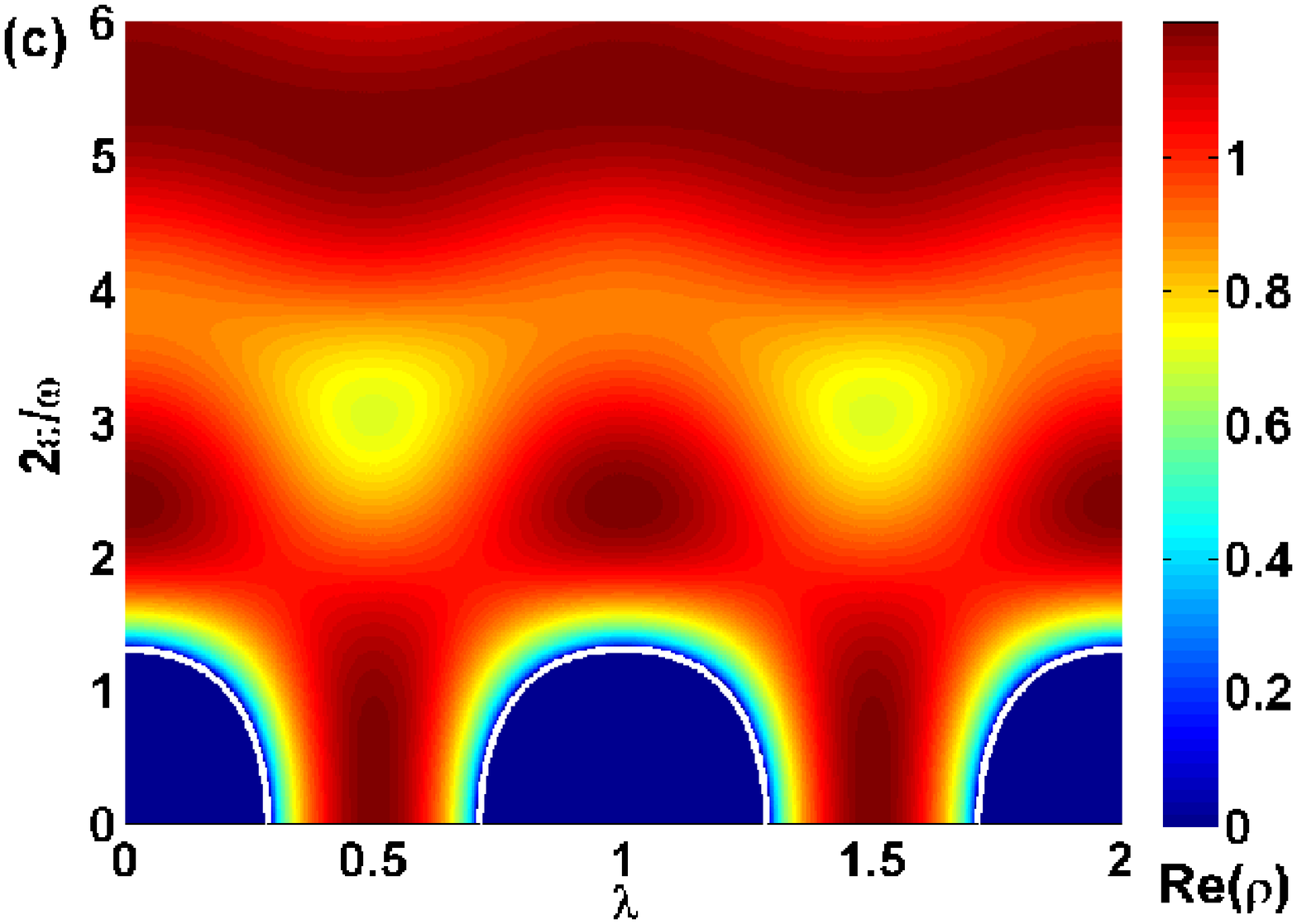}
\includegraphics[height=1.3in,width=2.2in]{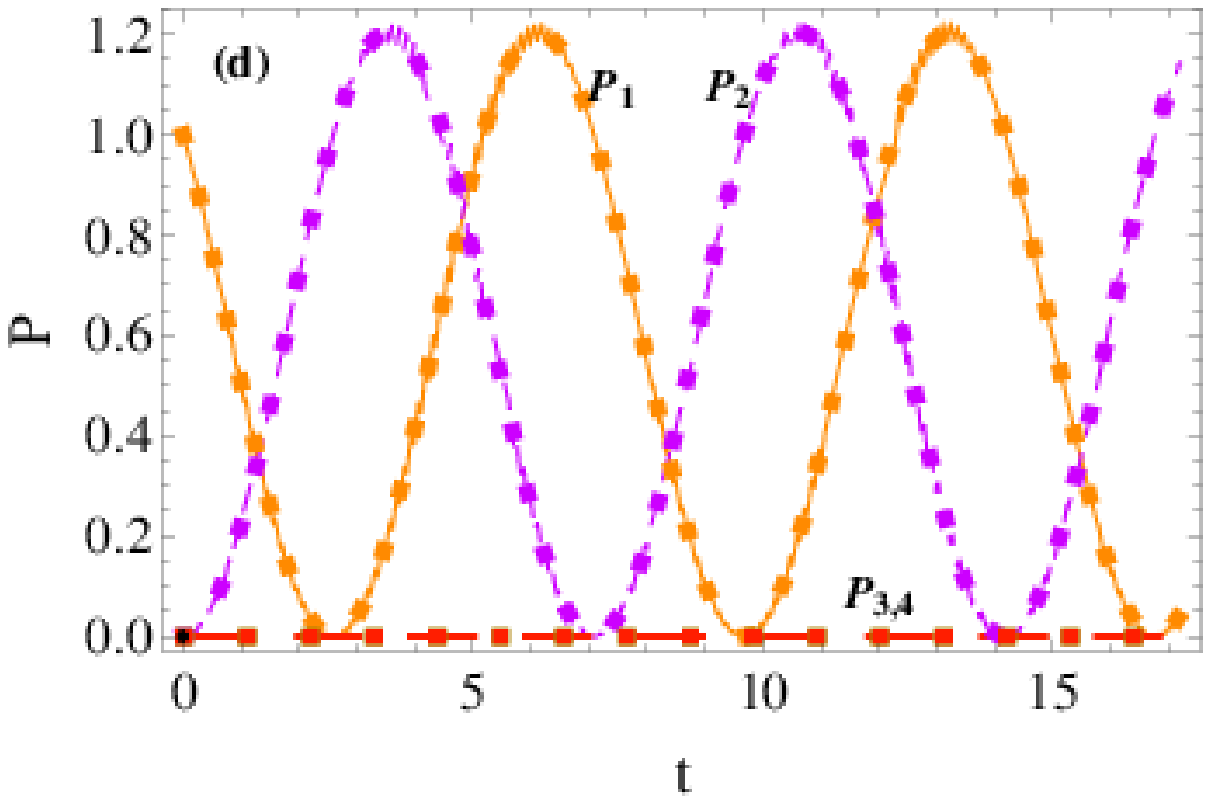}
\includegraphics[height=1.3in,width=2.2in]{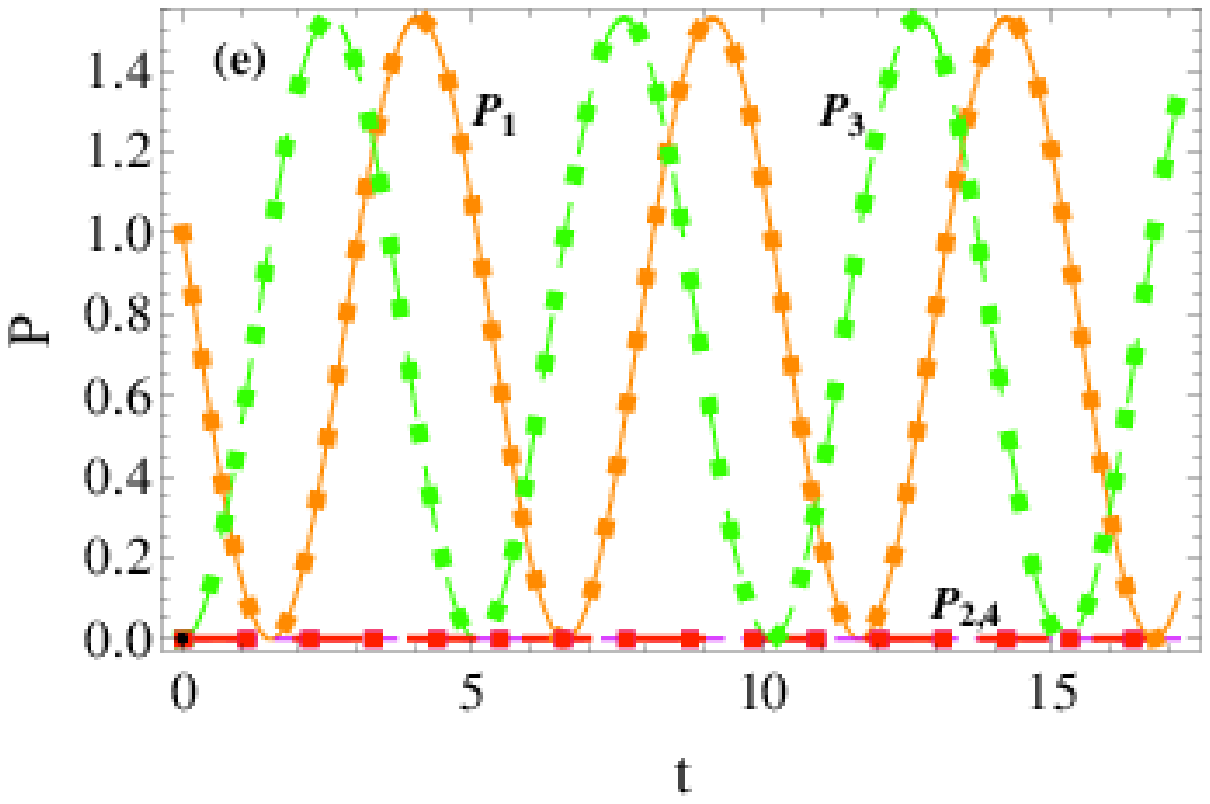}
\includegraphics[height=1.3in,width=2.2in]{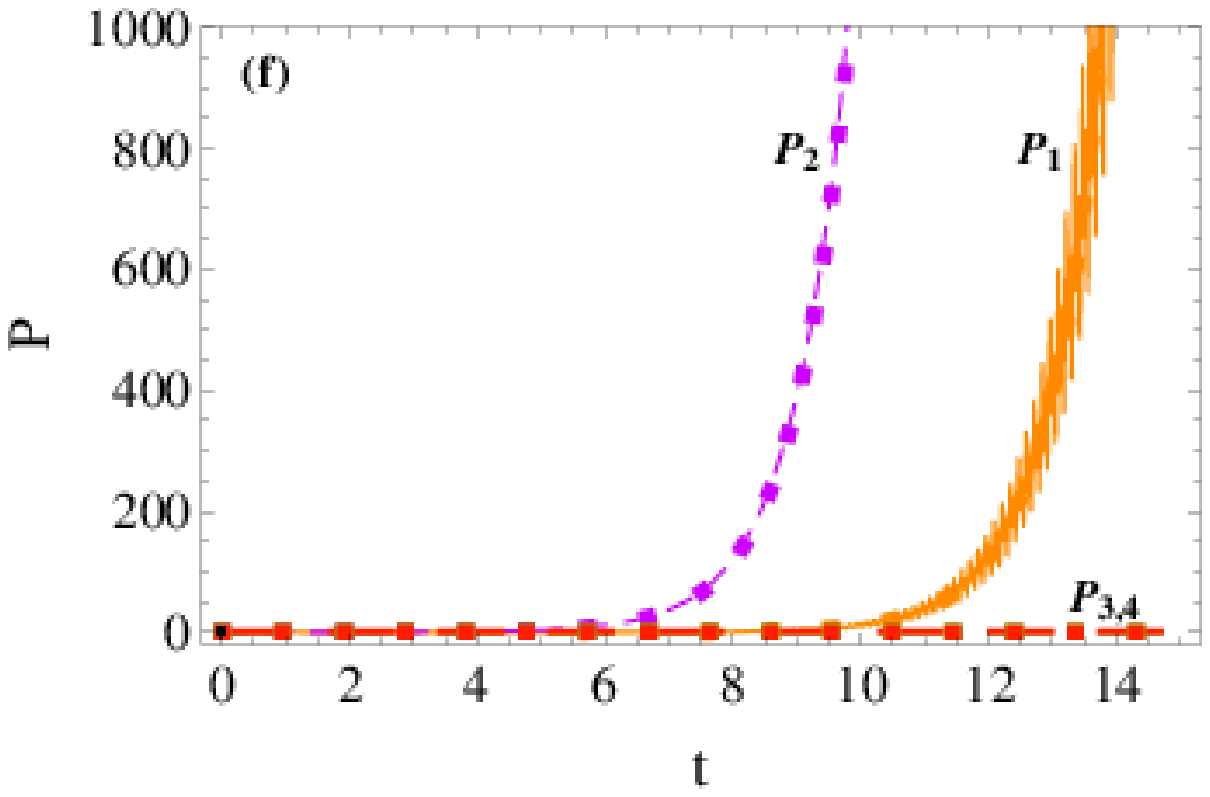}
\caption{\scriptsize{(Color online) Top panels depict the real part Re($\rho$) as a function
of $2\varepsilon/\omega$ and $\lambda$ for (a) $\beta=0.2$, (b) $\beta=0.45$, and (c) $\beta=0.6$. Here, Re($\rho$) corresponds to the imaginary part of complex quasienergy with $E_1=-E_2=-E_3=E_4=\frac{1}{2} i \rho$.
Bottom panels show the time-evolution curves of probabilities $P_{k}=|a_{k}(t)|^{2}$ given by the original model (3), for (d) $\beta=0.2$, $2\varepsilon/\omega=3$, $\lambda=0.5$; (e) $\beta=0.45$, $2\varepsilon/\omega=1$, $\lambda=1$; (f) $\beta=0.6$, $2\varepsilon/\omega=1$, $\lambda=0.5$, starting the system with a spin-up particle in the right well. The other parameters are chosen as $\nu=1$, $\Omega=100$, and $\omega=50$. Hereafter, the white curves are the boundary between stable and unstable parameter regions. Shown in Figs.~1 (d)-(f) are the numerical results of population $P_1=|a_1(t)|^2$ in state $|0,\uparrow\rangle$ (orange thin solid line), population $P_2=|a_2(t)|^2$ in state $|\downarrow,0\rangle$ (purple short-dashed line), population $P_3=|a_3(t)|^2$ in state $|\uparrow,0\rangle$ (green thin long-dashed line), and population $P_4=|a_4(t)|^2$ in state $|0,\downarrow\rangle$ (red thick long-dashed line). The squares denote the analytical correspondences obtained from the effective model (4), unless it is specially indicated. Specially, the white curves in Figs. 1 (a)-(c) correspond for the degenerate points (exceptional points) of quasienergies with $E_1=E_2=E_3=E_4=0$. All parameters adopted in these figures are dimensionless.}}
\end{figure*}

In Figs.~2 (a)-(c), we show the real part Re($\rho$) as a function of $\beta$ and $\lambda$ with (a) $2\varepsilon/\omega=5.1356$, (b) $2\varepsilon/\omega=2.405$, and (c) $2\varepsilon/\omega=1.5$, for the case of $\Omega/\omega=2$. Throughout this paper, the other two parameters are fixed as $v=1,\omega=50$ unless otherwise specified.
Figs.~2 (a) and (b) describe the stability  diagrams for the non-spin-flipping and spin-flipping tunneling respectively, which are associated with the adopted driving parameter $2\varepsilon/\omega=5.1356$ and $2\varepsilon/\omega=2.405$. The former ratio $2\varepsilon/\omega=5.1356$ infers $\mathcal{J}_{\frac{\Omega}{\omega}}(2\varepsilon/\omega)=\mathcal{J}_{2}(5.1356)=0$ and $\mathcal{J}_{0}(5.1356)\neq 0$, which leads to the quantum tunneling without spin flipping, while  the latter ratio $2\varepsilon/\omega=2.405$ gives $\mathcal{J}_{0}(2\varepsilon/\omega)=\mathcal{J}_{0}(2.405)=0$ and $\mathcal{J}_{2}(2.405)\neq 0$, which instead leads to the quantum tunneling with only spin flipping.
As such, when the system parameters are selected in the discrete stable region of Figs.~2 (a) and (b), stable generalized Rabi oscillations without and with spin flipping (see Figs. 2~(d) and (e)) will arise respectively.
In Fig.~2 (c), the ratio $2\varepsilon/\omega=1.5$ implies $\mathcal{J}_{0}(1.5)\neq 0$ and $\mathcal{J}_{2}(1.5)\neq 0$ such that the spin-conserving and spin-flipping tunnelings
will coexist. For the former two special case, where $\mathcal{J}_{2}(2\varepsilon/\omega)=0$ or $\mathcal{J}_{0}(2\varepsilon/\omega)=0$, there exist a sequence of crossings between the white boundary curve and the transverse axis. As can be seen from Figs.~2 (a)-(b), the crossings for the spin-conserving (spin-flipping) case are precisely located at $\lambda=m+1/2$ $(\lambda=m)$ respectively.
At these crossing points, where $J_0=J_2=0$, the quantum spin tunneling is not allowed and the system is stable if and only if $\beta_l=\beta_r=\beta=0$.  Indeed, the stable dynamics at these crossing points
characterizes the well-known CDT phenomenon in the Hermitian system, which is not shown here. For the general case of $\mathcal{J}_{0}(2\varepsilon/\omega)\neq 0$ and $\mathcal{J}_{2}(2\varepsilon/\omega)\neq 0$, from Fig.~2 (c), it can be seen that the white boundary curve does not cross with the transverse axis and the stable parameter region has been split into two parts.
The lower part corresponds to a continuous stable parameter region, whose width $d$ equals to the maximal safe value (the upper limit) of $\beta$ for the existence of the continuous stable parameter region and is determined essentially by the minimum value of $\sqrt{J_0^2+J_{\frac{\Omega}{\omega}}^2}$. In our case, the width $d$ is given by $d=\beta_{max}=|J_2|=\nu \mathcal{J}_2 (1.5)\approx 0.232088$; it corresponds to an upper limit in the sense that in the region under this limit, the system is stable for the whole range of parameters in the $(\lambda,\beta)$ space. The other (upper) part is associated with some discrete stable regions with the gain-loss strength $d < \beta < \nu \mathcal{J}_0 (1.5)$. Obviously, when the gain-loss strength is greater than the maximum value of $\sqrt{J_0^2+J_{\frac{\Omega}{\omega}}^2}$, namely, $\beta>|J_0|=\nu \mathcal{J}_0 (1.5)\approx 0.5118$, the system is always unstable.  As an example, we take $\beta=0.3 > d\approx 0.232088$ and $\lambda=0.8$ in the discrete stable parameter region to plot the time evolutions of probabilities as shown in Fig.~2 (f), and a stable generalized Rabi oscillation is presented accordingly. Besides, Figs.~2 (a)-(c) have confirmed, from a different angle, the fact that these stable parameter regions are shrunk with increasing the gain-loss strength $\beta$.

\begin{figure*}[htp]\centering
\includegraphics[height=1.3in,width=2.2in]{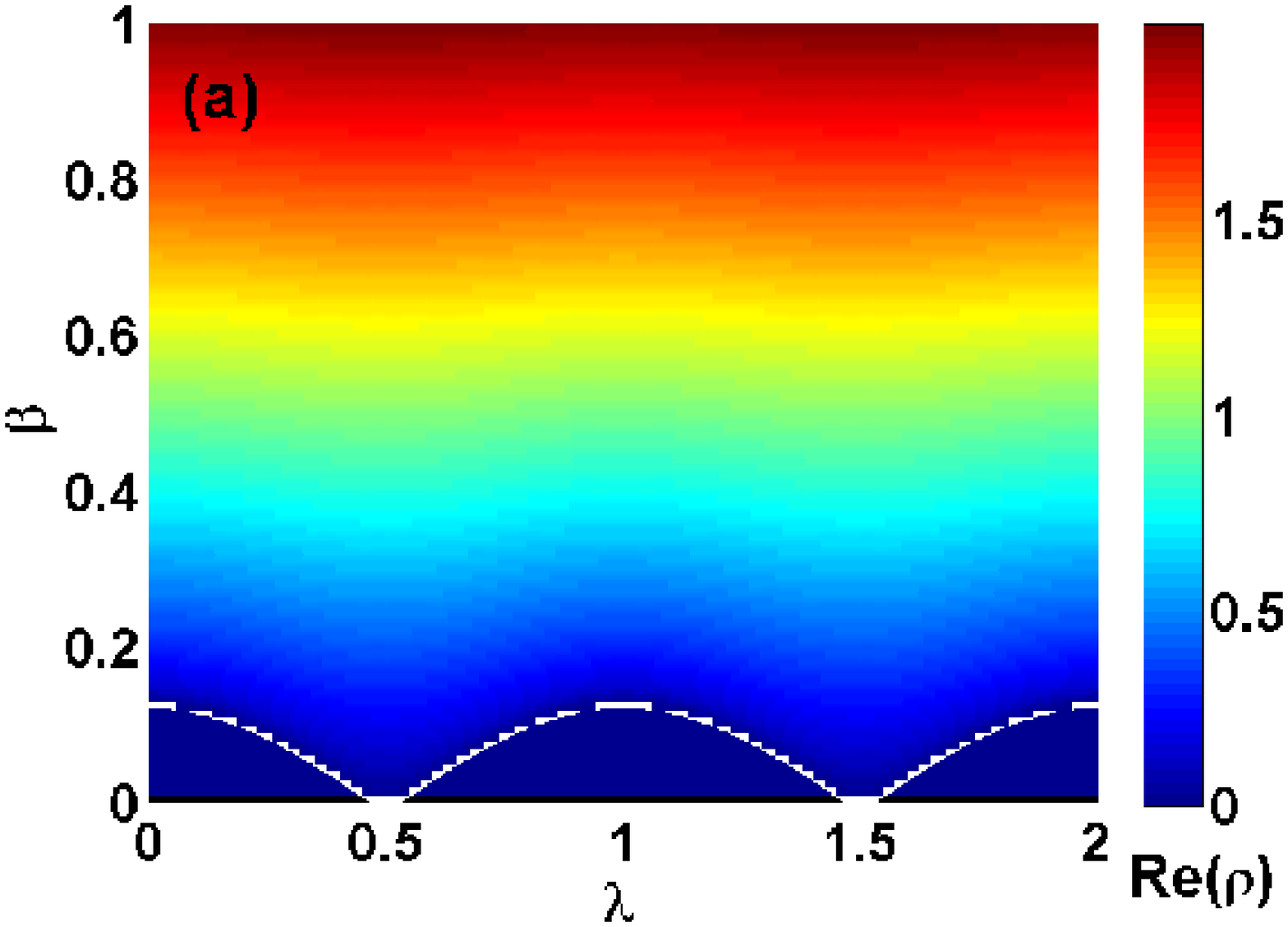}
\includegraphics[height=1.3in,width=2.2in]{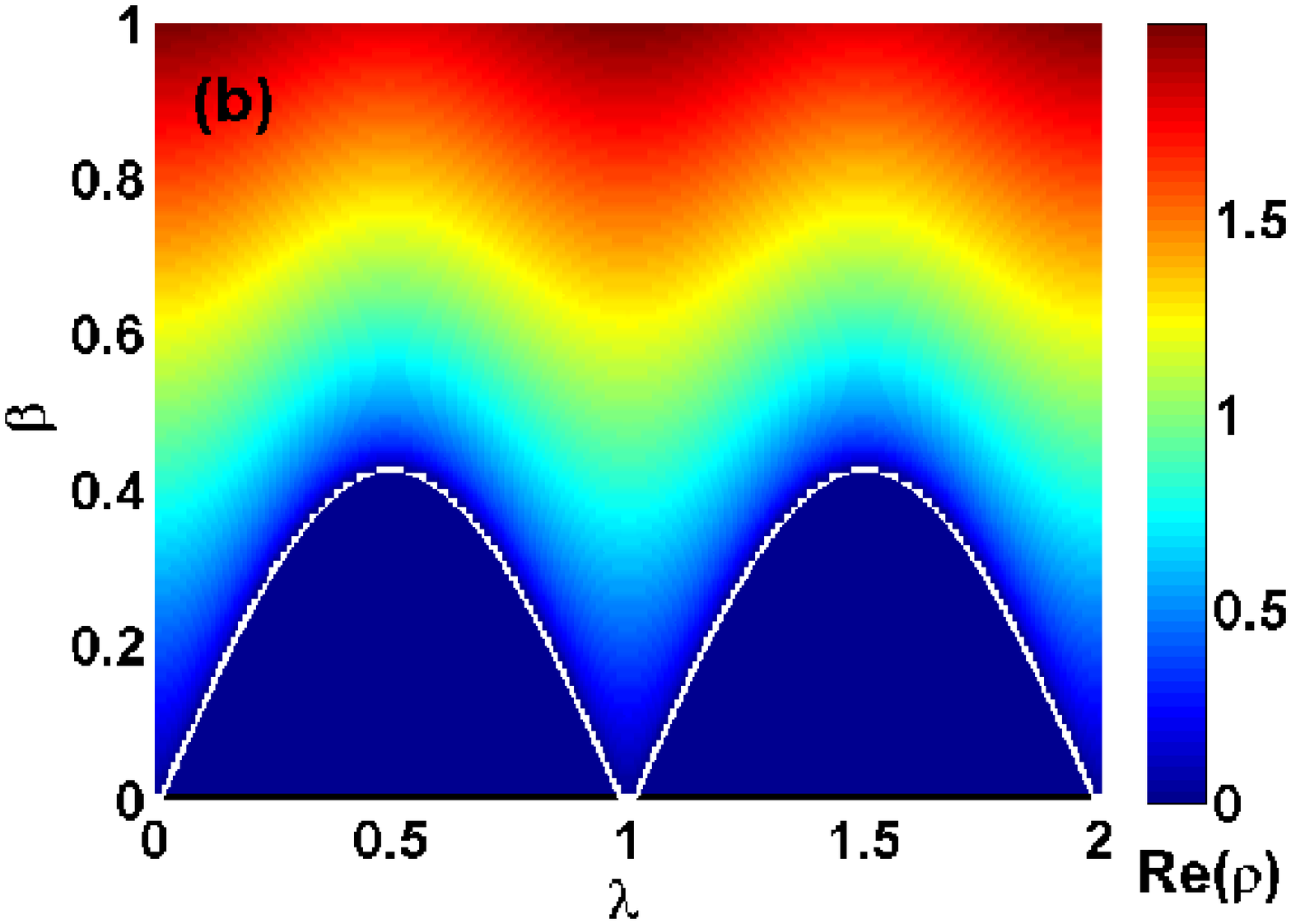}
\includegraphics[height=1.3in,width=2.2in]{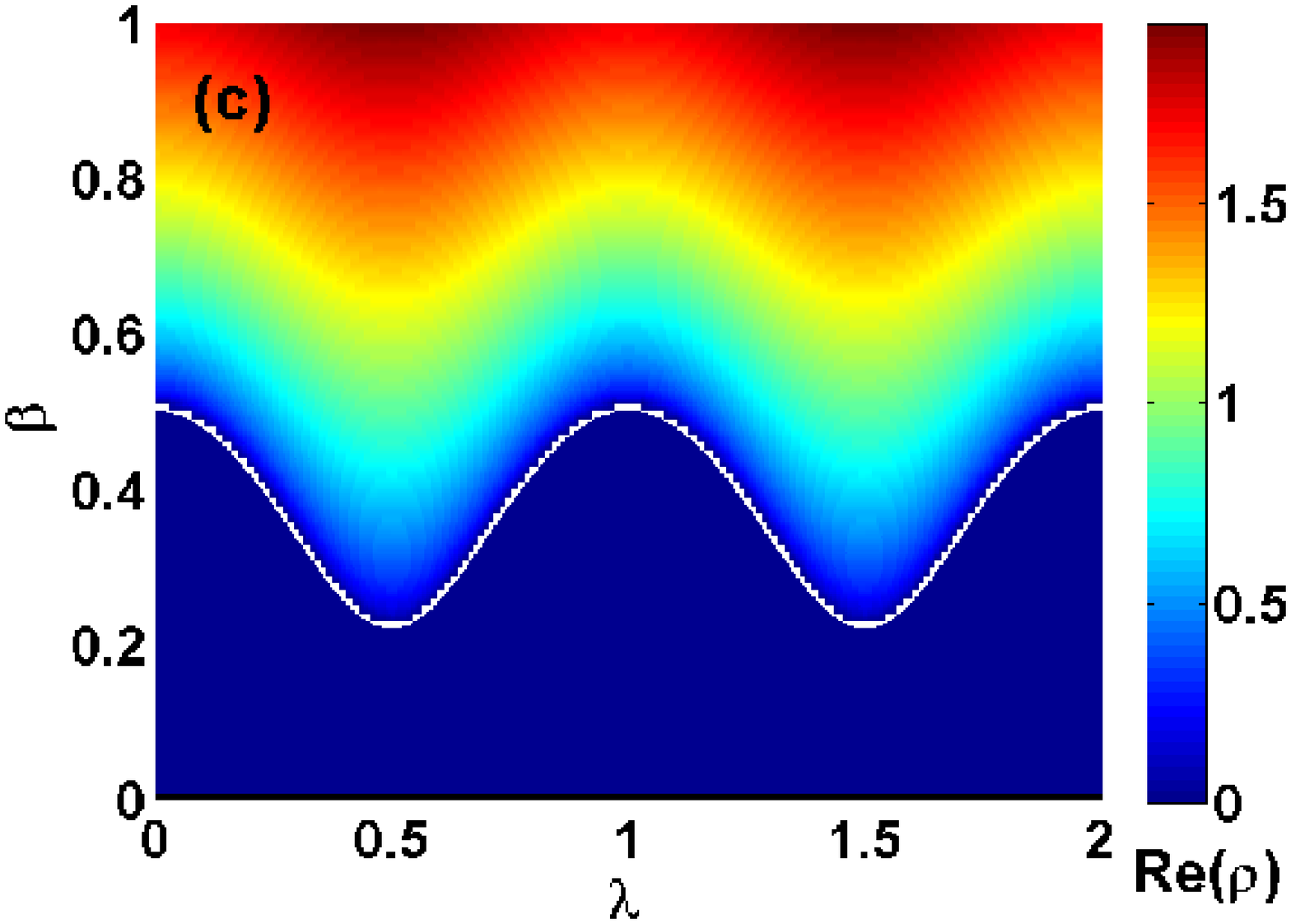}
\includegraphics[height=1.3in,width=2.2in]{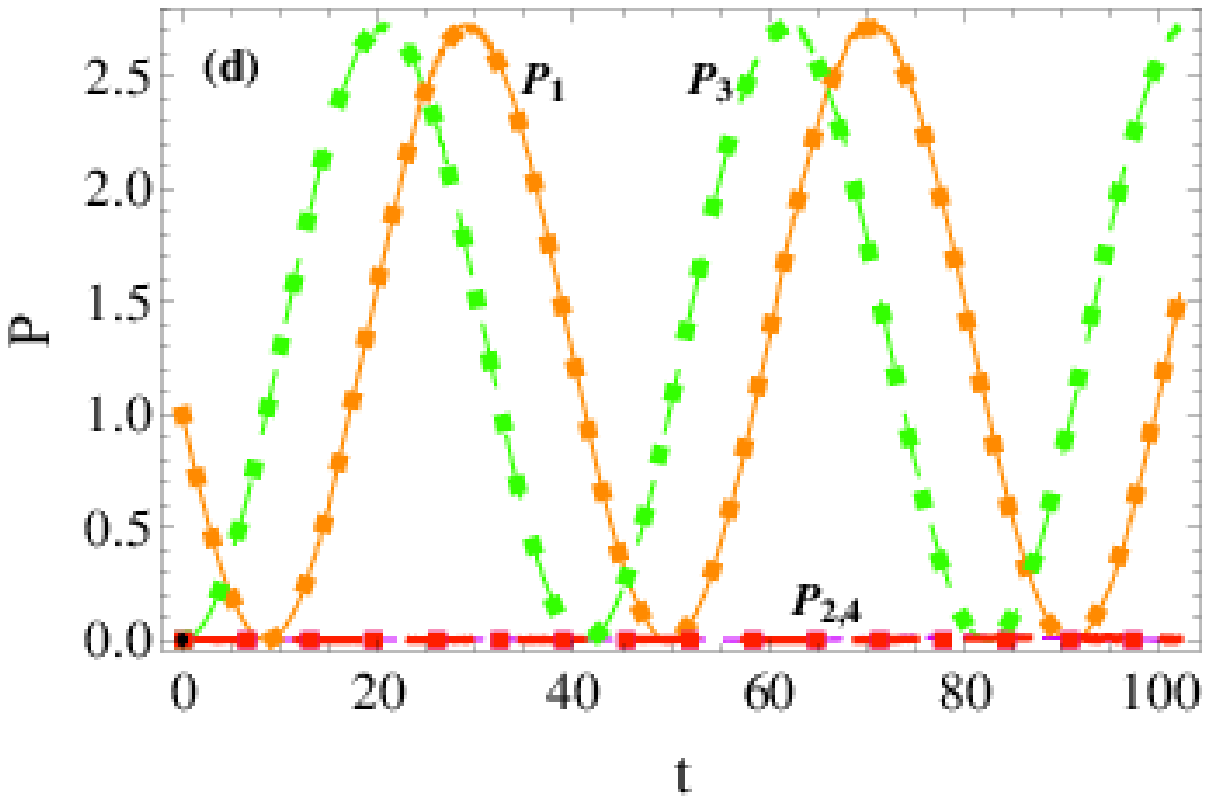}
\includegraphics[height=1.3in,width=2.2in]{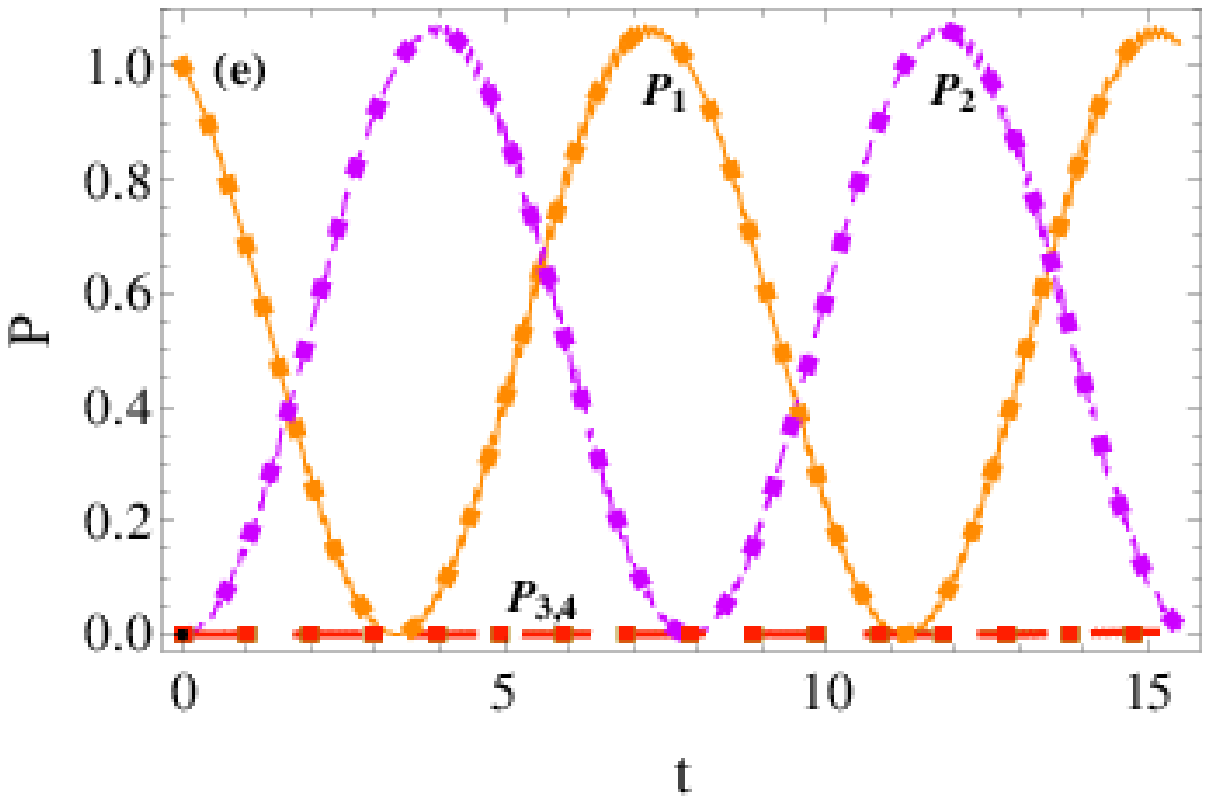}
\includegraphics[height=1.3in,width=2.2in]{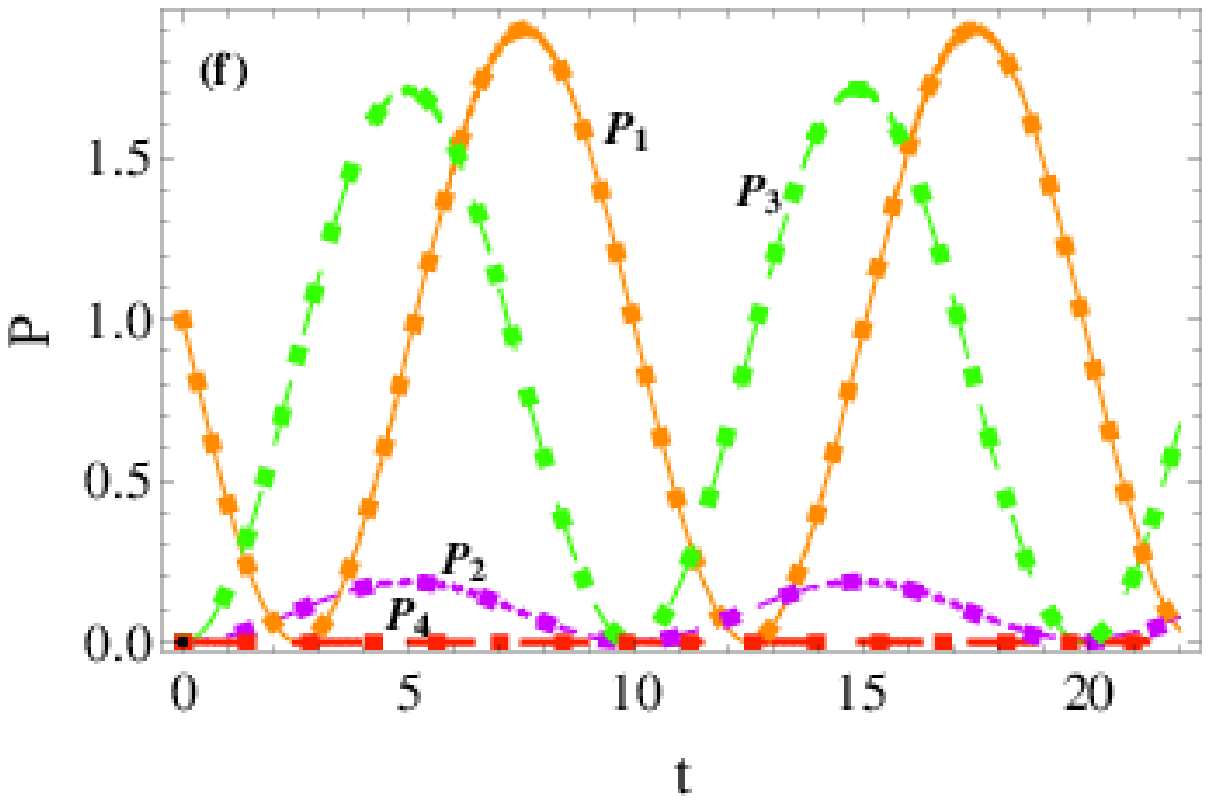}
\caption{\scriptsize{(Color online) Top panels depict the real parts Re($\rho$) as a function of
$\beta$ and $\lambda$ for (a) $2\varepsilon/\omega=5.1356$, (b) $2\varepsilon/\omega=2.405$, and (c) $2\varepsilon/\omega=1.5$. Here, Re($\rho$) corresponds to the imaginary part of complex quasienergy with $E_1=-E_2=-E_3=E_4=\frac{1}{2} i \rho$. The stable parameter regions are under the white boundary curves. Bottom panels show the time-evolution curves of probabilities $P_{k}=|a_{k}(t)|^{2}$ for (d) $2\varepsilon/\omega=5.1356$, $\beta=0.1$, $\lambda=0.1$; (e) $2\varepsilon/\omega=2.405$, $\beta=0.1$, $\lambda=0.4$; (f) $2\varepsilon/\omega=1.5$, $\beta=0.3$, $\lambda=0.8$, starting the system with a spin-up particle in the right well. The other parameters are the same as those of Fig.~1.}}
\end{figure*}

In Figs.~3 (a)-(c), we show the real part Re($\rho$) as a function of
$\beta$ and $2\varepsilon/\omega$ with (a) $\lambda=0.5$, (b) $\lambda=1$, and (c) $\lambda=1.7$, for the case of $\Omega/\omega=2$.  In Fig.~3 (a), the selected SO coupling strength $\lambda=0.5$ is associated with $J_{0}=0$, which means the quantum tunneling with spin flipping. In Fig.~3 (b), we consider the other special case with SO coupling strength $\lambda=1$, which corresponds with $J_{2}=0$ and thus describes the quantum tunneling without spin flipping. As expected, by taking system parameters in the discrete stable region of Figs.~3 (a) and (b) respectively, we can observe the stable generalized Rabi oscillation with and without spin flipping as shown in Figs.~3 (d) and (e) correspondingly. From  Figs.~3 (a)-(b). it is seen that the white boundary curve can cross with the transverse axis repeatedly. The positions of these crossing points correspond to the roots of $\mathcal{J}_2 (\frac{2\varepsilon}{\omega})=0$ for $\lambda=m+1/2$ (see Fig.~3 (a)) and of $\mathcal{J}_0(\frac{2\varepsilon}{\omega})=0$ for $\lambda=m$ (see Fig.~3 (b)). At these crossing points, we have $J_0=J_2=0$ and $\beta=0$, which means the occurrence of CDT (not shown in Fig.~3). In Fig.~3 (c), we discuss the general case with SO coupling strength $\lambda=1.7$,  and find  a continuous stable parameter region, due to the fact there is no intersection between the white boundary curve  with the transverse axis. Moreover, it is shown that the width of the continuous stable region decreases with the increase of $2\varepsilon/\omega$, which is consistent with the behaviors reflected perviously by Fig.~1(a). For the general case of $\lambda=1.7$ and with the parameter points taken in the stable region, when the ratio $2\varepsilon/\omega$ is changed to satisfy $\mathcal{J}_0(\frac{2\varepsilon}{\omega})=0$ or $\mathcal{J}_2 (\frac{2\varepsilon}{\omega})=0$, the stable generalized spin-flipping or non-spin-flipping Rabi oscillation will occur correspondingly; otherwise, the stable generalized spin-flipping and non-spin-flipping Rabi oscillation will simultaneously happen, as we can see from Fig.~3 (f).

\begin{figure*}[htp]\center
\includegraphics[height=1.3in,width=2.2in]{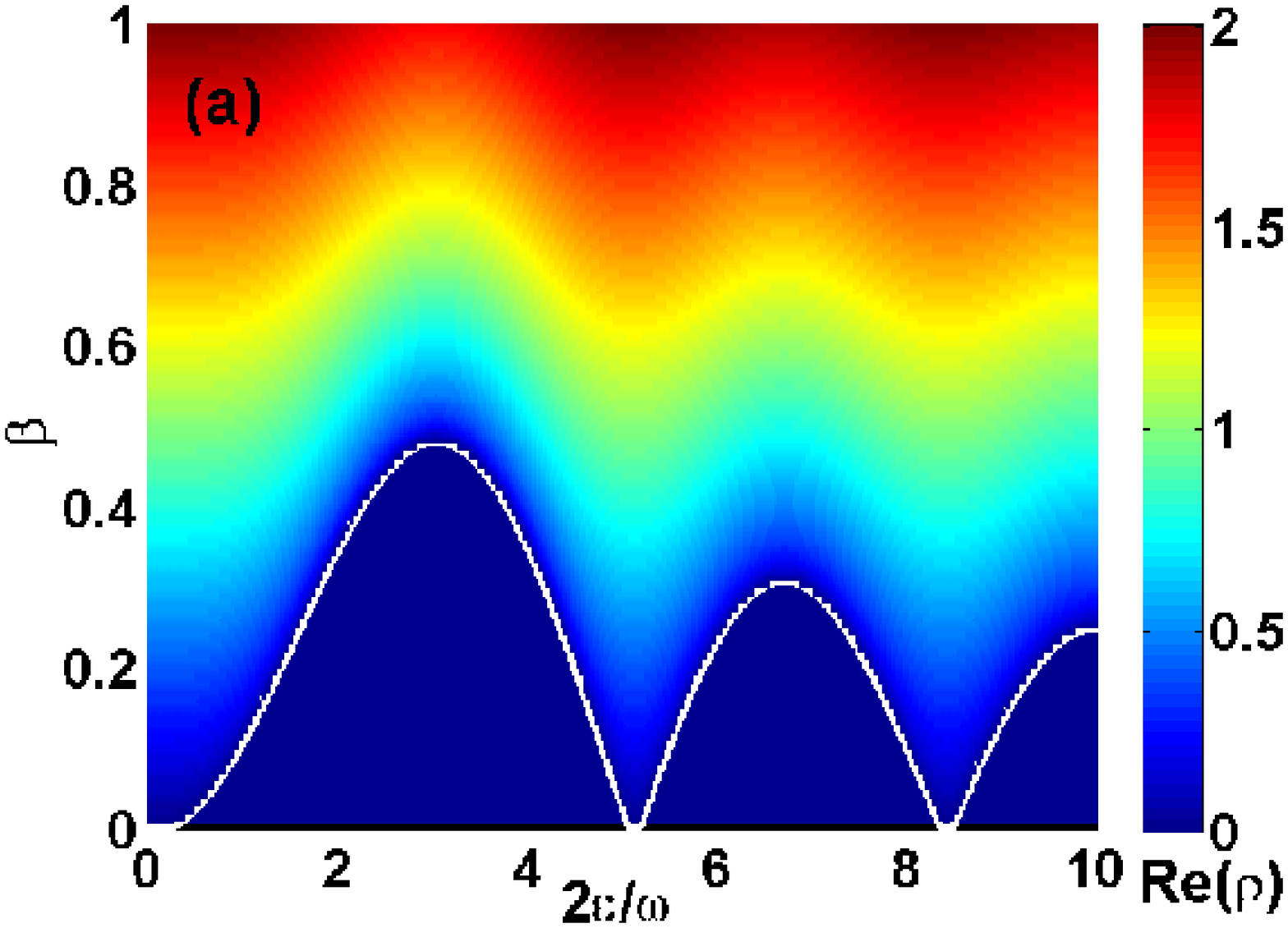}
\includegraphics[height=1.3in,width=2.2in]{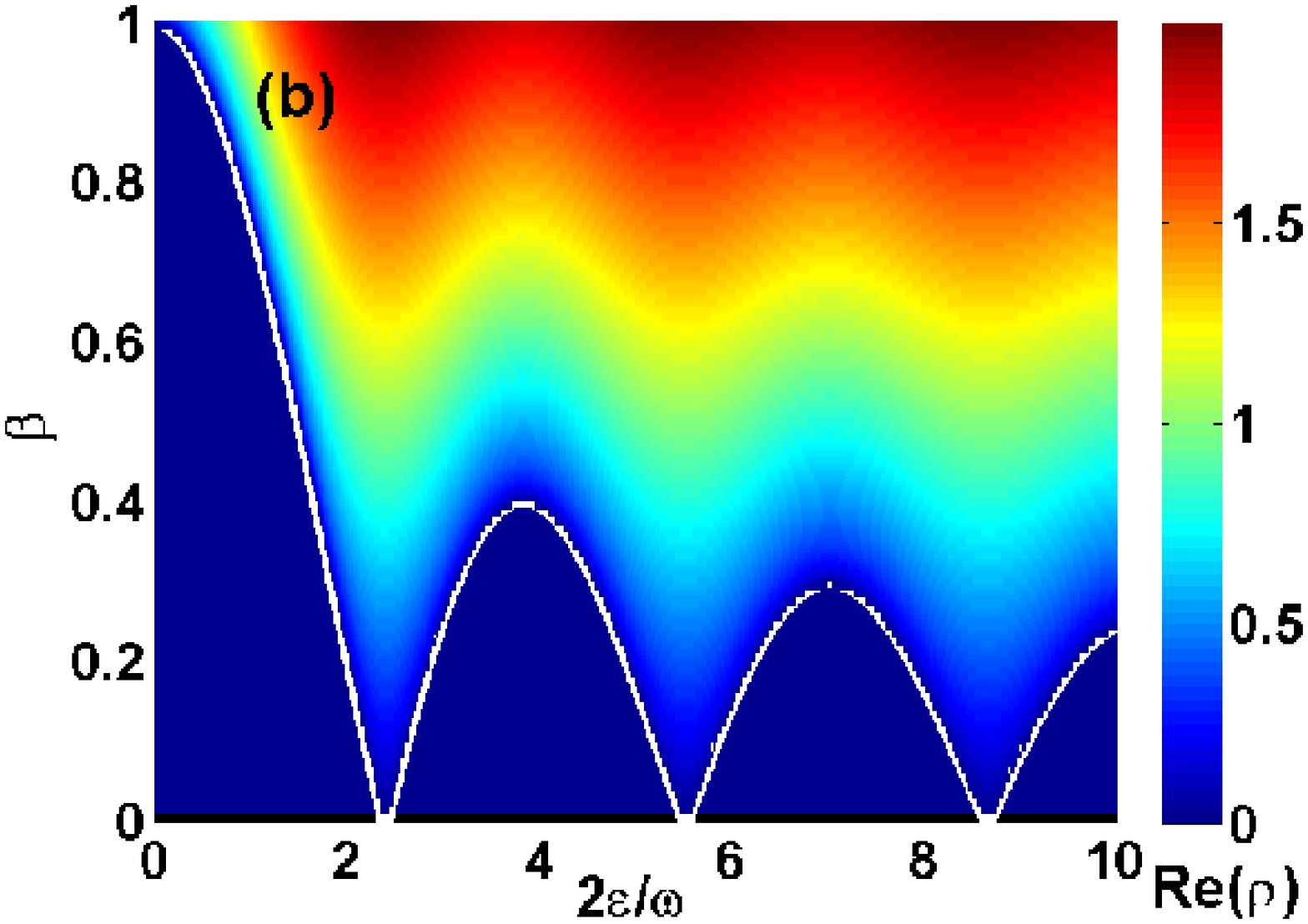}
\includegraphics[height=1.3in,width=2.2in]{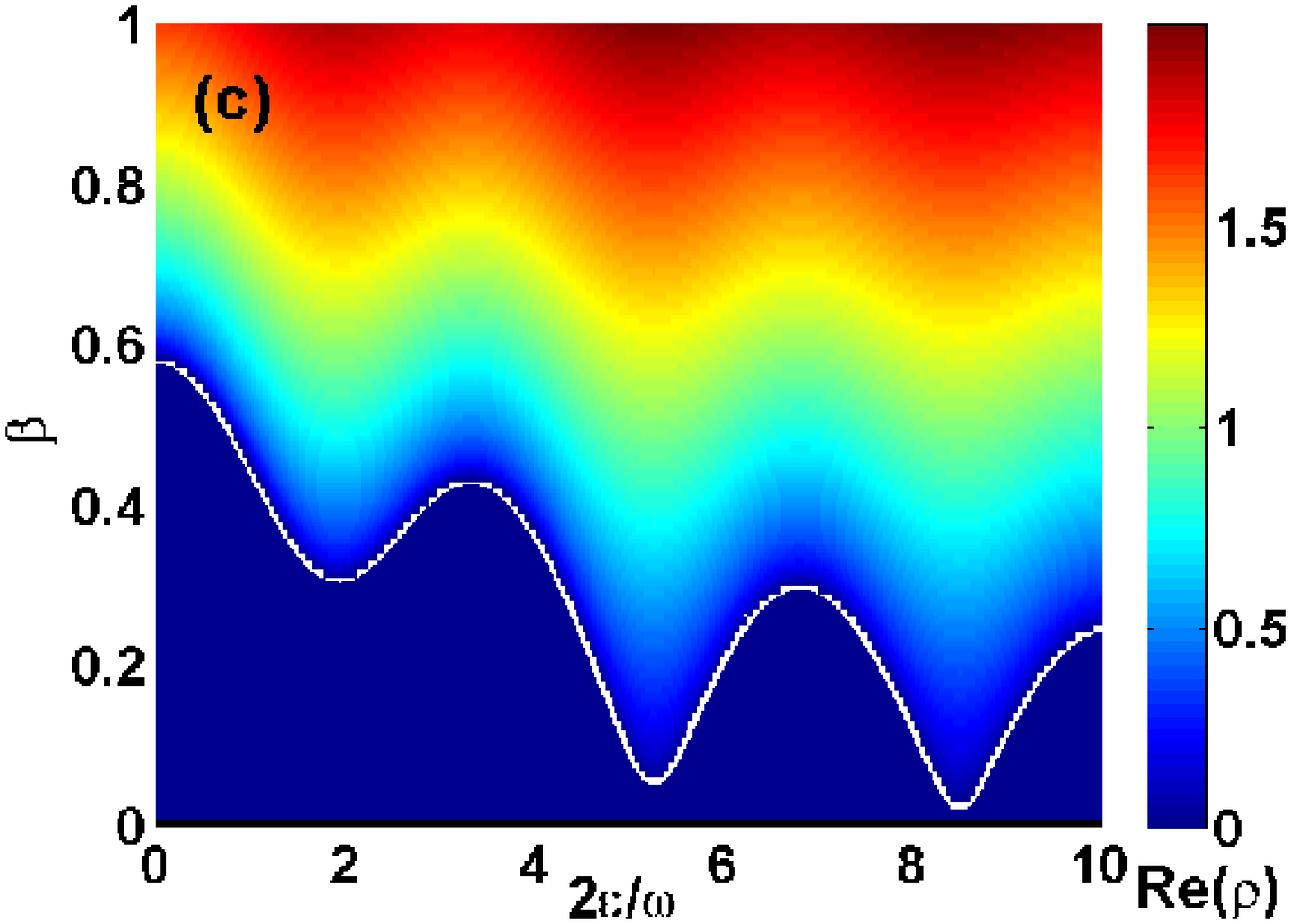}
\includegraphics[height=1.3in,width=2.2in]{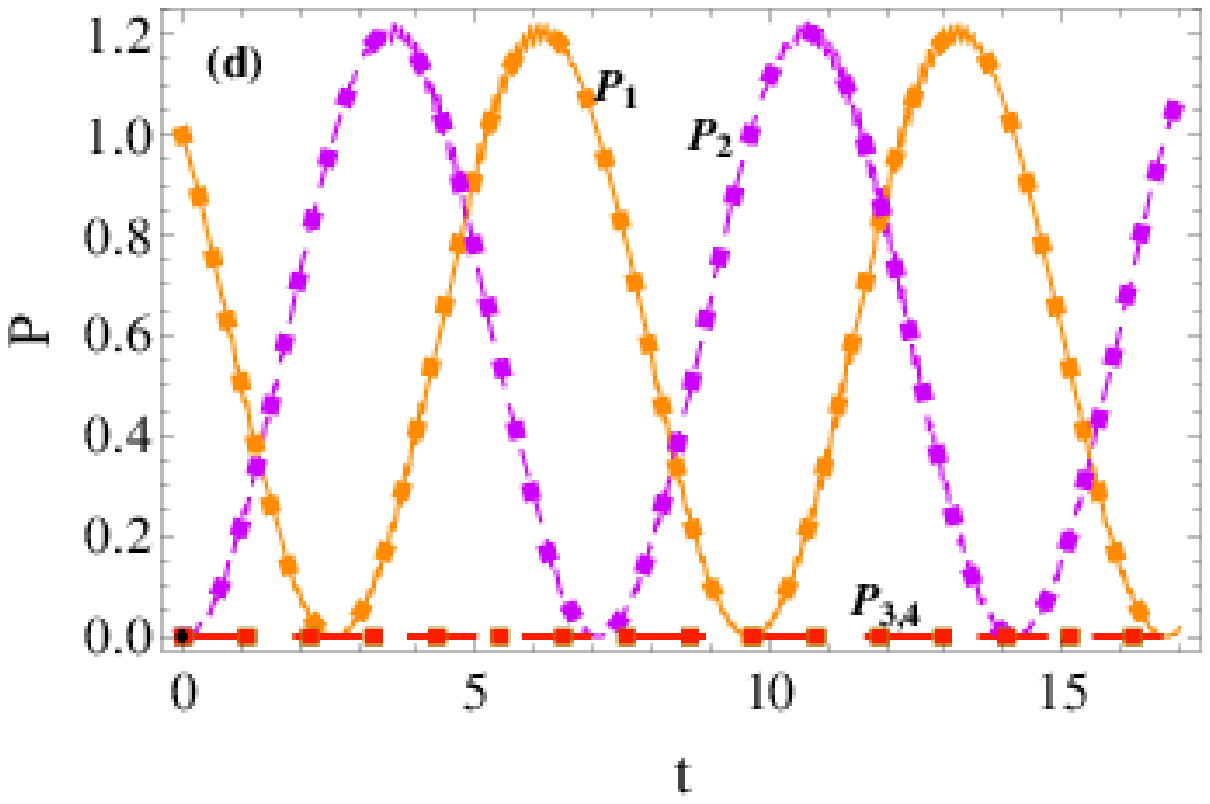}
\includegraphics[height=1.3in,width=2.2in]{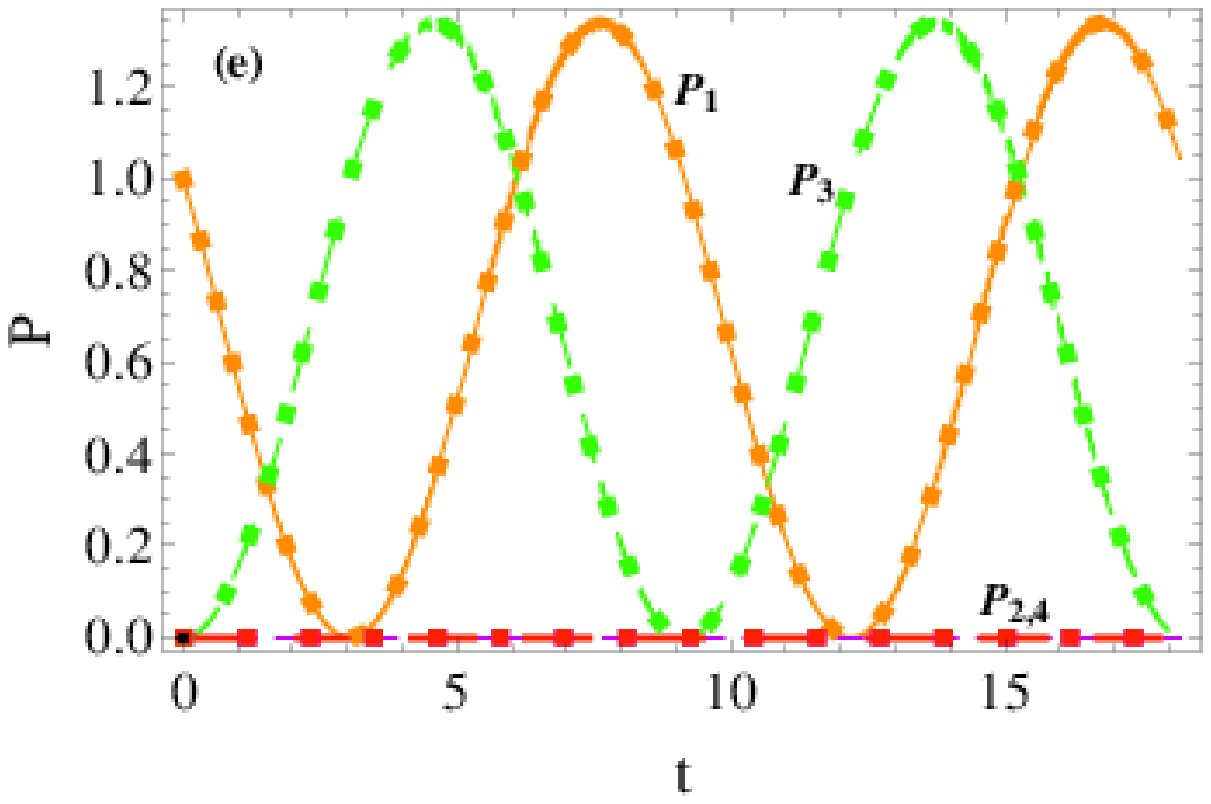}
\includegraphics[height=1.3in,width=2.2in]{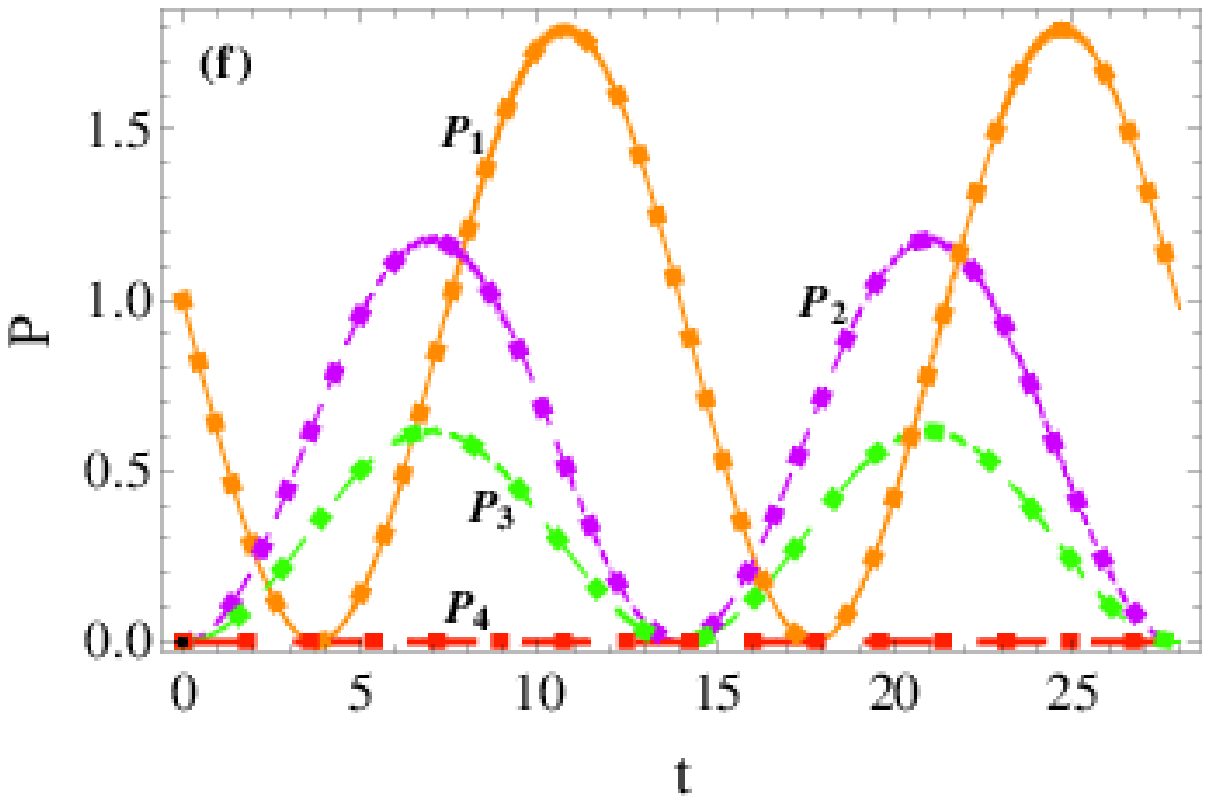}
\caption{\scriptsize{(Color online) Top panels depict the real part Re($\rho$) as a function of $\beta$ and $2\varepsilon/\omega$ for (a) $\lambda=0.5$, (b) $\lambda=1$ and (c) $\lambda=1.7$. Here, Re($\rho$) corresponds to the imaginary part of complex quasienergy with $E_1=-E_2=-E_3=E_4=\frac{1}{2} i \rho$. Bottom panels show the time-evolution curves of probabilities $P_{k}=|a_{k}(t)|^{2}$ with $\beta=0.2$, and (d) $\lambda=0.5$, $2\varepsilon/\omega=3$; (e) $\lambda=1$, $2\varepsilon/\omega=4$; (f) $\lambda=1.7$, $2\varepsilon/\omega=7$, starting the system with a spin-up particle in the right well. The other parameters are the same as those of Fig.~1.}}
\end{figure*}

(2) odd $\Omega/\omega$

When $\Omega/\omega$ is odd, we have $E_{1,2}=\pm \frac{1}{2} i \rho_{-}$ and $E_{3,4}=\mp \frac{1}{2} i \rho_{+}$ with $\rho_{\pm}=2 \sqrt{\beta^2-(|J_{0}|\pm |J_{\frac{\Omega}{\omega}}|)^2}$. Applying the condition $J_0=0$ or / and $J_{\frac{\Omega}{\omega}}=0$ will lead to the simple case: $\rho_{+}=\rho_{-}$, $E_1=E_4$, and $E_2=E_3$. We write $\rho_{+}+\rho_{-}$=Re($\rho_{+}+\rho_{-}$)+ \emph{i} Im($\rho_{+}+\rho_{-}$) and know that Re($\rho_{+}+\rho_{-}$)=0 will make Re($\rho_{-}$)=0 and  Re($\rho_{+}$)=0 hold simultaneously.
According to \emph{Case A} of stability analysis, we can deduce that when the system parameters satisfy Re($\rho_{+}+\rho_{-}$)=0, quasienergies $E_p$ become all real and the spin-dependent tunneling is stable. In fact, if Re($\rho_{-}$)=0 holds, both Re($\rho_{+}$)=0 and Re($\rho_{+}+\rho_{-}$)=0 will naturally hold.
As illustrated in Figs.~4-6, in all the stability diagrams with odd $\Omega/\omega$, the boundary between the stable (Re($\rho_{+}+\rho_{-}$)=0) and unstable (Re($\rho_{+}+\rho_{-}$)$\neq$ 0) regimes is displayed by the white curves, which can be obtained from the relation $\beta^2-(|J_{0}|-|J_{\frac{\Omega}{\omega}}|)^2=0$. In Figs.~4 (a)-(c), we take $\nu=1$, $\Omega=\omega=50$ (hence $\Omega/\omega=1$) to show Re($\rho_{+}+\rho_{-}$) as a function of
$2\varepsilon/\omega$ and $\lambda$, with different gain-loss strengths (a) $\beta=0.2$, (b) $\beta=0.45$, and (c) $\beta=0.6$. It is shown that only \emph{discrete} stable parameter regions exist for the case of odd $\Omega/\omega$. The basic explanation of this phenomenon is as follows: the value of $(|J_{0}|-|J_{\frac{\Omega}{\omega}}|)^2$ can become zero as either $2\varepsilon/\omega$ or $\lambda$ varies continuously, such that there does not exist a non-zero gain-loss coefficient $\beta$ below which $\rho_{-}$ is always a purely imaginary number (the system is stable) over the continuous range of  system parameters.
From Fig.~4 (a), we see that the sizes of discrete stable parameter regions decrease with the  increase of $2\varepsilon/\omega$. It can also be observed that for small driving parameters $2\varepsilon/\omega$, the stable spin-flipping tunneling with $\lambda=m+0.5$ (m=0, 1, ...) can not happen, whereas the stable generalized Rabi oscillation without spin flipping ($\lambda=m$) can happen, which is demonstrated by time-evolution
of the probabilities in Fig.~4 (d). By comparing Fig.~4 (b) with Fig.~4 (a),  we can see that the number of discrete stable parameter
regions  is getting smaller as the gain-loss strength increases.
From Fig.~4 (b), we also find that the adjustable values of  $2\varepsilon/\omega\in [1.03109, 2.67221]$ corresponding to the stable spin-flipping tunneling with $\lambda=m+0.5$ are not greater than all the adjustable values $2\varepsilon/\omega\in [0, 1.60947]$ corresponding to the stable non-spin-flipping tunneling with $\lambda=m$. This result is distinct from that of Fig. 1 (b) in the even $\Omega/\omega$ case. In order to support this finding, for instance, we set $\lambda=1$ and $2\varepsilon/\omega=1.4 > 1.03109$ to illustrate the time evolution of stable non-spin-flipping tunneling, as shown in Fig.~4 (e).
In Fig.~4 (c) with $\beta=0.6$, we observe that the stable spin-flipping tunneling with $\lambda=m+0.5$ vanishes, but the stable generalized non-spin-flipping Rabi oscillation with $\lambda=m$ still exists (e.g., see the time-evolution of probabilities in Fig.~4 (f)). Based on these numerical results, a conclusion can be drawn: for odd $\Omega/\omega$, the stable spin-flipping tunneling  is suppressed preferentially as the gain-loss strength is increased. This conclusion is basically the same as that  reached for the case of even $\Omega/\omega$.

\begin{figure*}[htp]\center
\includegraphics[height=1.3in,width=2.2in]{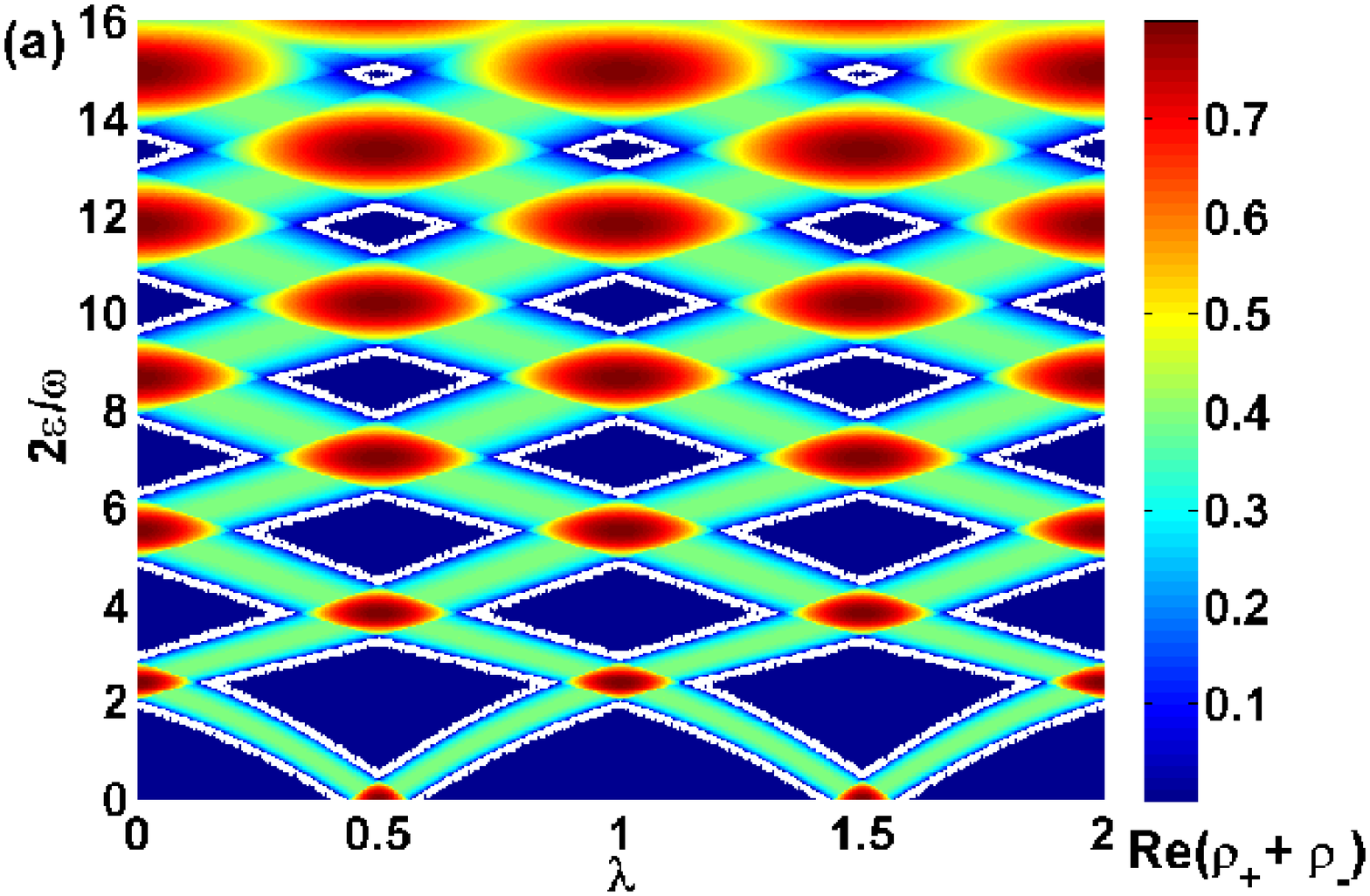}
\includegraphics[height=1.3in,width=2.2in]{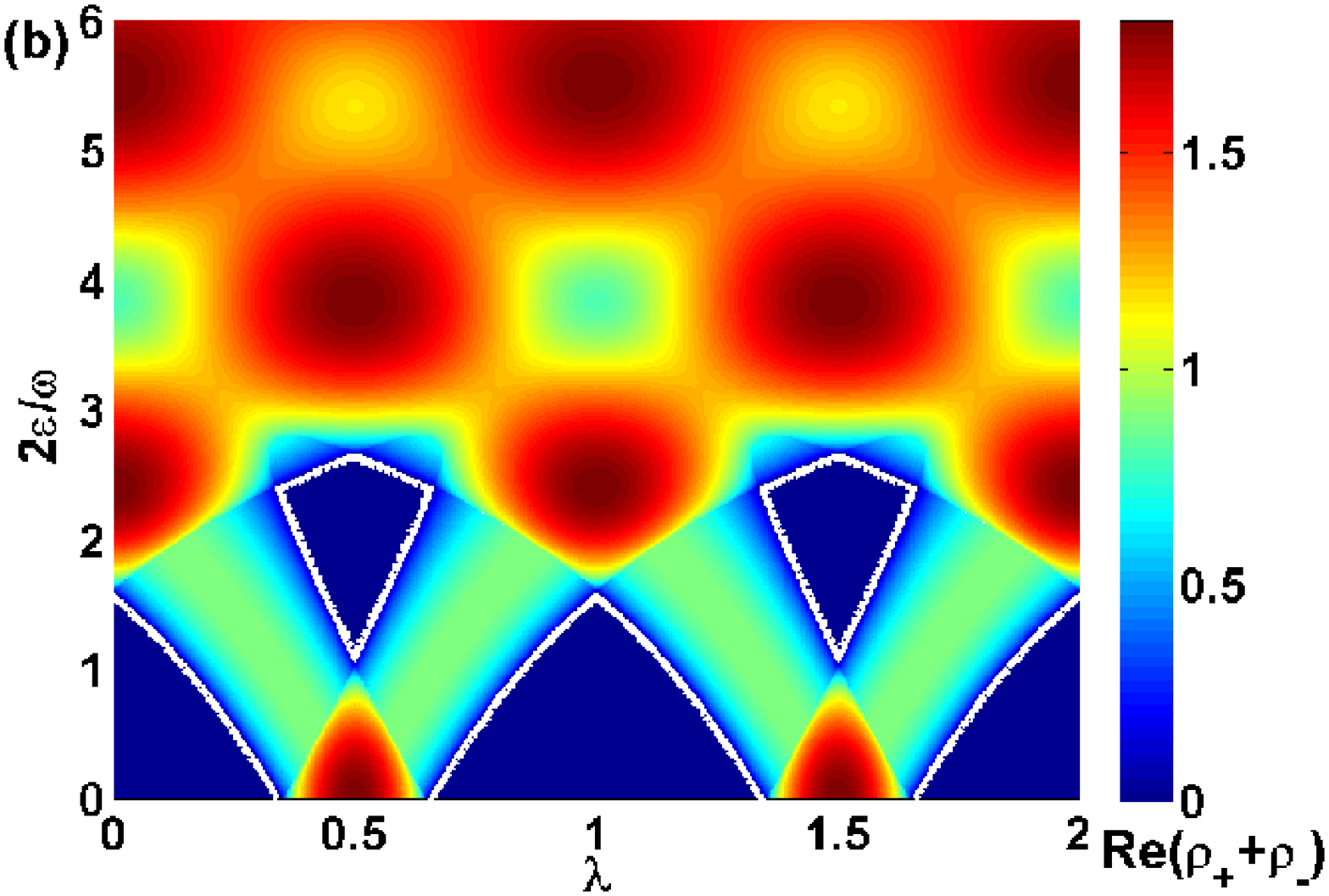}
\includegraphics[height=1.3in,width=2.2in]{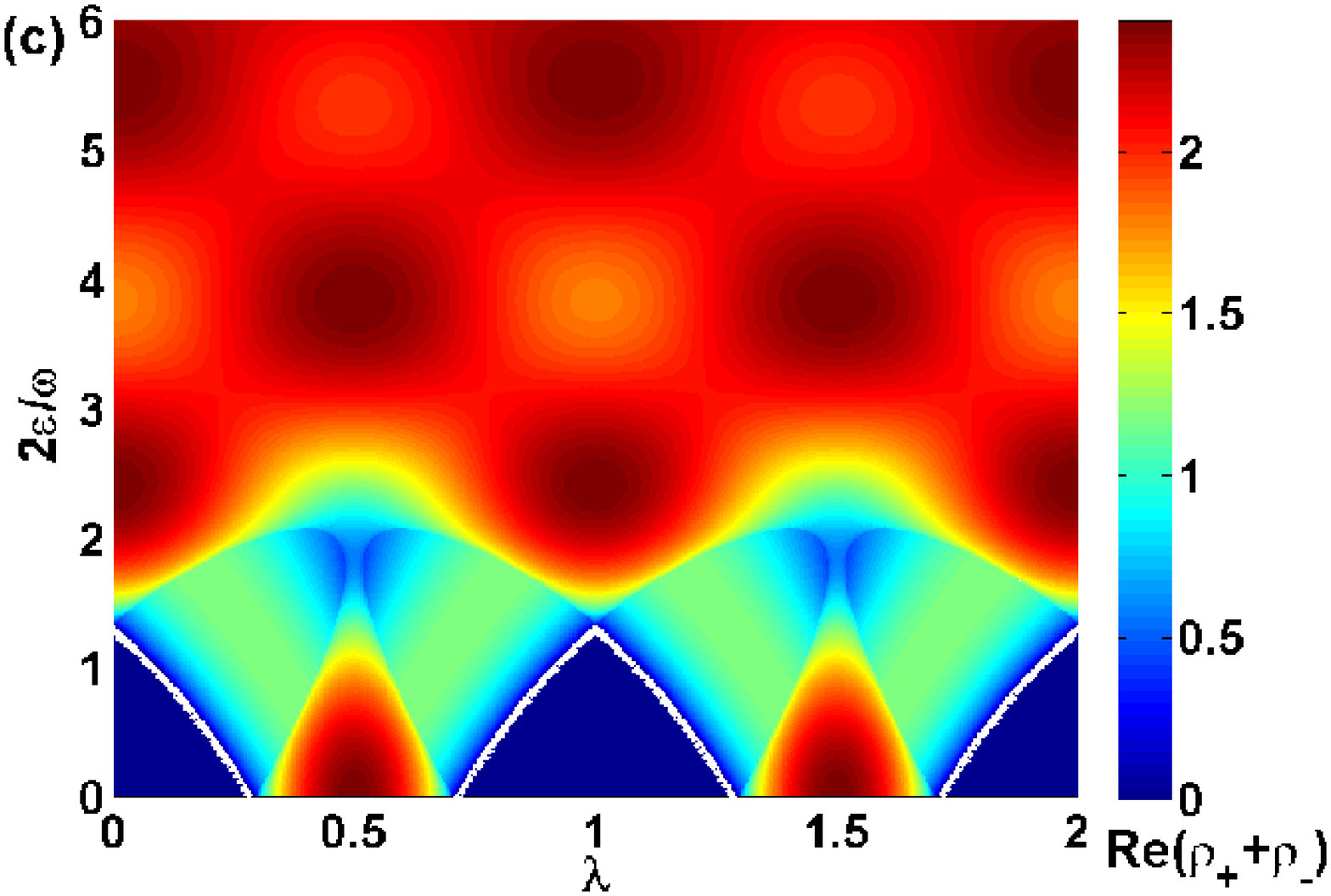}
\includegraphics[height=1.3in,width=2.2in]{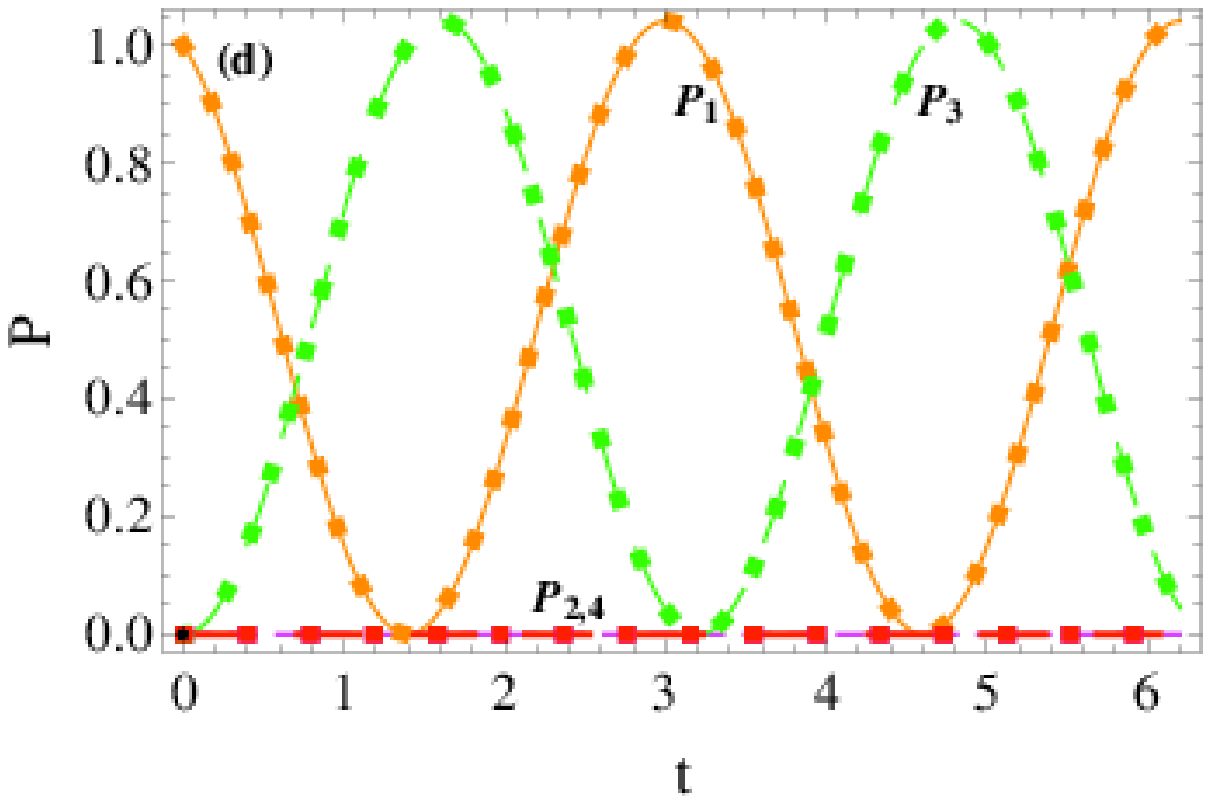}
\includegraphics[height=1.3in,width=2.2in]{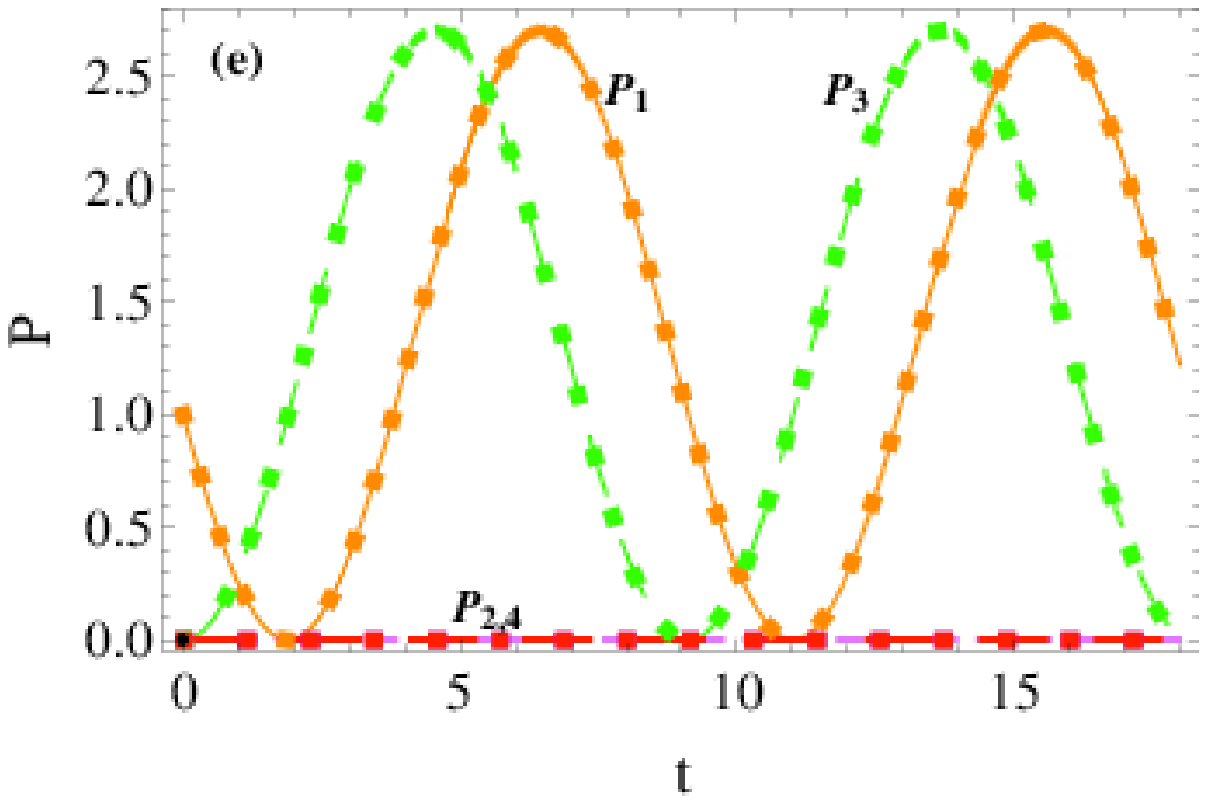}
\includegraphics[height=1.3in,width=2.2in]{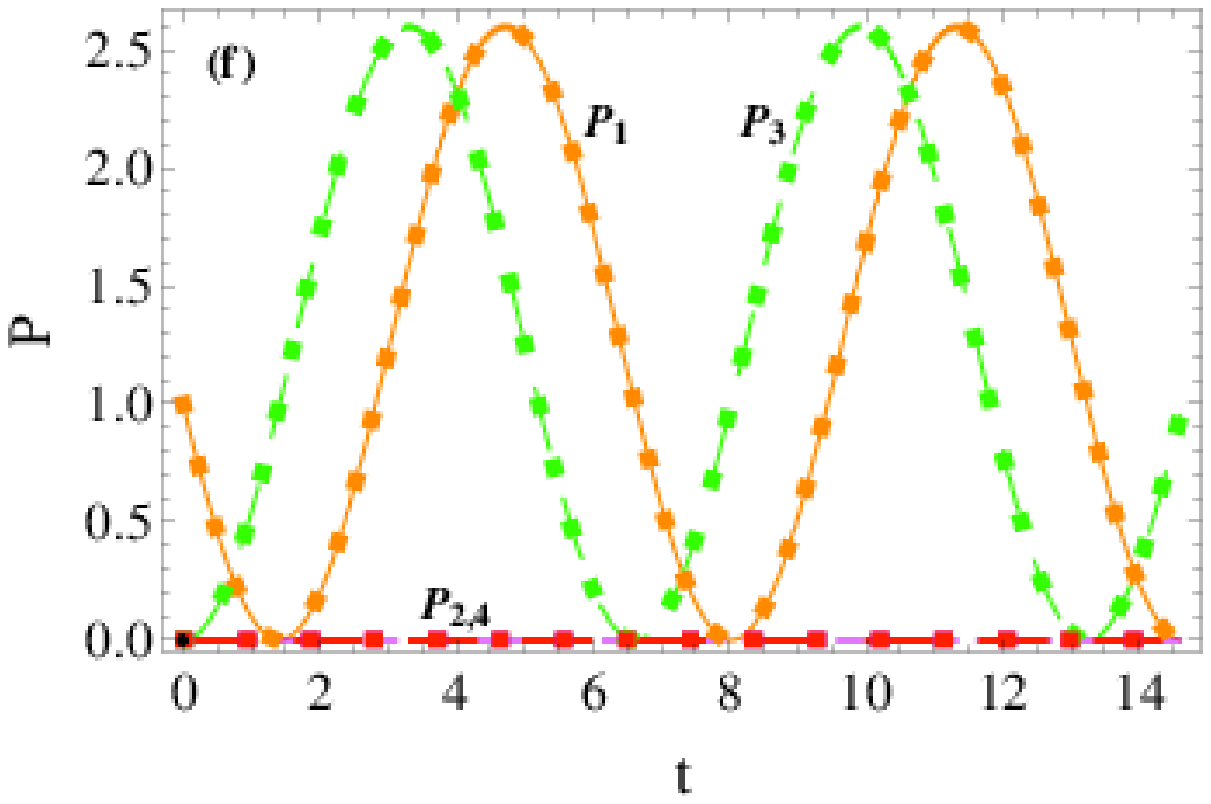}
\caption{\scriptsize{(Color online) Top panels depict the real part Re($\rho_{+}+\rho_{-}$) as a function of $2\varepsilon/\omega$ and $\lambda$ for (a) $\beta=0.2$, (b) $\beta=0.45$ and (c) $\beta=0.6$. Here, Re($\rho_{+}+\rho_{-}$) is associated with the imaginary part of complex quasienergy with $E_{1,2}=\pm \frac{1}{2} i \rho_{-}$ and $E_{3,4}=\mp \frac{1}{2} i \rho_{+}$. Bottom panels show the time-evolution curves of probabilities $P_{k}=|a_{k}(t)|^{2}$ for  $\lambda=1$, and (d) $\beta=0.2$, $2\varepsilon/\omega=0$; (e) $\beta=0.45$, $2\varepsilon/\omega=1.4$; (f) $\beta=0.6$, $2\varepsilon/\omega=1$, starting the system with a spin-up particle in the right well. The other parameters are chosen as $\nu=1$ and $\Omega=\omega=50$. In (a)-(c), the white curves are the boundary between stable (Re($\rho_{+}+\rho_{-}$)=0) and unstable (Re($\rho_{+}+\rho_{-}$)$\neq$ 0) regimes.}}
\end{figure*}

In Figs.~5 (a)-(c), we show the real part Re($\rho_{+}+\rho_{-}$) as a function of
$\beta$ and $\lambda$ with (a) $2\varepsilon/\omega=3.8317$, (b) $2\varepsilon/\omega=2.4048$, and (c) $2\varepsilon/\omega=1$, for the case of $\Omega/\omega=1$. Similarly, only
discrete stable parameter regions are observed. As the ratio $2\varepsilon/\omega=3.8317$ is taken to satisfy $\mathcal{J}_{1}(3.8317)=0$ ($J_1=0$), Fig.~5 (a) indicates the stability diagram for non-spin-flipping tunneling, where
stable generalized Rabi oscillation without spin flipping(e.g., see Fig.~5 (d)) is observed when the parameters are suitably chosen. In this case, the maximum adjustable value of gain-loss strength $\beta$ for achieving stable non-spin-flipping tunneling with $\lambda=m$ is given by $\beta_{max}=|J_0|=\nu |\mathcal{J}_0(3.8317)|\approx 0.40276$. In Fig.~5 (b), the ratio $2\varepsilon/\omega=2.4048$ satisfying $\mathcal{J}_{0}(2.4048)=0$ ($J_0=0$) is associated with the stability diagram for spin-flipping tunneling, and stable generalized Rabi oscillation with spin flipping (e.g., see Fig.~5 (e)) is presented for the appropriately chosen parameters. The maximum adjustable value of gain-loss strength $\beta$ for realizing stable spin-flipping quantum tunneling with $\lambda=m+0.5$ is $\beta'_{max}=|J_{\frac{\Omega}{\omega}}|=\nu |\mathcal{J}_{1}(2.4048)|\approx 0.51915$. As can be seen in Fig.~5 (c), for the general case of $2\varepsilon/\omega=1$, where  $\mathcal{J}_{0}(2\varepsilon/\omega)\neq 0$ and $\mathcal{J}_{1}(2\varepsilon/\omega)\neq 0$,  the discrete stable parameter region can be divided into two categories, corresponding to the two situations in which the biggest adjustable values of $\beta$ for realizing stable spin tunneling are located at $\lambda=m$ and $\lambda=m+0.5$ respectively. With the gain-loss strength $\beta$ taken in the stable parameter regions of Fig.~5 (c) and the SO coupling strength taken of $\lambda=m$ or $\lambda=m+0.5$, stable generalized Rabi oscillation without spin flipping or with spin flipping will occur respectively.  As excepted, when other values of $\lambda$ are selected in the stable parameter regions, stable generalized spin-flipping and non-spin-flipping Rabi oscillation will simultaneously occur, as illustrated in  Fig.~5 (f)).

\begin{figure*}[htp]\center
\includegraphics[height=1.3in,width=2.2in]{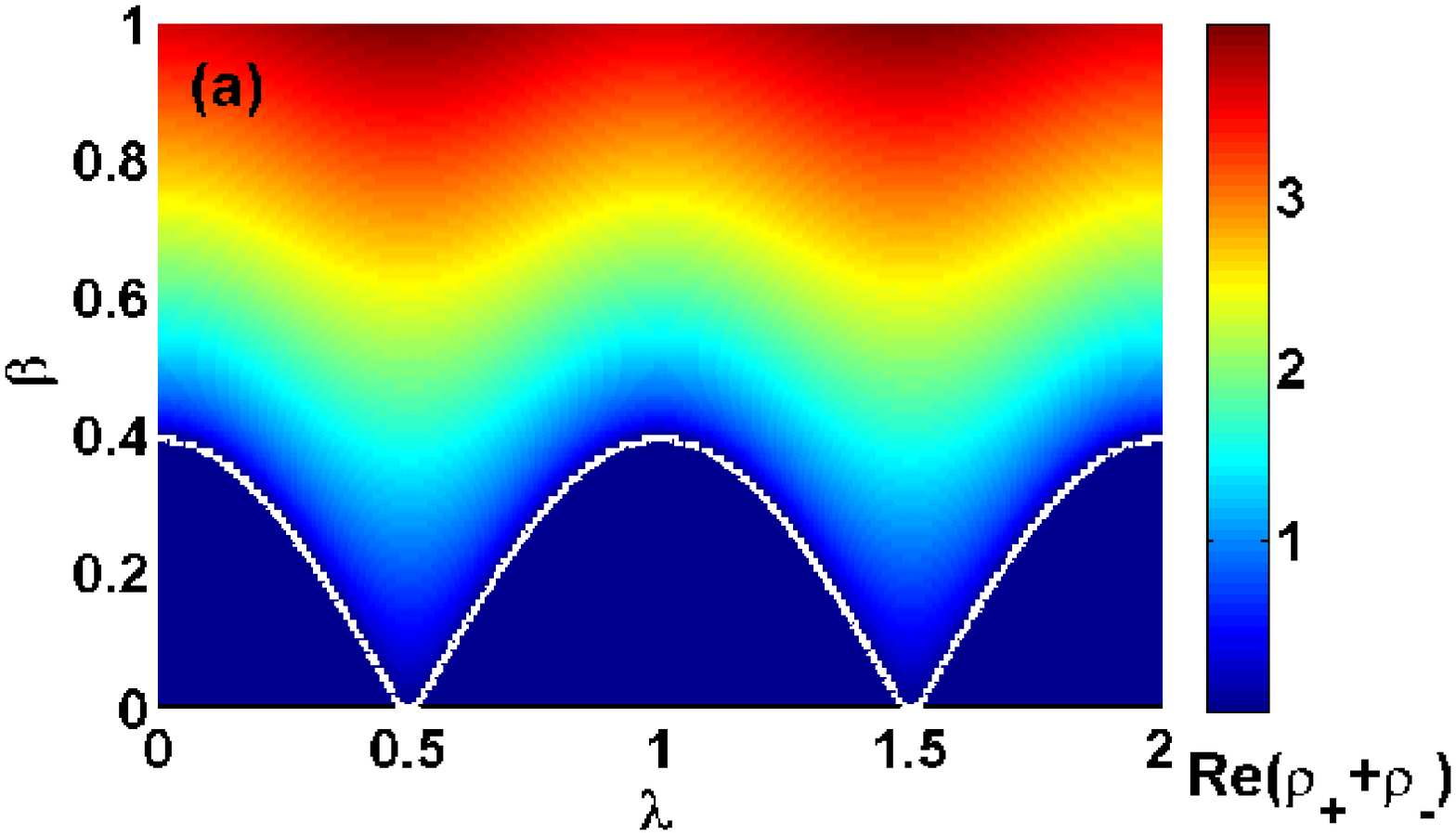}
\includegraphics[height=1.3in,width=2.2in]{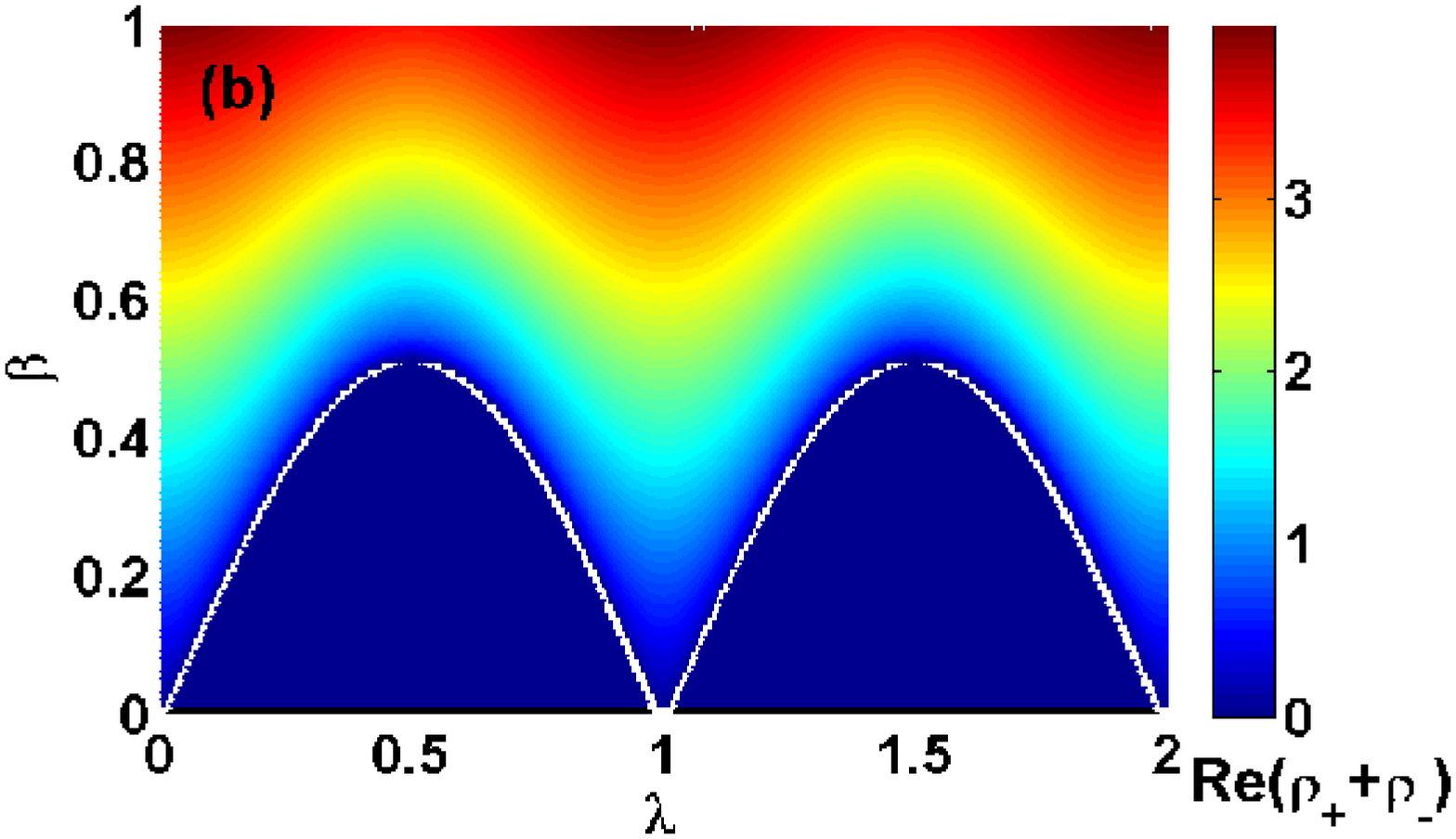}
\includegraphics[height=1.3in,width=2.2in]{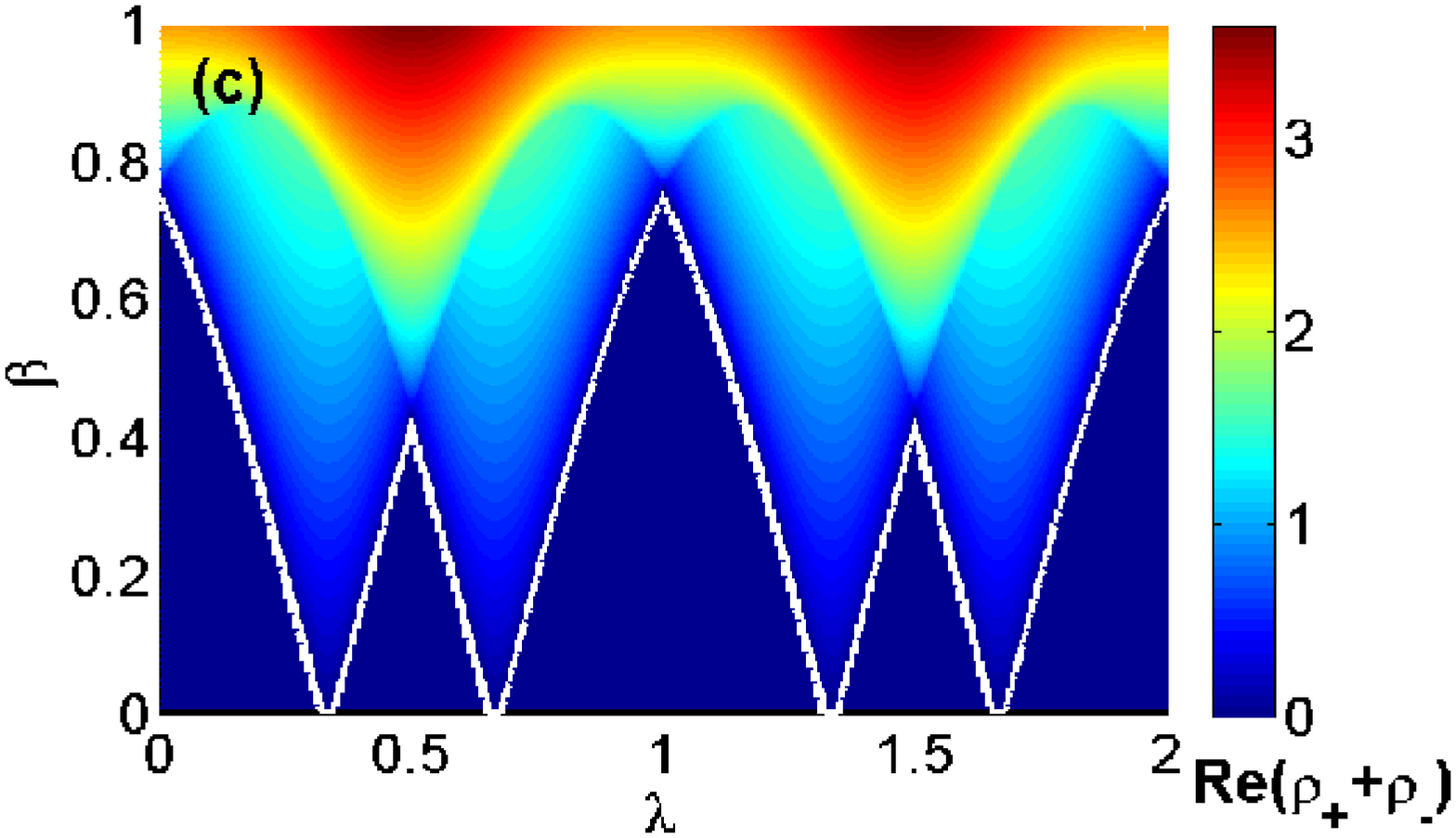}
\includegraphics[height=1.3in,width=2.2in]{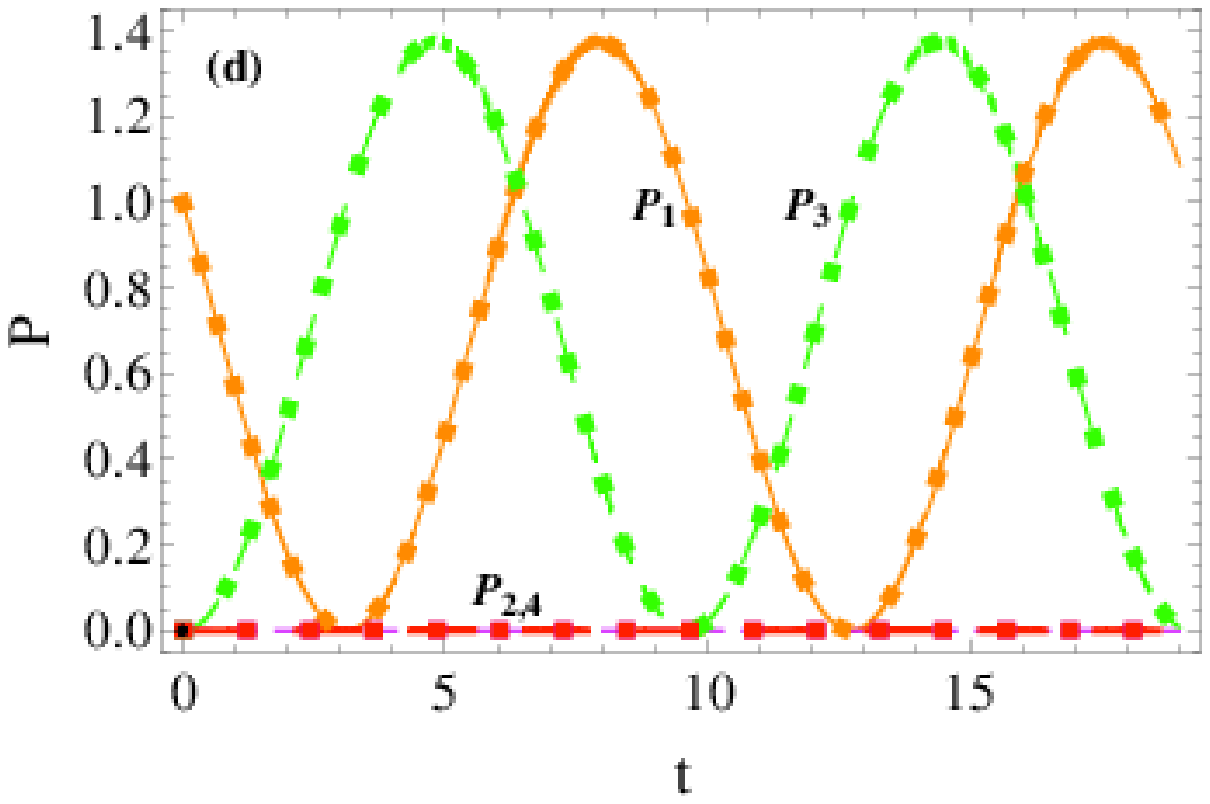}
\includegraphics[height=1.3in,width=2.2in]{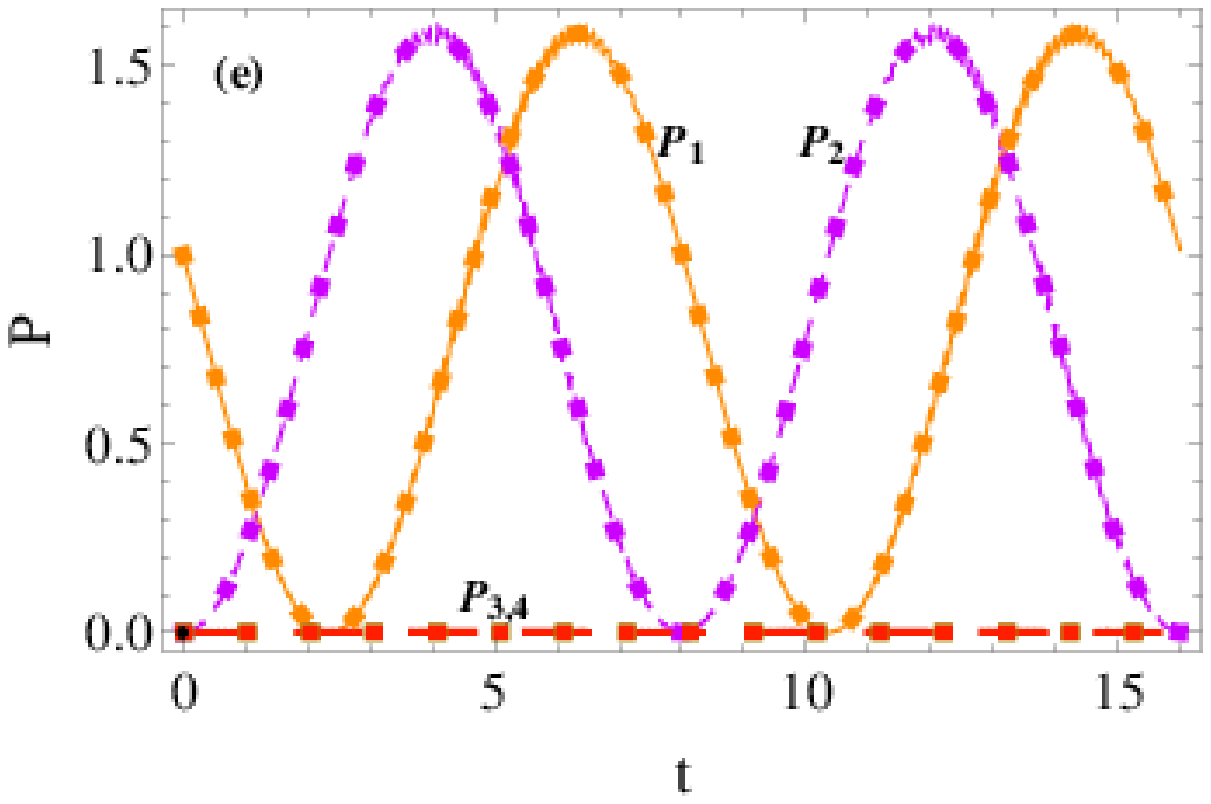}
\includegraphics[height=1.3in,width=2.2in]{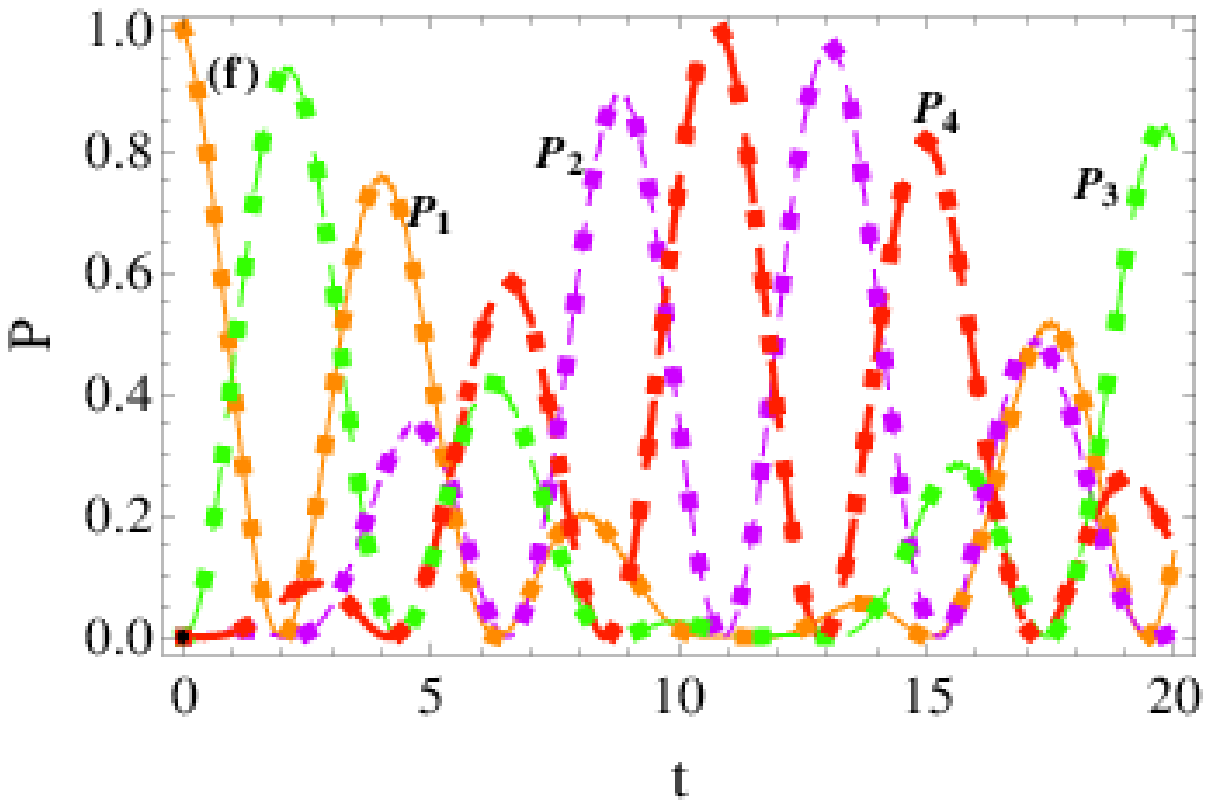}
\caption{\scriptsize{(Color online) Top panels depict the real part Re($\rho_{+}+\rho_{-}$) as a function of $\beta$ and $\lambda$ for (a) $2\varepsilon/\omega=3.8317$, (b) $2\varepsilon/\omega=2.4048$ and (c) $2\varepsilon/\omega=1$. Here, Re($\rho_{+}+\rho_{-}$) is associated with  the imaginary part of complex quasienergy with $E_{1,2}=\pm \frac{1}{2} i \rho_{-}$ and $E_{3,4}=\mp \frac{1}{2} i \rho_{+}$. Bottom panels show the time-evolution curves of probabilities $P_{k}=|a_{k}(t)|^{2}$ for (d) $2\varepsilon/\omega=3.8317$, $\beta=0.2$, $\lambda=0.9$; (e) $2\varepsilon/\omega=2.4048$, $\beta=0.3$, $\lambda=0.4$; (f) $2\varepsilon/\omega=1$, $\beta=0.1$, $\lambda=0.1$, starting the system with a spin-up particle in the right well. The other parameters are the same as those of Fig.~4. In (a)-(c), the white curves are the boundary between Re($\rho_{+}+\rho_{-}$)=0 and Re($\rho_{+}+\rho_{-}$)$\neq$ 0. The stable parameter region is under the white curve.}}
\end{figure*}

In Figs.~6 (a)-(b), we show the real part Re($\rho_{+}+\rho_{-}$) as a function of
$\beta$ and $2\varepsilon/\omega$ with (a) $\lambda=0.5$ and (b) $\lambda=1.7$, for the case of  $\Omega/\omega=1$. As before, only the \emph{discrete} stable parameter regions are found. In Fig.~6 (a), we take the SO coupling strength $\lambda=0.5$ which is associated with $J_0=0$. When the system parameters are chosen in the discrete stable regions of Fig.~6 (a), stable generalized Rabi oscillation with spin flipping can be observed (e.g., see Fig.~6 (c)). In Fig.~6 (a), the crossing points of the white boundary curve and transverse axis are precisely aligned with the roots of $\mathcal{J}_{1}(\frac{2\varepsilon}{\omega})=0$, at which CDT will occur (not shown). In Fig.~6 (b), for the general case of $\lambda=1.7$, with the parameters in the discrete stable parameter regions and $2\varepsilon/\omega$  obeying $\mathcal{J}_{0}(\frac{2\varepsilon}{\omega})=0$ or $\mathcal{J}_{1}(\frac{2\varepsilon}{\omega})=0$, the stable generalized Rabi oscillation with spin flipping or without spin flipping will occur  respectively. In Fig.~6 (d), we have shown the stable generalized Rabi oscillation without spin flipping for the suitable chosen parameters.
Meanwhile, the adjustable values of $\beta$ for realizing stable spin tunneling are shown to exhibit a sequence of peaks. These peaks are situated at the zeros of $\mathcal{J}_{1}(\frac{2\varepsilon}{\omega})$ and $\mathcal{J}_{0}(\frac{2\varepsilon}{\omega})$ alternately, which stand for the non-spin-flipping tunneling and  spin-flipping tunneling.
The situation for $\lambda=1$ is similar to the even $\Omega/\omega$ case as shown in Fig.~3 (b) and it is not discussed here.

\begin{figure}[htp]\center
\includegraphics[height=1.3in,width=1.6in]{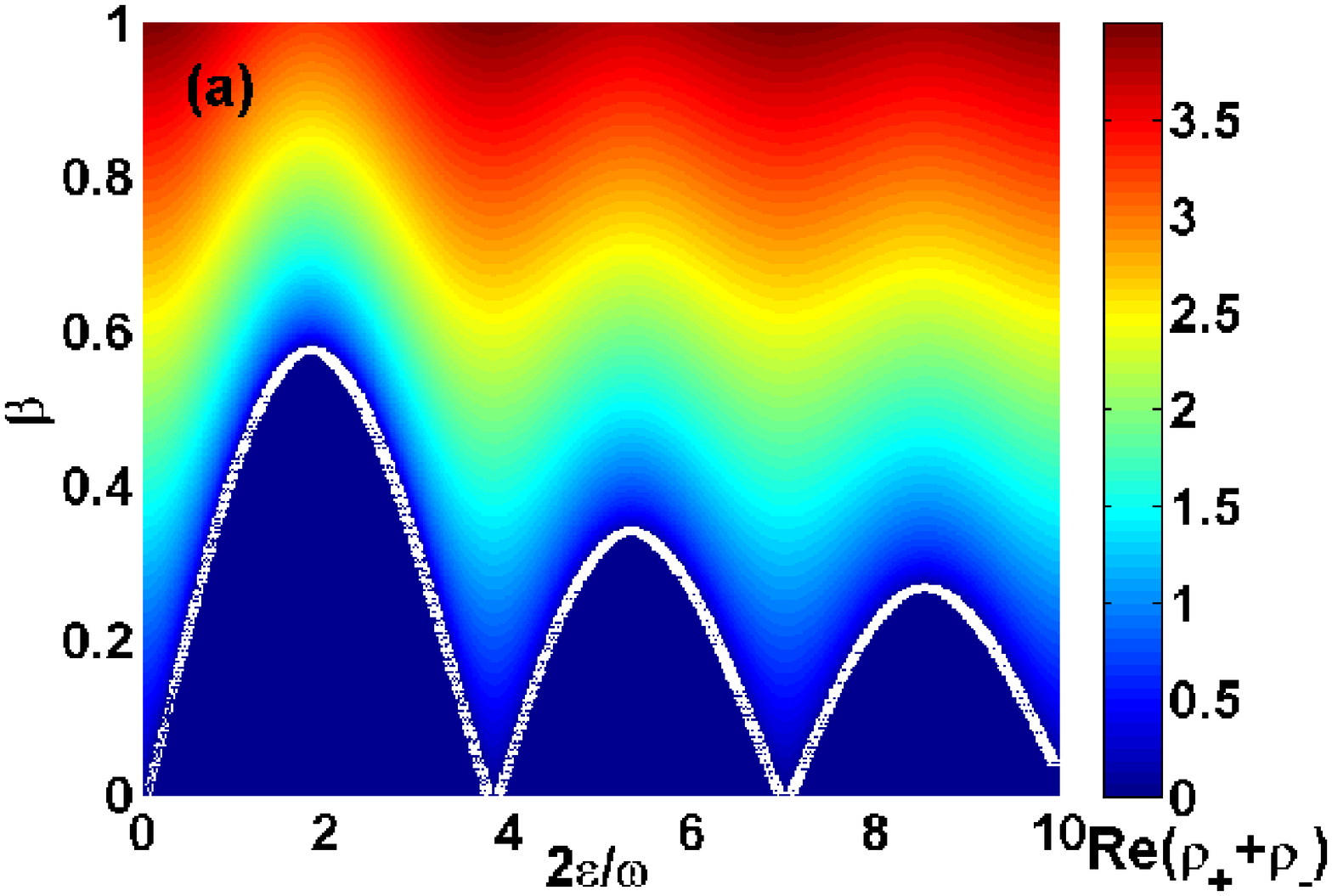}
\includegraphics[height=1.3in,width=1.6in]{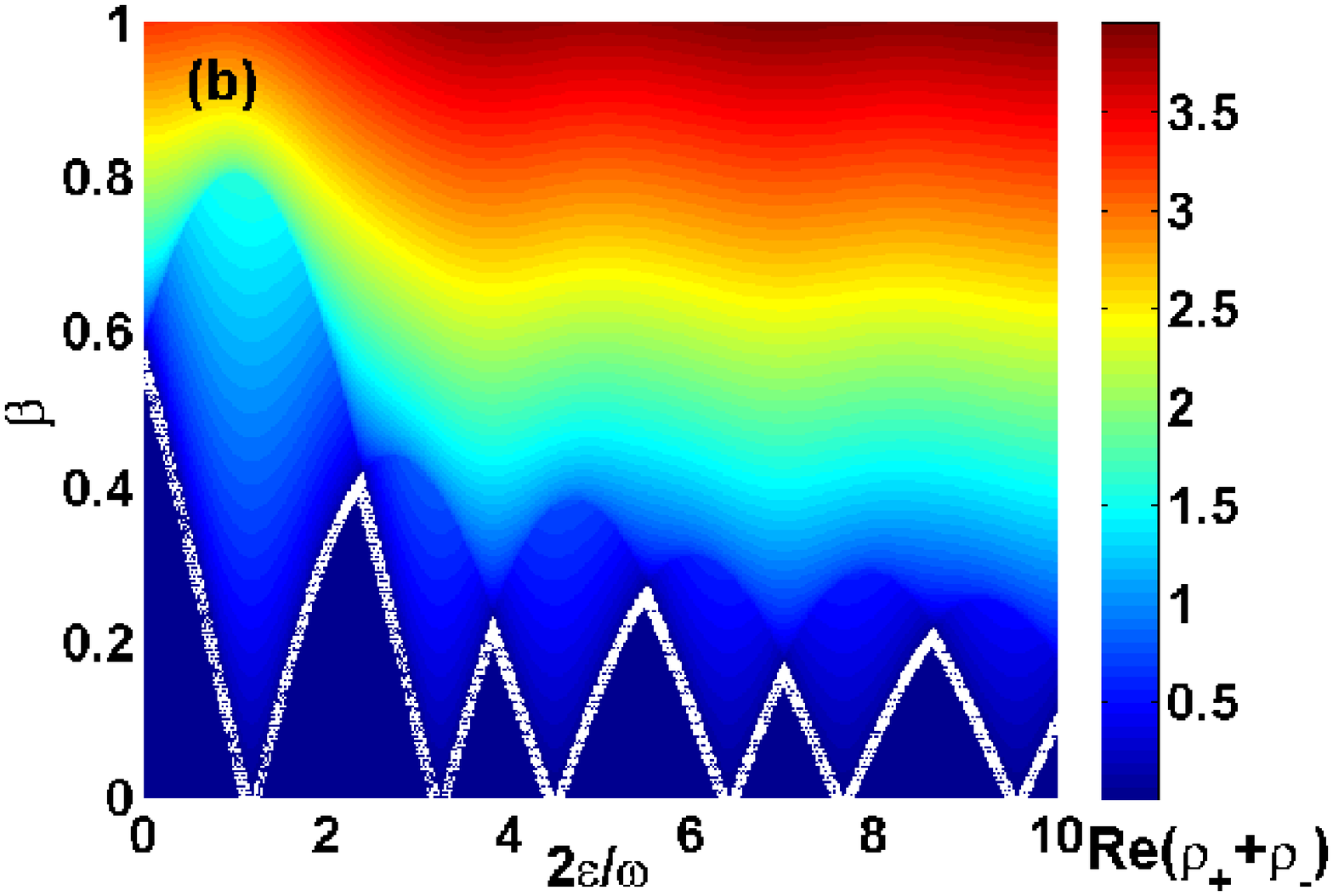}
\includegraphics[height=1.3in,width=1.6in]{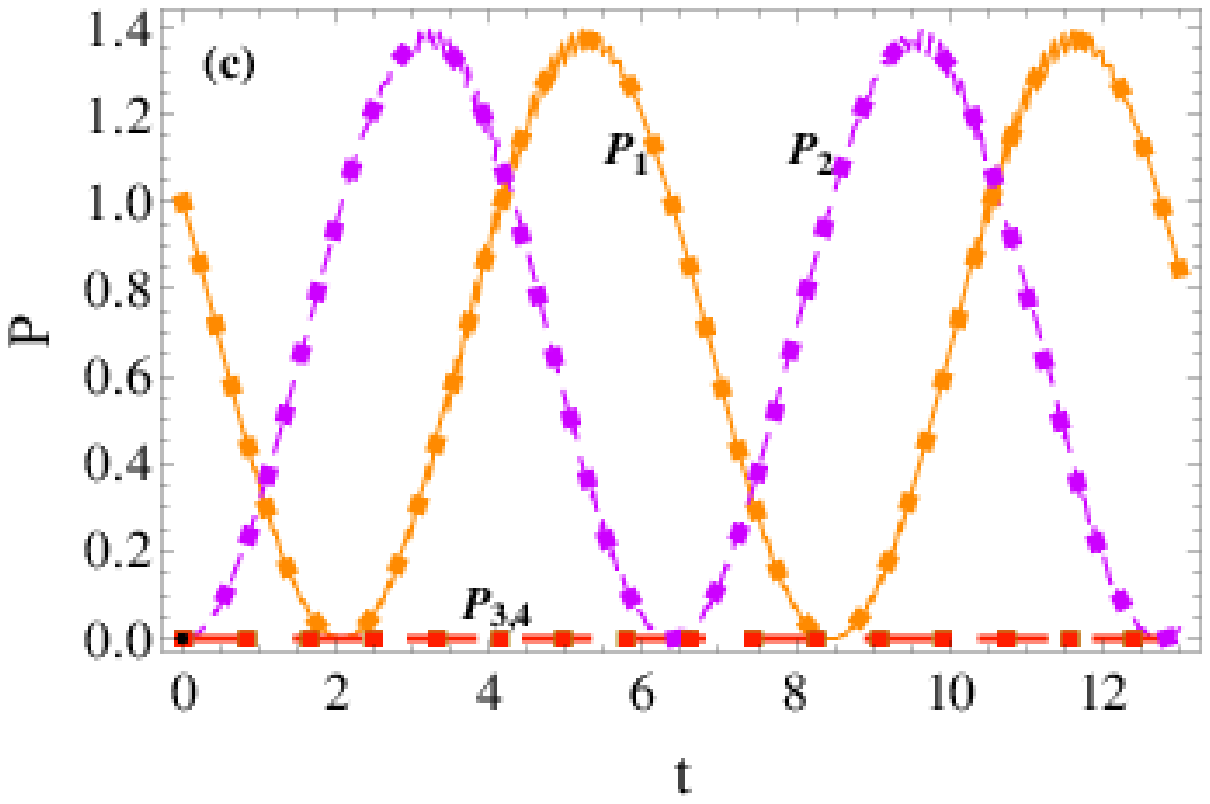}
\includegraphics[height=1.3in,width=1.6in]{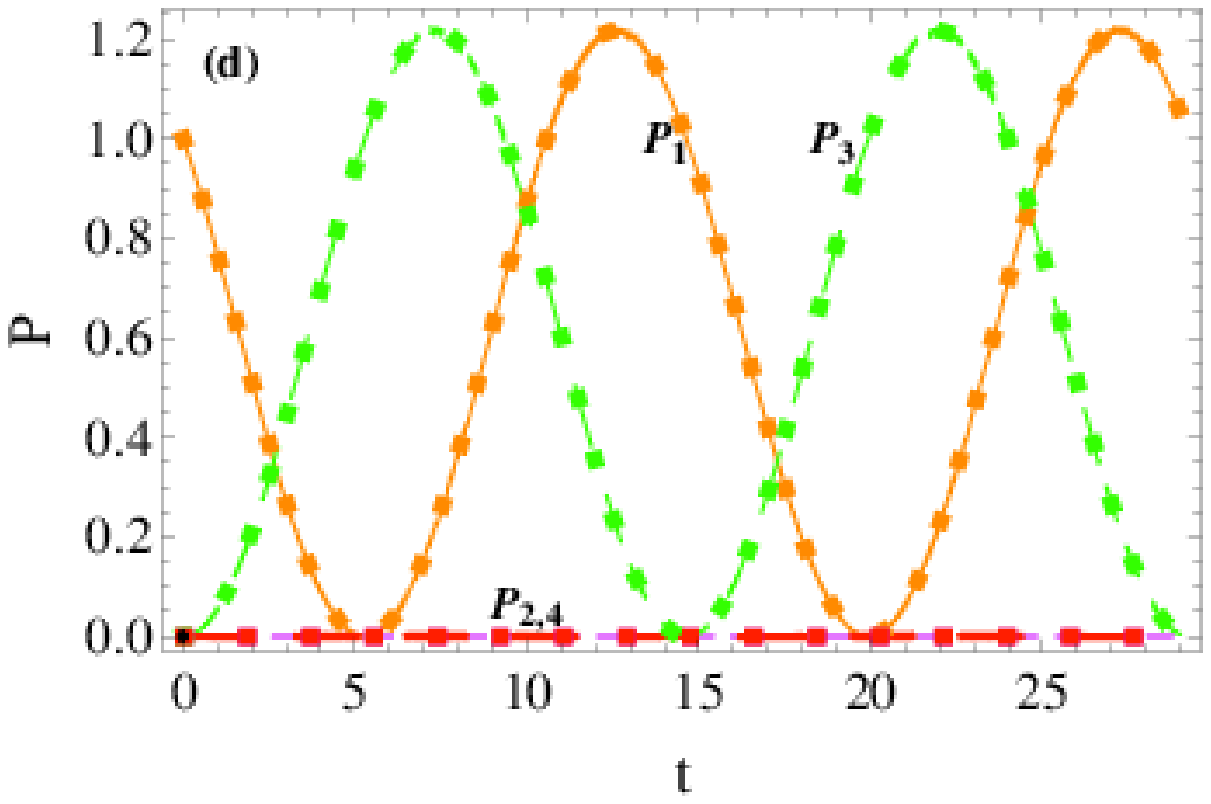}
\caption{\scriptsize{(Color online) Top panels the real part Re($\rho_{+}+\rho_{-}$) as a function of $\beta$ and $2\varepsilon/\omega$ for (a)
$\lambda=0.5$ and (b) $\lambda=1.7$. Here, Re($\rho_{+}+\rho_{-}$) is associated with the imaginary part of complex quasienergy with $E_{1,2}=\pm \frac{1}{2} i \rho_{-}$ and $E_{3,4}=\mp \frac{1}{2} i \rho_{+}$. Bottom panels show the time-evolution curves of probabilities $P_{k}=|a_{k}(t)|^{2}$ for (c) $\lambda=0.5$, $\beta=0.3$, $2\varepsilon/\omega=2$; (d) $\lambda=1.7$, $\beta=0.1$, $2\varepsilon/\omega=3.8317$, starting the system with a spin-up particle in the right well. In (a)-(c), the white curves are the boundary between stable (Re($\rho_{+}+\rho_{-}$)=0) and unstable (Re($\rho_{+}+\rho_{-}$)$\neq$ 0) regimes. The other parameters are the same as those of Fig.~4.}}
\end{figure}

\emph{2. Stability analysis under unbalanced gain and loss}

 Now we turn to the unbalanced gain-loss situation, where the loss (gain) coefficients of two wells do not take the same values, namely, $\beta_r\neq \beta_l$. In such situation,  we will perform  stability analysis based on \emph{Case B} with $\Omega/\omega$ being even and odd respectively. To guarantee the system's stability, we find that the spin-boson loss must be stronger than the gain, namely, $\beta_r > \beta_l$\cite{lunt96}.

(1) even $\Omega/\omega$

When $\Omega/\omega$ is even, the quasienergies become $E_1=E_4=\frac{1}{2} i (\beta_l-\beta_r+\rho')$ and $E_2=E_3=\frac{1}{2} i (\beta_l-\beta_r-\rho')$ with $\rho'=\sqrt{(\beta_l+\beta_r)^2-4 J_0^2-4 J_{\frac{\Omega}{\omega}}^2}$.
By adjusting the system parameters to satisfy the balance between the effective coupling strengths and the gain-loss coefficients, $\beta_r \beta_l= J_0^2+ J_{\frac{\Omega}{\omega}}^2$, we can simplify the quasienergies as:  $E_1=E_4=0$ and $E_2=E_3=i (\beta_l-\beta_r)$. According to \emph{Case B} of stability analysis, for the gain-loss coefficients obeying  $\beta_r > \beta_l$, we find that the considered system is stable under the balance (equilibrium) condition of $\beta_r \beta_l= J_0^2+ J_{\frac{\Omega}{\omega}}^2$. To verify the  analytical results, by fixing the parameters $\Omega=100$, $\omega=50$, $\nu=1$, $\lambda=1/3$, $\varepsilon=75$, and adjusting the dissipation coefficients $\beta_r, \beta_l$ to meet the above balance condition, from the accurate model (3), we
plot the time-evolution curves of all probabilities $P_k=|a_k(t)|^2$ and $P=\sum_{k=1}^{4} P_k$, as shown in Fig.~7. It is clearly seen that all the probabilities tend to stable values after a period of time and the system is stable\cite{xiao85, zhou384}.  In Figs.~7 (a)-(b), we adopt the same parameters and yet different initial states.  As shown in Fig.~7 (a), for the particle initialized in the right (loss) well, the total probability of the particle decays quickly in the initial time interval, then increases to approach a constant
value of 0.4. If we instead start the particle in the left (gain) well, the total probability monotonically increases and finally reaches a higher stable value of 1.9.
In Fig.~7 (c), when the special ratio $\beta_r/\beta_l=3$ is set, we can see
that the total probability of the particle initially prepared in the right well monotonically
tends to the steady value of 1 after a temporary decay, and the system is then stabilized.
In the limit of $t\rightarrow \infty$, the special ratio between the two dissipation
parameters, $\beta_r/\beta_l=3$, can make the total probability evolve in time toward unity, namely, $P(\infty)=1$, a result which is in analogy to the non-Hermitian cold atomic system without SO coupling\cite{xiao85}. This nontrivial result can be demonstrated analytically from the general non-Floquet solution (7) with a given initial condition (not shown here).

\begin{figure*}[htp]\center
\includegraphics[height=1.3in,width=2.2in]{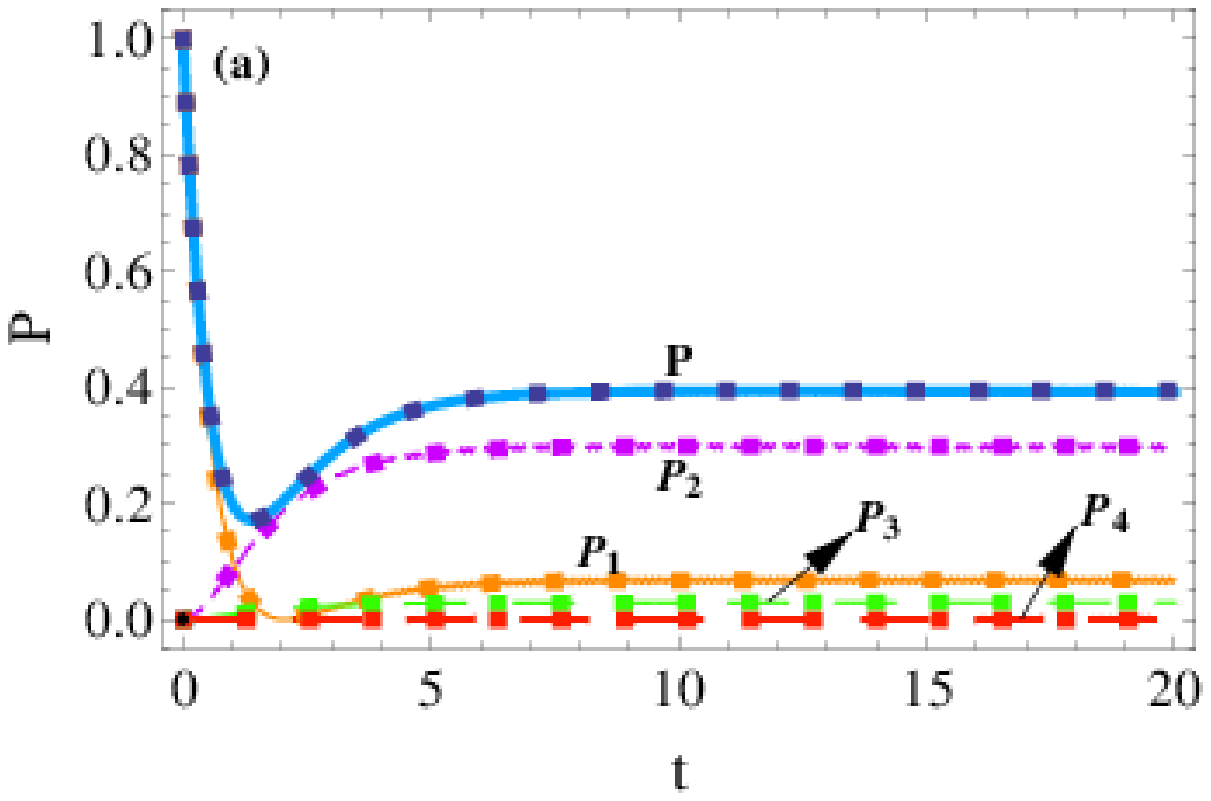}
\includegraphics[height=1.3in,width=2.2in]{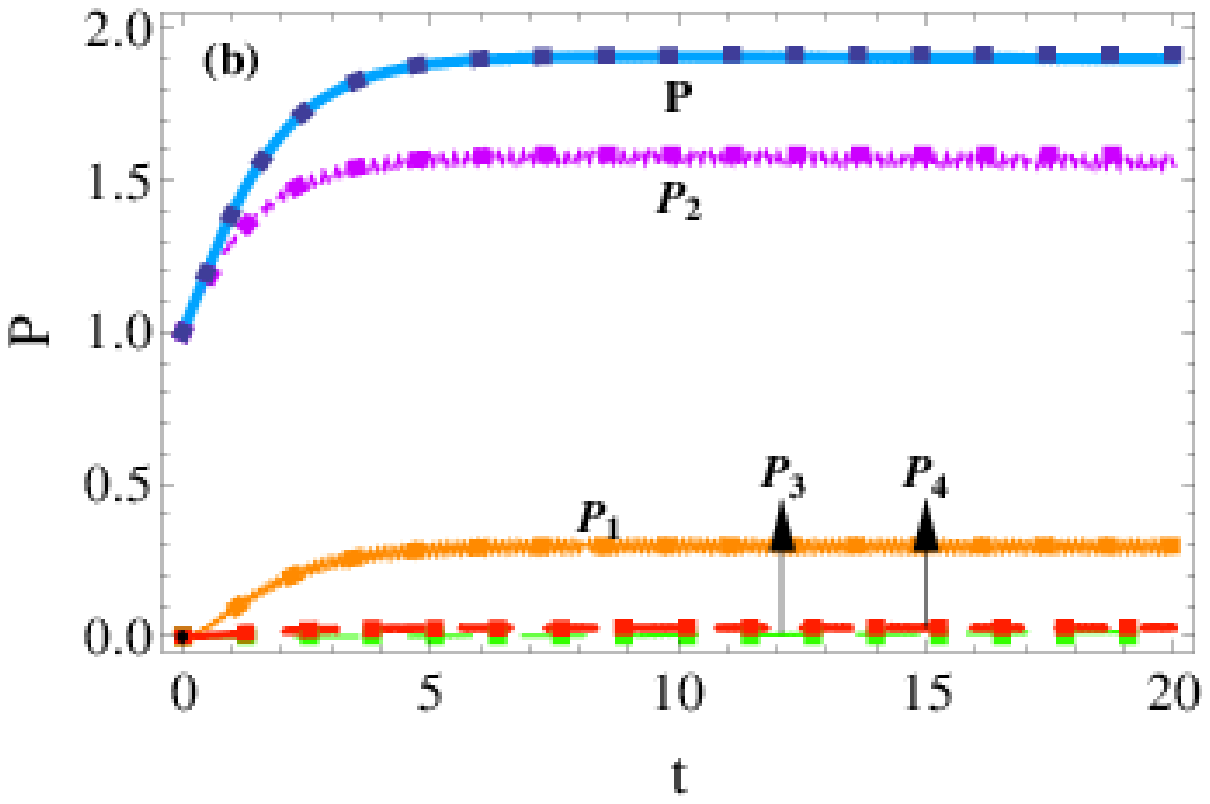}
\includegraphics[height=1.3in,width=2.2in]{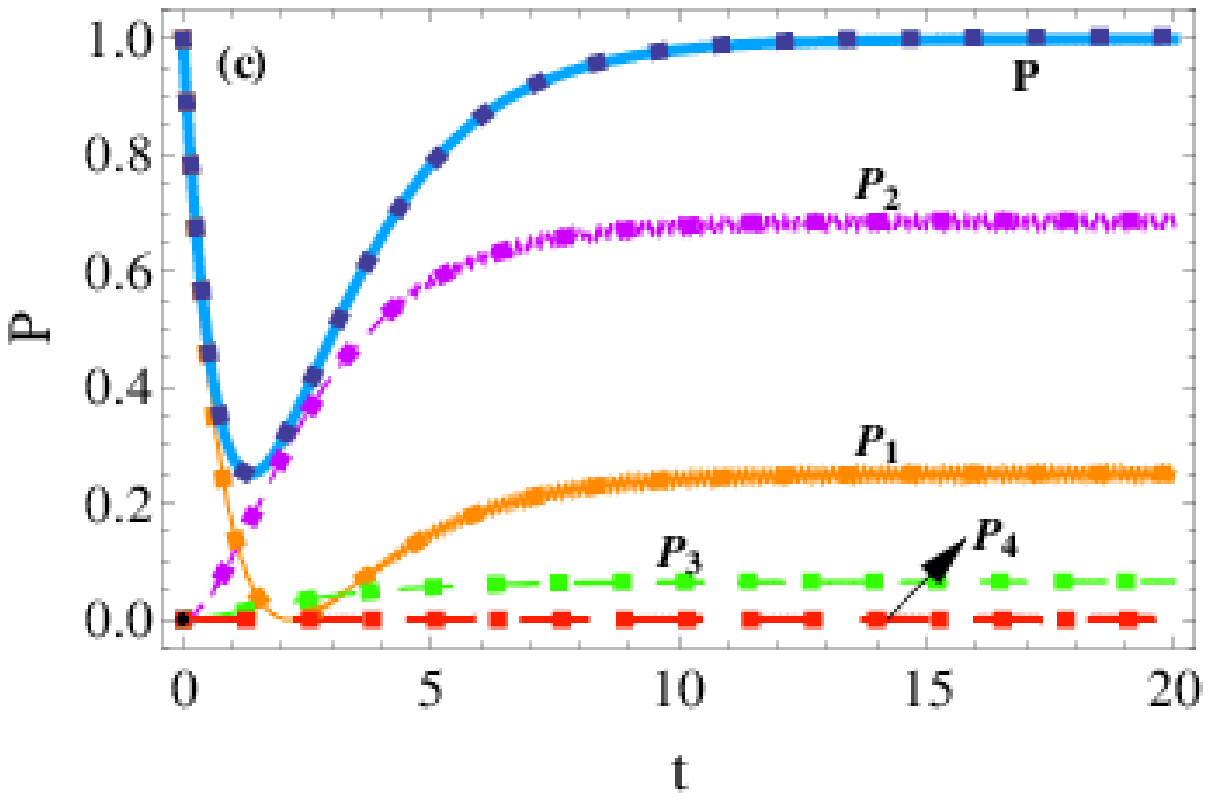}
\caption{\scriptsize{(Color online) Time evolution  of the probabilities $P_k (t)$ and total probability $P$ with different initial conditions and different dissipation coefficients. (a) $a_1 (0)=1, a_k(0)=0 (k\neq 1)$, $\beta_r=0.9706$ and $\beta_l=0.2$; (b) $a_2 (0)=1, a_k(0)=0 (k\neq 2)$, $\beta_r=0.9706$ and $\beta_l=0.2$; (c) $a_1 (0)=1, a_k(0)=0 (k\neq 1)$, $\beta_r=3* 0.254427$ and $\beta_l=0.254427$. The other parameters are chosen as $\nu=1$, $\Omega=100$, $\omega=50$, $\lambda=1/3$, and $\varepsilon=75$. Hereafter, the blue solid lines  label the numerical results of the total population $P=\sum_{k=1}^{4} P_k$ obtained from the original model (3) and the black squares denote the analytical correspondences given by the effective model (4).}}
\end{figure*}

(2) odd $\Omega/\omega$

When $\Omega/\omega$ is odd, the quasienergies become $E_{1,2}=\frac{1}{2} i (\beta_l-\beta_r \pm \rho'_{-})$ and
$E_{3,4}=\frac{1}{2} i (\beta_l-\beta_r \mp \rho'_{+})$ with $\rho'_{\pm}=\sqrt{(\beta_l+\beta_r)^2-4 (|J_0| \pm |J_{\frac{\Omega}{\omega}}|)^2}$. According to \emph{Case B} of the stability analysis, we obtain some stable conditions which are distinguished as the following two categories.

\emph{Category 1. two of the quasienergies are 0, and the imaginary parts of the other two are less than 0.}

(i) $J_{\frac{\Omega}{\omega}}=0$ and $\beta_r \beta_l=J_0^2$

\begin{figure}[htp]\center
\includegraphics[height=1.3in,width=1.6in]{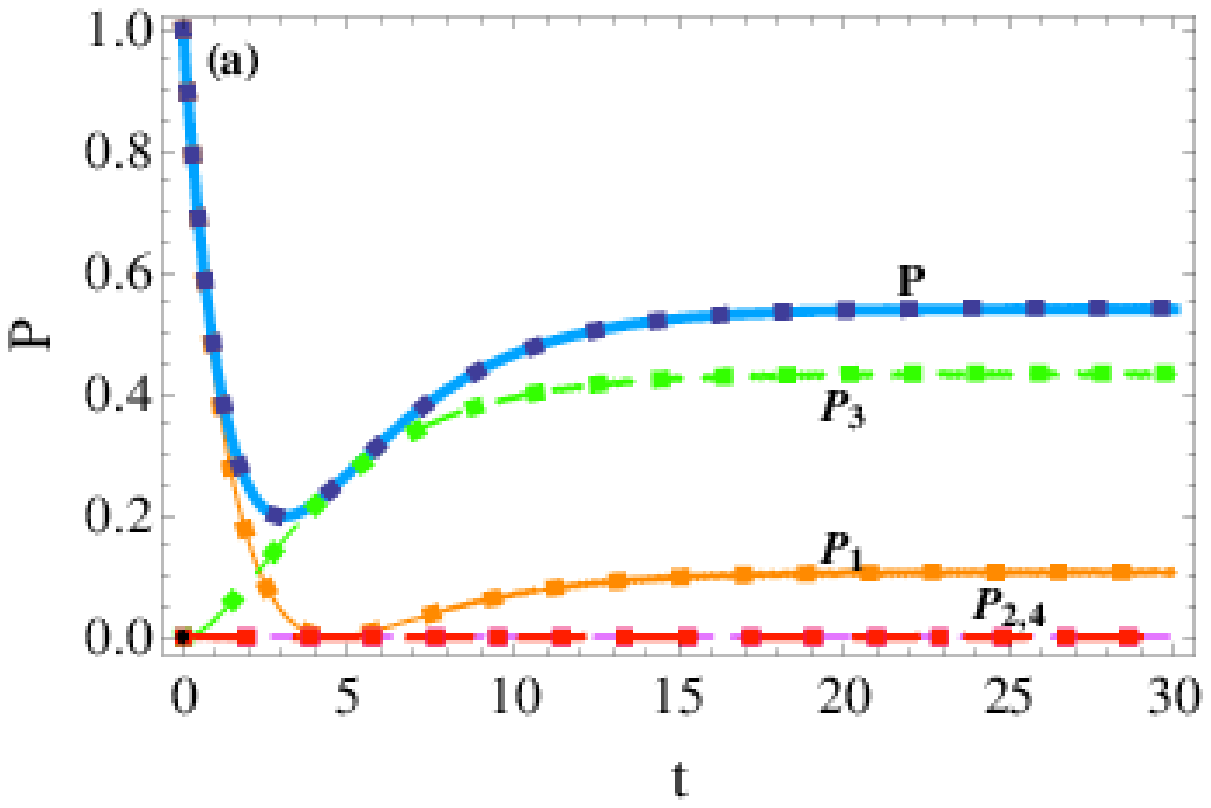}
\includegraphics[height=1.3in,width=1.6in]{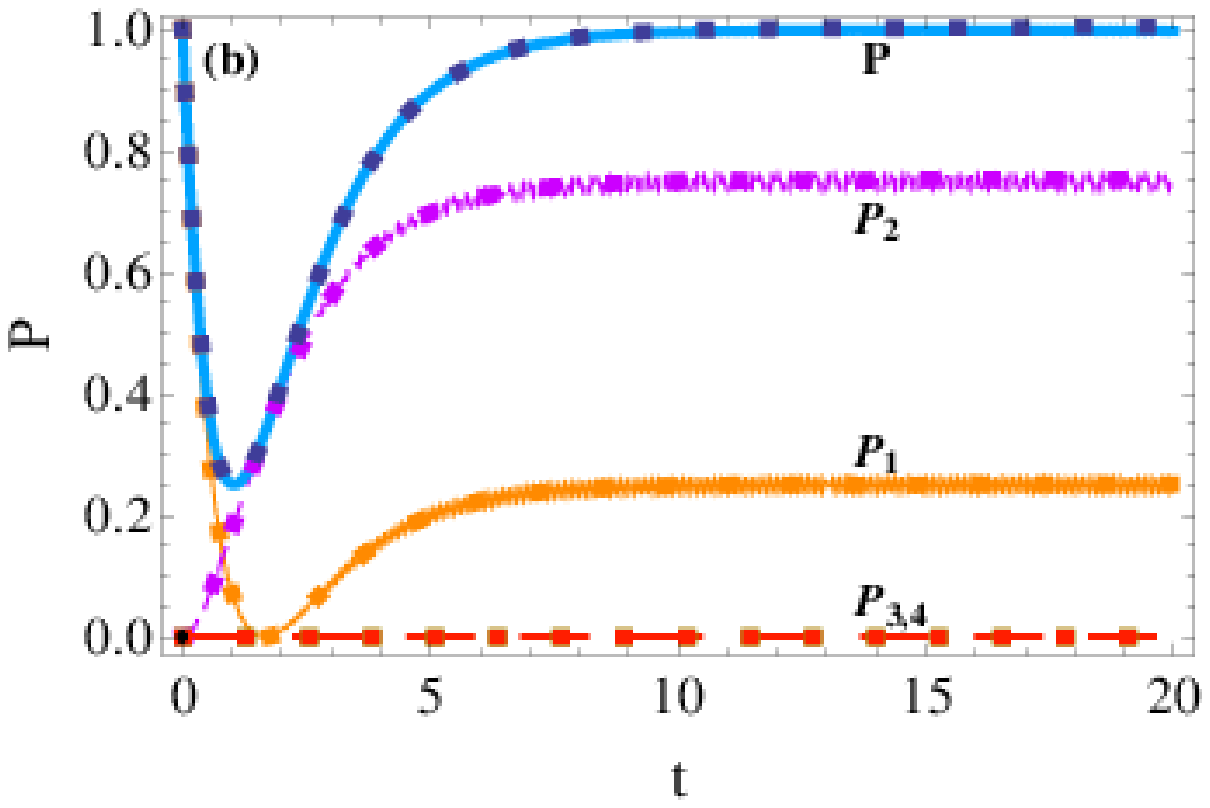}
\includegraphics[height=1.3in,width=1.6in]{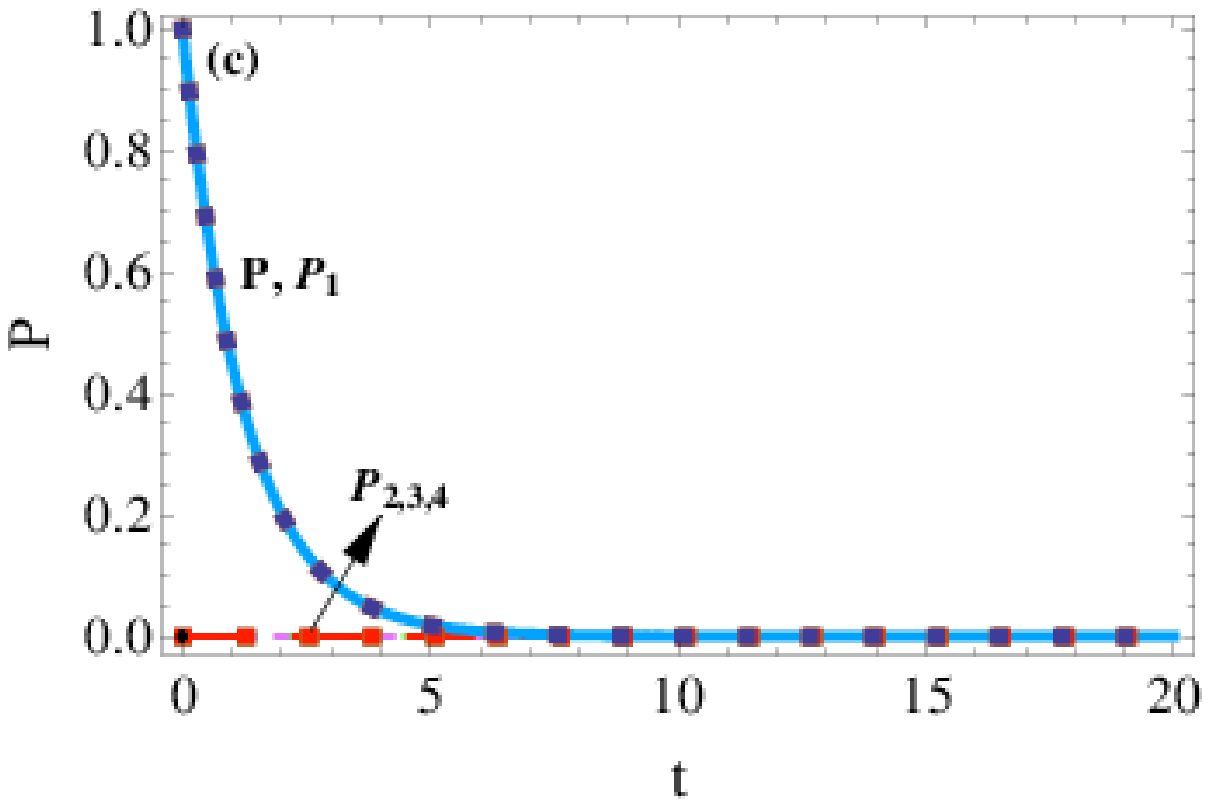}
\includegraphics[height=1.3in,width=1.6in]{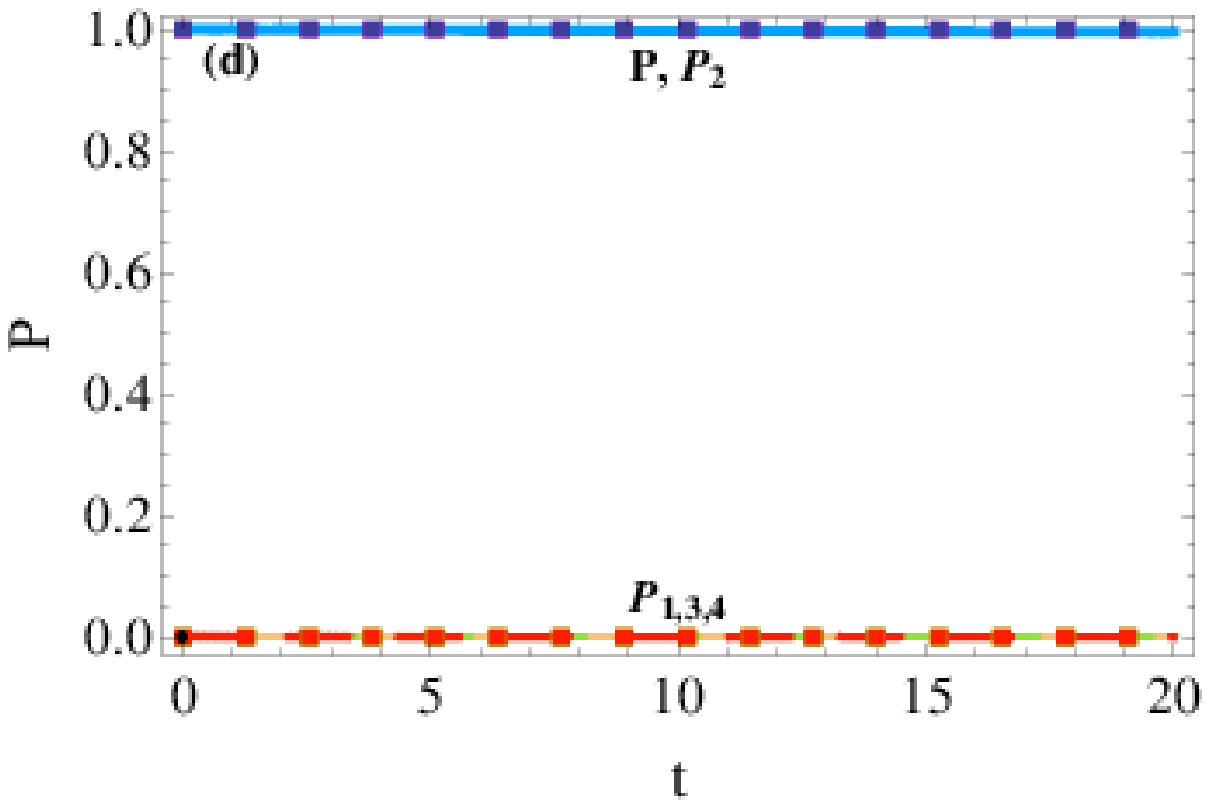}
\caption{\scriptsize{(Color online) Time evolution of the probabilities $P_k (t)$ and total probability $P$, with (a) $\lambda=1/3$, $\beta_r=0.405538$, $\beta_l=0.1$, and $\varepsilon=3.8317*50/2$; (b) $\lambda=1/2$, $\beta_r=3*0.332971$, $\beta_l=0.332971$, and $\varepsilon=2.001*50/2$; (c)  $\lambda=1$, $\beta_r=0.4$, $\beta_l=0$, and $\varepsilon=2.4048*50/2$; (d) $\lambda=1$, $\beta_r=0.4$, $\beta_l=0$, and $\varepsilon=2.4048*50/2$. In (a)-(c), we assume that the system is initialized with a spin-up particle in the right well; In (d), we start a spin-down particle in the left well without gain ($\beta_l=0$). The other parameters are chosen as $\nu=1$ and $\Omega=\omega=50$.}}
\end{figure}
When the stable condition $J_{\frac{\Omega}{\omega}}=0$ and $\beta_r \beta_l=J_0^2$ is fulfilled, the quasienergies can be simplified as $E_1=E_4=0$ and $E_2=E_3=i (\beta_l-\beta_r)$, which thus produces stable non-spin-flipping tunneling  for $\beta_r > \beta_l$. To verify this analytical result, we set the parameters $\Omega=\omega=50$ ($\Omega/\omega=1$), $\nu=1$, $\lambda=1/3$, $\varepsilon=3.8317*50/2$, $\beta_r=0.405538$ and $\beta_l=0.1$ to satisfy the above stable condition (i), and from equation (3) we plot the time-evolution curves of probabilities for the initial spin-up atom occupying the right well, as shown in Fig.~8 (a). It can be seen that the total probability of the final-state will approach 0.55 after a sufficiently large time. Here, we observe that the stable non-spin-flipping  tunneling occurs because the driving parameters $2\varepsilon/\omega=3.8317$ satisfy $\mathcal{J}_1(2\varepsilon/\omega)=0$.

(ii) $J_{0}=0$ and $\beta_r \beta_l=J_{\frac{\Omega}{\omega}}^2$

When the stable condition $J_{0}=0$ and $\beta_r \beta_l=J_{\frac{\Omega}{\omega}}^2$ is fulfilled, the quasienergies can also be written as $E_1=E_4=0$ and $E_2=E_3=i (\beta_l-\beta_r)$, thereby producing stable spin-flipping tunneling  for $\beta_r > \beta_l$. Taking the parameters to satisfy the above stable condition (ii) as: $\Omega=\omega=50$, $\nu=1$, $\lambda=1/2$, $\varepsilon=2.001*50/2$, $\beta_r=3*0.332971$ and $\beta_l=0.332971$, from equation (3) we have plotted the time-evolution curves of probabilities with the spin-up atom initially localized in the right well, as shown in Fig.~8 (b). Obviously, as time increases to infinity, namely, $t\rightarrow \infty$, the final total probability tends to 1, which also seems  reminiscent of the previous result  with $\beta_r/\beta_l=3$ studied in the non-Hermitian bosonic junction without considering SO coupling\cite{xiao85}. Here, we observe that the stable spin-flipping  tunneling occurs because the SO coupling strength $\lambda=1/2$ is associated with $J_{0}=0$.

(iii) $J_{0}=J_{\frac{\Omega}{\omega}}=0$ and $\beta_l=0$

When $J_{0}=J_{\frac{\Omega}{\omega}}=0$ and $\beta_l=0$ are set, we have the quasienergies: $E_1=E_4=0$ and $E_2=E_3=-i \beta_r$. In such a case, the particle will be remain frozen in the initially occupied well. Thus, under the stable condition (iii) $J_{0}=J_{\frac{\Omega}{\omega}}=0$ and $\beta_l=0$, we find that the system's stability depends on its initial condition: when the initial state is state $|0, \uparrow\rangle$ or $|0, \downarrow\rangle$ or a superposition state of $|0, \uparrow\rangle$ and $|0, \downarrow\rangle$, that is, the system is initialized with the particle in the right (loss) well, the total probability of system will exponentially decay to 0; when the initial state is state $|\downarrow, 0\rangle$ or $|\uparrow, 0\rangle$ or a superposition state of $|\downarrow, 0\rangle$ and $|\uparrow, 0\rangle$, that is, the system is initialized with the particle in the left (zero gain) well, conventional CDT will occur. For example, we take the parameters to satisfy the above stable condition (iii) as: $\Omega=\omega=50$, $\nu=\lambda=1$, $\varepsilon=2.4048*50/2$, $\beta_r=0.4$, and $\beta_l=0$, and from equation (3) we have plotted the time-evolution curves of probabilities with the particle initially localized in the right ($P_1 (0)=1$) and left well ($P_2 (0)=1$), as shown in
 Fig.~8 (c) and in Fig.~8 (d) respectively.
It is clearly seen that for the former initial condition ($P_1 (0)=1$), the total probability of system (here $P=P_1$) exponentially decays to 0, for the latter ($P_2 (0)=1$), however, the initial state is kept and conventional CDT occurs.

\emph{Category 2. one of the quasienergies is 0, and the imaginary parts of the others are less than 0.}

(i) $(|J_{0}|-|J_{\frac{\Omega}{\omega}}|)^2=\beta_r \beta_l$ and $(\beta_l+\beta_r)^2<4 (|J_0|+ |J_{\frac{\Omega}{\omega}}|)^2$

\begin{figure}[htp]\center
\includegraphics[height=1.3in,width=1.6in]{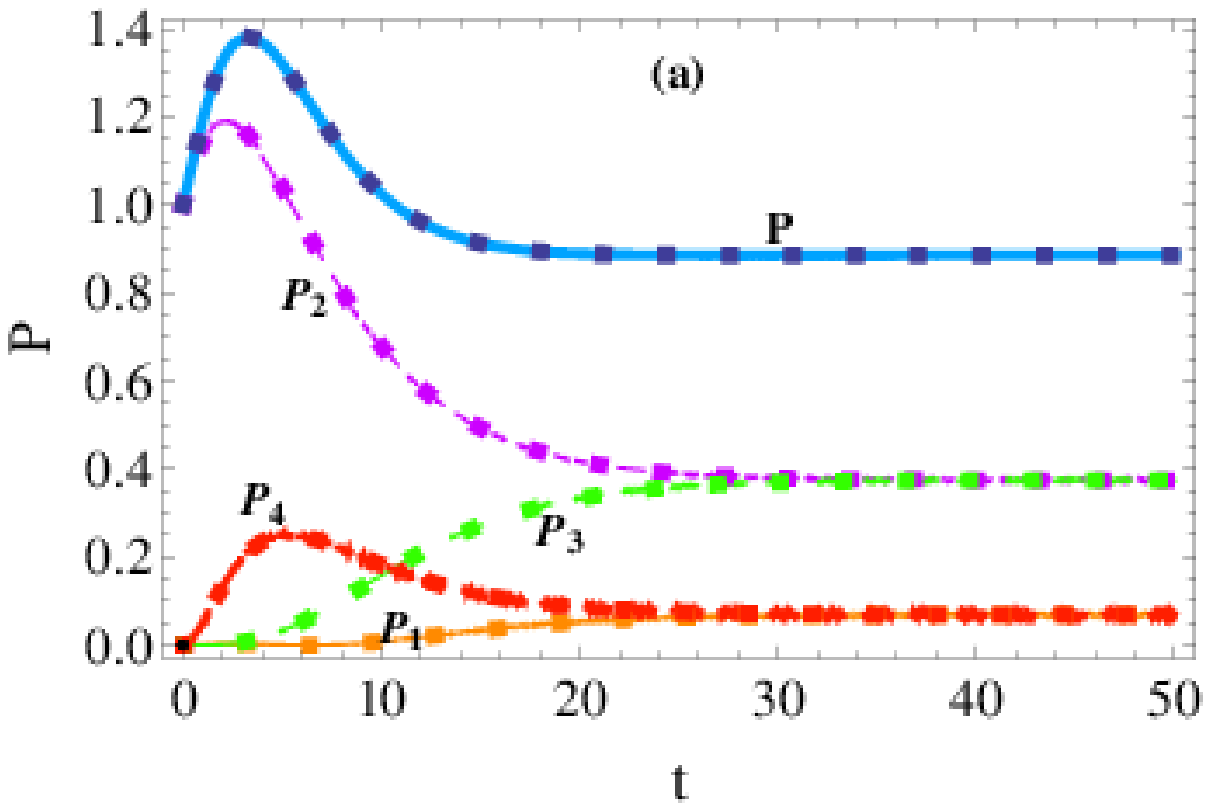}
\includegraphics[height=1.3in,width=1.6in]{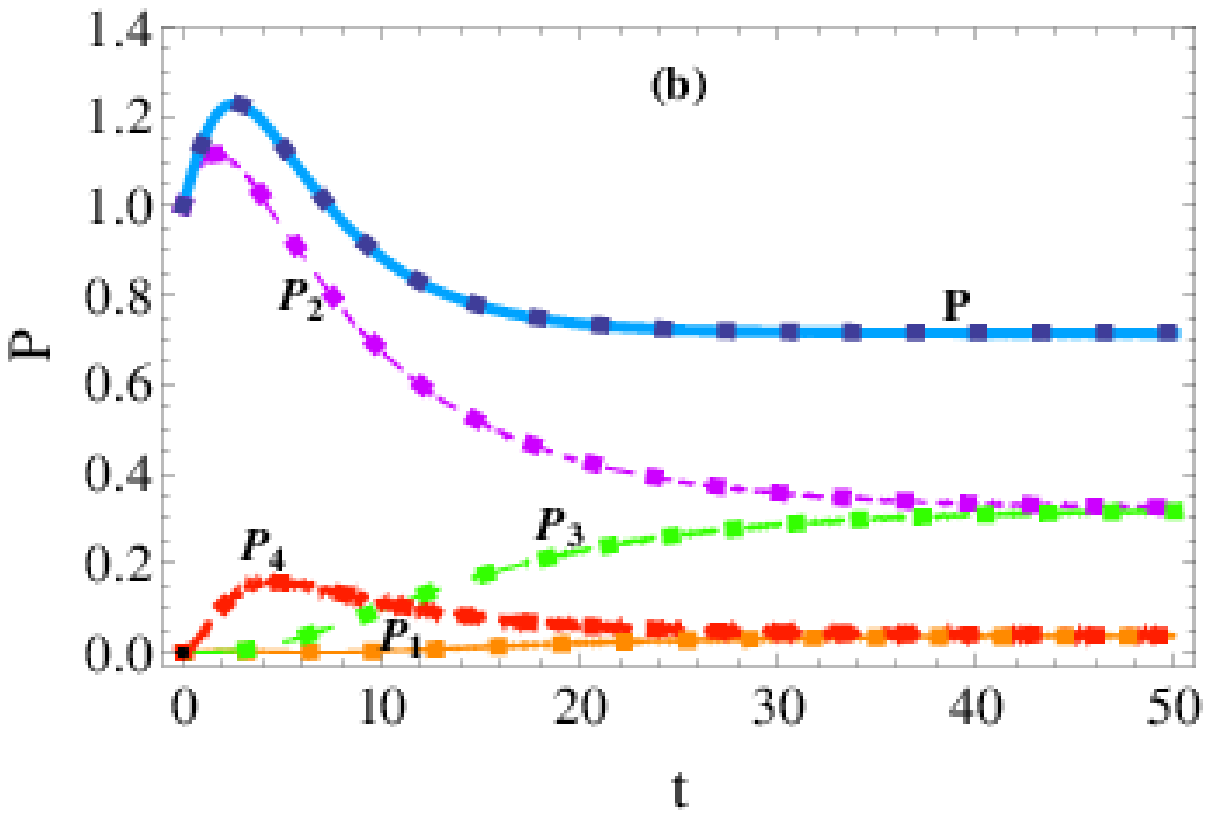}
\caption{\scriptsize{(Color online) Time-evolution curves of the probabilities $P_k (t)$ and total probability $P$ for $\lambda=1/4$, $\varepsilon=100$, and (a) $\beta_r=0.54816$, $\beta_l=0.1$; (b) $\beta_r=0.685209$, $\beta_l=0.08$, starting the system with a spin-down particle in the left well. The other parameters are chosen as $\nu=1$ and $\Omega=\omega=50$.}}
\end{figure}

When the system parameters satisfy the stable conditions $(|J_{0}|-|J_{\frac{\Omega}{\omega}}|)^2=\beta_r \beta_l$ and $(\beta_l+\beta_r)^2<4 (|J_0|+ |J_{\frac{\Omega}{\omega}}|)^2$, the quasienergies becomes $E_1=0$, $E_2=i (\beta_l-\beta_r)$, $E_3=\frac{1}{2} i (\beta_l-\beta_r-\rho'_{+})$, and $E_4=\frac{1}{2} i (\beta_l-\beta_r+\rho'_{+})$ with $\rho'_{+}$ being a purely imaginary number, and thus the system is stable for $\beta_r > \beta_l$. In Fig.~9 (a), we take the parameters to match the stable conditions  $(|J_{0}|-|J_{\frac{\Omega}{\omega}}|)^2=\beta_r \beta_l$ and $(\beta_l+\beta_r)^2<4 (|J_0|+ |J_{\frac{\Omega}{\omega}}|)^2$ as follows: $\Omega=\omega=50$, $\nu=1$, $\lambda=1/4$, $\varepsilon=100$, $\beta_r=0.54816$, and $\beta_l=0.1$. Adopting such parameters, we illustrate the time evolution
of all the probabilities with a spin-down particle initially localized in the left well, as illustrated in Fig.~9 (a).
Obviously, after a sufficiently long time, the final total probability tends to a steady value of 0.88.

(ii) $(|J_{0}|-|J_{\frac{\Omega}{\omega}}|)^2=\beta_r \beta_l$ and $0 \leq \rho'_{+} < \beta_r-\beta_l$

When the system parameters satisfy the stable conditions $(|J_{0}|-|J_{\frac{\Omega}{\omega}}|)^2=\beta_r \beta_l$ and $0 \leq \rho'_{+} < \beta_r-\beta_l$, the quasienergies becomes $E_1=0$, $E_2=i (\beta_l-\beta_r)$, $E_3=\frac{1}{2} i (\beta_l-\beta_r-\rho'_{+})$, and $E_4=\frac{1}{2} i (\beta_l-\beta_r+\rho'_{+})$, and thus stable spin-dependent tunneling can be achieved  for $\beta_r > \beta_l$. In Fig.~9 (b), we  take the parameters to match the stable conditions of  $(|J_{0}|-|J_{\frac{\Omega}{\omega}}|)^2=\beta_r \beta_l$ and $0 \leq \rho'_{+} < \beta_r-\beta_l$ as follows: $\Omega=\omega=50$, $\nu=1$, $\lambda=1/4$, $\varepsilon=100$, $\beta_r=0.685209$, and $\beta_l=0.08$. With such parameters, we have shown in Fig.~9 (b) the time evolution
of all the probabilities with a spin-down particle initially localized in the left well. It can be seen that the final total probability approaches a steady value of 0.72 after a sufficiently large time.

\section{conclusion and outlook}

In summary, we have studied the stabilization of spin tunneling of a single SO-coupled atom placed in a periodically driven non-Hermitian double-well potential. By use of the high-frequency approximation, we derive the analytical expressions of quasienergies and Floquet states of the non-Hermitian system, which can give us the direct information about the system's stability. The main results are summarized as follows, which are numerically confirmed by monitoring the time-evolution of the accurate model. When the loss (gain) coefficients of two wells take the same values, $\beta_r=\beta_l=\beta$, we find that as the gain-loss strength is increased, the stable spin-flipping tunneling is preferentially suppressed and the stable parameter regions will become narrower. In such balanced gain-loss situation, when $\Omega/\omega$ is even, \emph{continuous} stable parameter regions can emerge;  when $\Omega/\omega$ is odd, nevertheless, only \emph{discrete} stable parameter regions are found. On the other hand, when the loss (gain) coefficients of two wells does not take the same values, the parametric equilibrium conditions for realizing  stable spin tunneling are presented whether $\Omega/\omega$ is even or odd. These results may be relevant to quantum control of the spin-dependent tunneling dynamics
in the realistic dissipative systems and promise some potential applications in the design of novel spintronics device.

Finally, we shall briefly discuss some possible experimental schemes for observing our theoretical predictions. A promising candidate for an experimental realization of our model
is a SO-coupled Bose-Einstein condensate in an optical double-well
potential where particles
are injected into one well and removed from the other. The localized loss and gain have already been
realized experimentally: localized loss can be created by a focused
electron beam\cite{barontini110, gericke4, wurtz103, labouvie116}  and gain by pumping atoms into the trapped
condensate from a physically separate cloud\cite{robins4, doring79}. However, in real cold-atomic experiment the perfect control of localized loss and gain is a very demanding task, especially because of the challenging realization
of localized gain. Recently, other  schemes have been proposed for  realization of balanced gain and loss in cold-atomic system. In these schemes, injecting particles (gain) can be realized by embedding
a double-well system in a multi-well
potential or an optical lattice loaded with ultra-cold atoms, which severs as particle reservoirs and permits
a steady current of particles into the embedded double-well subsystem\cite{kreibich87, kreibich90, kreibich93, kogel99}. In this way, the subsystem
can behave exactly as a non-Hermitian double-well system with balanced gain and loss contributions.
It has been demonstrated that balanced gain and loss in a cold-atomic system can be realized even for the non-$\mathcal{PT}$-symmetric potentials\cite{lunt96, Altinisik100}, which relaxes the required symmetries of the complex potential and thus is beneficial in real experimental implementations.
In order to test our theoretical results, one may first prepare the SO-coupled bosonic gases by the Raman-laser coupling schemes\cite{lin471}, and
then load them into the experimental setups as proposed in Refs.~\cite{kreibich87, kreibich90, kreibich93, kogel99}, where the  embedded double-well system should be periodically driven with the well-established shaking techniques.

In our theoretical model, we only consider the simple case where the contact interaction between the atoms is neglected. Experimentally,
the interatomic interaction in ultracold atomic gases can be tuned using Feshbach resonances\cite{chin82, cazalilla83} over a wide parameter range, which  of course allows for zero interaction to be achieved. When the impact of weak interatomic interaction is considered, the system will be more complex and the corresponding physical effect will be much richer, which is worth being studied in the future.

\section*{ACKNOWLEDGMENTS}

This work was supported by the Scientific Research Fund of Hunan Provincial Education Department under Grant No. 18C0027, the National Natural Science Foundation of China under Grants 11975110, 11747034, and the Hunan Provincial Natural Science Foundation of China under Grant Nos. 2017JJ3208, 2019JJ40060. Xiaobing Luo was also supported by the Scientific and Technological Research Fund of Jiangxi Provincial Education Department (numbers
GJJ180559), and Open Research Fund Program of the State Key Laboratory of Low-Dimensional
Quantum Physics (KF201903).

\end{document}